\documentclass[aps,twocolumn,showpacs,pra,groupedaddress,superscriptaddress,nofootinbib]{revtex4-1}
\usepackage{graphicx,amsmath,amssymb,txfonts,float}
\usepackage{subfigure,hyperref,bbm,times}
\usepackage[T1]{fontenc}
\usepackage{braket}
\usepackage{epsfig}
\usepackage{color}
\usepackage{graphicx}
\usepackage{dcolumn}
\usepackage{bm}

\newcommand{\scal}[2]{\langle#1|#2\rangle}
\providecommand{\openone}{\leavevmode\hbox{\small1\kern-3.8pt\normalsize1}}

\usepackage{soul}

\hypersetup{
   colorlinks=true,
   linkcolor=blue,
}

\graphicspath{{figure/}}

\begin{document}

\title{Dynamics of spatially indistinguishable particles and entanglement protection}

\author{Farzam Nosrati}
\email{farzam.nosrati@unipa.it}
\affiliation{Dipartimento di Ingegneria, Universit\`{a} di Palermo, Viale delle Scienze, Edificio 9, 90128 Palermo, Italy}
\affiliation{INRS-EMT, 1650 Boulevard Lionel-Boulet, Varennes, Qu\'{e}bec J3X 1S2, Canada}

\author{Alessia Castellini}
\affiliation{Dipartimento di Fisica e Chimica - Emilio Segr\`e, Universit\`a di Palermo, via Archirafi 36, 90123 Palermo, Italy}

\author{Giuseppe Compagno}
\affiliation{Dipartimento di Fisica e Chimica - Emilio Segr\`e, Universit\`a di Palermo, via Archirafi 36, 90123 Palermo, Italy}

\author{Rosario Lo Franco}
\email{rosario.lofranco@unipa.it}
\affiliation{Dipartimento di Fisica e Chimica - Emilio Segr\`e, Universit\`a di Palermo, via Archirafi 36, 90123 Palermo, Italy}
\affiliation{Dipartimento di Ingegneria, Universit\`{a} di Palermo, Viale delle Scienze, Edificio 6, 90128 Palermo, Italy}

\begin{abstract}
We provide a general framework which allows one to obtain the dynamics of $N$ noninteracting spatially indistinguishable particles locally coupled to separated environments. The approach is universal, being valid for both bosons and fermions and for any type of system-environment interaction. It is then applied to study the dynamics of two identical qubits under paradigmatic Markovian noises, such as phase damping, depolarizing and amplitude damping. We find that spatial indistinguishability of identical qubits is a controllable intrinsic property of the system which protects quantum entanglement against detrimental noise.
\end{abstract}

\maketitle

\section{Introduction}

Dealing with identical particles, as building blocks of quantum networks, has been subject of debate because of the difficulty in characterizing their physical quantum correlations \cite{vogel2015PRA,Shi2003PRA,Li2001PRA,Paskauskas2001PRA,cirac2001PRA,zanardiPRA,tichyFort,
tichy2011essential,bose2013,sasaki2011PRA,ghirardi2002,balachandranPRL,plenio2014PRL,benatti2014review,
benattiOSID2017,franco2016quantum,compagno2018dealing,duzzioniPRA,bordone2011,morris2019,benatti2020entanglement}.
In quantum mechanics, particle identity refers to particles of the same species having the same intrinsic properties (such as mass, charge, spin) \cite{cohen2006quantum}, while indistinguishability can be given a continuous degree related to a given set of quantum measurements \cite{nosrati2019control}. In contrast, nonidentical (distinguishable) particles are individually addressable bodies of a composite system, where individual operations on a given particle can be performed by means of local operations and classical communication (LOCC) \cite{horodecki2009quantum, chitambar2014everything}. As a direct consequence of the addressability notion (particle-locality), the dynamical treatment of a composite quantum system of distinguishable particles is well understood \cite{nielsen2010quantum,breuer2002theory,aolitareview}. This is not the case when the system is made of indistinguishable particles, for which the concept of addressability is lost. 

A particle-based no-label approach has been introduced to characterize states composed of identical particles, where the many-body system state is considered as a whole state vector of single-particle states \cite{franco2016quantum,compagno2018dealing}. Employing this approach together with the operational framework of spatially localized operations and classical communication (sLOCC) \cite{franco2018indistinguishability}, one can directly access exploitable entanglement between internal degrees of freedom of indistinguishable particles having arbitrary spatial configurations of wave functions. Experimental evidence of spatial indistinguishability as a direct resource for remote entanglement has been recently reported \cite{sunetalexp,barros2019entangling}. The no-label approach has been compared with other methods in both first and second quantization \cite{chin2019entanglement, lourencco2019entanglement, chin2019reduced} and has been applied to analyze the Hanbury Brown-Twiss effect with wave packets \cite{Quanta66} and the quantum entanglement in one-dimensional systems of anyons \cite{mani2020quantum}. Thanks to the sLOCC framework within the no-label approach, spatial indistinguishability has been proven to be quantifiable by an entropic-informational measure \cite{nosrati2019control} which plays a direct role in activating quantum-enhanced information processing \cite{sunetalexp,PhysRevA.65.062305, PhysRevA.68.052309, castellini2019activating,benatti2014dissipative, PhysRevA.100.012308}. A remaining open question is how indistinguishable particles, under general conditions of spatial overlap, interact with local environments and, as a result, how their system state dynamically evolves. 
In fact, some reported studies about the effect of environmental noise on identical particles have been so far limited to particles occupying specific modes of a system \cite{argentieri2011,Marzolino2013,bordone2016}, due to the lack of a method to treat a continuous degree of spatial indistinguishability. The no-label approach plus sLOCC framework appear to be the ideal tools to fill this gap in the context of open quantum systems. 

Any realistic quantum system inevitably interacts with the surrounding environment, leading to the decoherence process \cite{breuer2002theory,Zurek2003}. Quantum resources, like coherence and entanglement, are more fragile than the classical ones since even a small perturbation would drive the whole system towards decoherence phenomena \cite{peres1984stability,PhysRevLett.86.2490}. Also, the decay rate increases as the number of degrees of freedom of the quantum system increases. As a consequence of local Markovian environments, for example, the entanglement between two separated (distinguishable) qubits completely disappears at a finite time \cite{yu2004finite,Yu598, almeida2007environment, laurat2007heralded}. This is why the preservation of quantum resources is an important challenge in quantum information and computation science. It is known that quantum entanglement can revive as a result of local non-Markovian (quantum or classical) environments \cite{aolitareview,bellomo2007non,bellomo2008entanglement,lofrancoreview,
lofrancoClassical,darrigo2012AOP,adeline2014,lofranco2012PRA,
LoFrancoNatCom,trapaniPRA,devegaRMP,xuPRL}. However, the degree of entanglement decreases and eventually vanishes after a certain critical time. Other possible protection strategies have been purposed, such as quantum Zeno effect \cite{maniscalco2008protecting}, structured environments \cite{mazzola2009sudden, bellomo2008entanglementtr,Man2015a}, distillation protocols \cite{Bennett1996}, decoherence-free subspaces \cite{zanardi1997noiseless, lidar1998decoherence}, dynamical decoupling and control techniques \cite{Viola1998,viola2005random,franco2014preserving,orieux2015experimental, damodarakurup2009experimental,cuevas2017cut,mortezapour2018protecting}, quantum error corrections \cite{preskill1998reliable, knill2005quantum, shor1995scheme} and topological properties \cite{kitaev2003fault,freedman2003topological}. Some of these techniques rely on harnessing many system parameters, whose number increases as the number of particles increases. Also, they require near-perfect suppression of experimental imperfections \cite{suter2016colloquium}. Finding intrinsic properties of the system which can naturally shield entanglement from noise during the dynamics is highly desirable.

In the present study, we accomplish a twofold purpose: quantum dynamics of indistinguishable subsystems (particles) and quantumness protection. In fact, we first provide a dynamical framework for noninteracting identical particles with an arbitrary amount of spatial overlap which are locally coupled to separated environmental noise sources. Our derivation, valid for both bosons and fermions, sheds light on the microscopic processes of $N$ spatially indistinguishable particles that interact with $M$ independent environments. Afterward, we apply the procedure to study the entanglement dynamics of two identical (bosonic and fermionic) qubits under general conditions of spatial overlap in Markovian environments at zero temperature under the effect of the phase damping process, depolarizing process and amplitude damping process. We show that spatial indistinguishability plays an important role in preserving entanglement against detrimental noise.   

The paper is organized as follows. In Sec.~\ref{General-Framework}, we supply the general procedure to deal with the dynamics of noninteracting spatially overlapping identical particles coupled to separated environments. The entanglement dynamics is reported in Sec.~\ref{PhaseD} for the case of phase damping, in Sec.~\ref{depol} for the depolarizing noise, in Sec.~\ref{AM} for the amplitude damping process. Finally, in Sec.~\ref{Conclusion} we summarize our results.

\section{General framework}\label{General-Framework}

In this section, we introduce the general framework to treat the dynamics of identical particles under arbitrary conditions of spatial overlap in the case of localized separated environments. We recall that this is a typical situation in composite physical systems where the noise sources are localized in space. While nonidentical (distinguishable) particles individually interact with their own environment \cite{lofrancoreview,yu2004finite}, when particles are identical and spatially overlapping, they can no longer be individually addressed by the local environment, so a dedicated treatment is needed. 

In the following we provide this treatment by adopting the no-label approach to identical particles \cite{franco2016quantum,compagno2018dealing} together with the sLOCC measurements \cite{franco2018indistinguishability}, which are performed to access the indistinguishability-enabled quantumness contained in the evolved distributed resource state of the system.

\subsection{System-environment Hamiltonian}

A global state composed of $N$ identical particles, taken as the set of one-particle states $\ket{\Psi}:=\ket{\phi_1, \phi_2, \ldots,\phi_N }$ \cite{franco2016quantum}, must be considered as an holistic indivisible entity. In this notation, as usual, each $\phi_i$ contains all the degrees of freedom characterizing the particle. According to the standard definition \cite{cohen2006quantum}, a single-particle operation $\hat{A}$ linearly acts on a $N$-particle state one at a time, namely: $\hat{A}\ket{\Psi}:=\sum_i \ket{\phi_1,\ldots, \hat{A}\phi_i, \ldots ,\phi_N}$. On the other hand, a single-particle quantum operation can be physically restricted to act locally on a given region of space where the identical particles of the composite system can be found with a given probability. Therefore, a suitable definition of a localized single-particle operation acting on a global state of identical particles with arbitrary spatial wave functions is needed to encompass the locality of the quantum operator. 

\textbf{Definition.} \textit{A spatially localized single-particle operator $\hat{S}_X$, performed on a region of space $X$, acts on a $N$-particle state of identical particles $\ket{\Psi}=\ket{\phi_1, \phi_2, \ldots,\phi_N}$ as follows}
\begin{equation} \label{Spatial_Op}
   \hat{S}_X\ket{\Psi}:=\sum_i |\bra{X}\phi_i\rangle|
   \ket{\phi_1, \ldots,\hat{S}\phi_i,\ldots,\phi_N },
\end{equation}
{\it where $|\bra{X}\phi_k\rangle|$ is the absolute value of the probability amplitude of finding a particle in the region $X$, which weights the strength of the operation as physically desired}. 

The operator $\hat{S}$ is any operator on internal degrees of freedom of single particle. Notice that, if the region $X$ represents all the space where the particles of the system can be found, the operator $\hat{S}_X$ reduces to the usual single-particle operator $\hat{A}$ above, as expected. The definition of Eq.~(\ref{Spatial_Op}) shall be a central ingredient to describe the time evolution of spatially overlapping identical particles that interact with localized environments.

\begin{figure}[t] 
\includegraphics[width=0.49\textwidth]{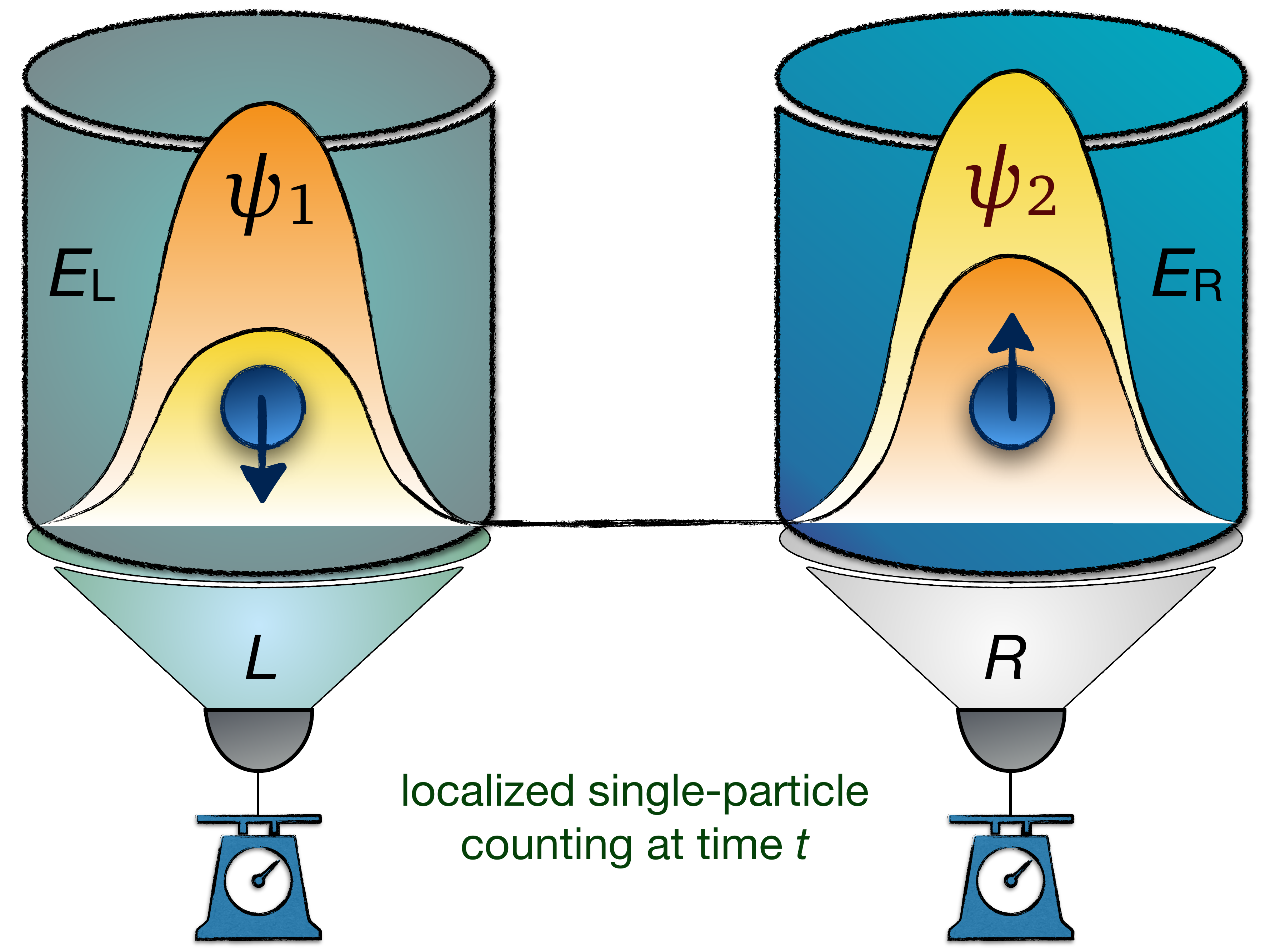}
\caption{\textbf{Sketch of the open quantum system.} Two noninteracting identical spin $\frac{1}{2}$-like subsystems (qubits), with spatial wave functions $\psi_1$ and $\psi_2$, locally interact with separated environments $E_L$ and $E_R$ placed in the spatial regions $L$ and $R$, respectively. At time $t$, single-particle local counting is performed (sLOCC measurement).}
\label{fig1}
\end{figure}

For the sake of simplicity and without loss of generality, we limit our analysis to two identical two-level (spin $\frac{1}{2}$-like) particles (qubits) coupled to two separated bosonic environments $E_L$ and $E_R$, as depicted in Fig.~\ref{fig1}, localized in separated sites $L$ and $R$, respectively. The elementary two-particle system state $\ket{\Psi}_S=\ket{\phi_1,\phi_2}$ is characterized by the set of single-particle states $\{\ket{\phi_i}=\ket{\psi_i s_i},\ i=1,2\}$, where $\psi_i$ is the spatial wave function with associated pseudospin state $s_i$ of basis $\{\uparrow, \downarrow\}$. Since the system-environment interaction occurs locally in two distant space regions, any environmental action on the internal degrees of freedom of the particles has to take into account the locality of the interaction. So, indicating with $g_{kL}$ and $g_{kR}$ the coupling constants between the system and the environmental mode $k$ in regions $L$ and $R$, respectively, the general system-environment interaction Hamiltonian can be written as
\begin{equation} \label{Hamiltonin}
     H_{I}:=\sum_k \left(g_{kL}\hat{E}_{kL}\hat{\sigma}_L+ g_{kR}\hat{E}_{kR}\hat{\sigma}_R\right)+h.c.,
\end{equation}
where $\hat{E}_{kX}$ ($X=\{L,R\}$) is some environmental operator in terms of annihilation or creation operators of local mode $k$, $\hat{\sigma}_{X}$ is a generic localized pseudospin single-particle operator and \textit{h.c.} indicates Hermitian conjugation. The explicit form of $\hat{\sigma}_{X}$ clearly depends on the type of local system-environment interaction and its action on the system state follows the definition of Eq.~(\ref{Spatial_Op}). The explicit types of environmental and pseudospin operators are assigned by the given physical process, as we shall see in the applications of the following sections. To achieve a general treatment of the system dynamics, it is useful to see how the interaction Hamiltonian above acts on a global system-environment elementary state vector $\ket{\Psi}_{SE}=\ket{\phi_1,\phi_2}\otimes\ket{\xi_L,\xi_R}$, where $\ket{\xi_X}$ represents a global pure state of the modes of the local environment $E_X$. By using Eqs.~(\ref{Spatial_Op}) and (\ref{Hamiltonin}), we have 
\begin{equation}
   \begin{split} \label{actionofHamiltonin}
        H_{I}\ket{\Psi}_{SE} &=\sum_k g_{kL}\ket{\hat{E}_{kL}\xi_L,\xi_R}\otimes|\bra{L}\psi_1\rangle|\ket{\psi_1\hat{\sigma}_L s_1,\psi_2s_2} \\
        &+\sum_k g_{kL}\ket{\hat{E}_{kL}\xi_L,\xi_R}\otimes|\bra{L}\psi_2\rangle|\ket{\psi_1s_1,\psi_2\hat{\sigma}_L s_2} \\
        &+\sum_k g_{kR}\ket{\xi_L,\hat{E}_{kR}\xi_R}\otimes|\bra{R}\psi_1\rangle|\ket{\psi_1
        \hat{\sigma}_R s_1,\psi_2s_2}\\
        &+\sum_k g_{kR}\ket{\xi_L,\hat{E}_{kR}\xi_R}\otimes|\bra{R}\psi_2\rangle|\ket{\psi_1s_1,\psi_2\hat{\sigma}_R s_2}\\
        &+ h.c.
   \end{split}
\end{equation}
From this equation, it is immediate to see that the interaction Hamiltonian of Eq.~(\ref{Hamiltonin}) can be recast as a sum of two terms
\begin{equation}\label{H1plusH2}
H_{I} = H_{I}^{(1)} +  H_{I}^{(2)},
\end{equation}
with
\begin{equation} \label{H1&H2}
H_{I}^{(j)} = \sum_k \sum_{X=L,R} g_{kX}^{(j)}\hat{E}_{kX}\hat{\sigma}_X^{(j)}+h.c.,\quad (j=1,2)
\end{equation}
where $\hat{\sigma}^{(j)}$, and thus $H_{I}^{(j)}$, is defined as applied only on the particle pseudospin at the $j$-th slot of the system state vector and $g_{kX}^{(j)}=g_{kX}|\bra{X}\psi_j\rangle|$ ($X=L,R$) is the effective coupling constant between the particle with spatial wave function $\psi_j$ and the mode $k$ of the environment $E_X$. The latter shows that the localized single-particle operator defined in Eq.~(\ref{Spatial_Op}) naturally leads to effective coupling constants $g_{kX}^{(j)}$ directly dependent on the probability amplitude of finding the particle in the regions $L$ and $R$ where the environment (noise source) is placed, as physically expected. Hereafter, the notation $\hat{O}^{(j)}$ shall indicate a single-particle operator which only acts on the particle state at the $j$-th position of the system state vector.

Having the above Hamiltonian model of the system-environment interaction, the global (unitary) time evolution operator is
\begin{equation}\label{timeevo}
    \hat{U}=e^{-\frac{i}{\hbar}\left(H_{I}^{(1)}+H_{I}^{(2)}\right)t}.
\end{equation} 
Looking at Eq.~(\ref{H1plusH2}), one can see an analogy with the usual total interaction Hamiltonian $H_\mathrm{tot}=H_A+H_B$ of two nonidentical (distinguishable) particles $A$ and $B$. However, while $H_A$ and $H_B$ always commute due to the individual addressability of the particles (particle-locality), this is not in general the case for $H_{I}^{(1)}$ and $ H_{I}^{(2)}$. Indeed, two important aspects need to be pointed out when one deals with an open quantum system of identical particles and separated environments:
\begin{itemize}
\item A unitary time evolution cannot generally be factorized, that is $\hat{U}\neq e^{-\frac{i}{\hbar}H_{I}^{(1)}t}e^{-\frac{i}{\hbar}H_{I}^{(2)}t}$ because of the nonzero commutative property of interaction Hamiltonians, $[H_{I}^{(i)},H_{I}^{(j)}]\neq 0$ for $i\neq j$. This property describes the process that particles can interact with the same environment and, as a result, collective effects of the environment show up. Hamiltonians $H_{I}^{(1)}$ and $ H_{I}^{(2)}$ commute only when the spatial wave functions of the identical particles are spatially separated in correspondence of each local environment, e.g. $\ket{\psi_1} = \ket{L}$, $\ket{\psi_2} = \ket{R}$ (or vice versa), as easily seen from Eq.~(\ref{H1&H2}). In this case, in fact, the identical particles are distinguishable by their position and individually interact with their own environment. 

\item On the other hand, one can write the interaction Hamiltonian of Eq.~(\ref{H1plusH2}) as a sum of two terms corresponding to the two interaction spatial regions $H_{I}=H_{IL}+H_{IR}$, where each $H_{IX}=H_{IX}^{(1)}+H_{IX}^{(2)}$ (see Eqs.~(\ref{actionofHamiltonin}) and (\ref{H1plusH2})). Since each environmental operator $\hat{E}_X$ is localized and independent, one has $[H_{IL},H_{IR}]= 0$ and the global time evolution of Eq.~(\ref{timeevo}) can be written as product of two (localized) operations $\hat{U}=e^{-\frac{i}{\hbar}H_{IL}t}e^{-\frac{i}{\hbar}H_{IR}t}=\hat{U}_L\otimes\hat{U}_R$. However, each operator $H_{IX}$ (and thus $\hat{U}_X$) does not address in general an individual particle of the system state vector due to the spatial overlap (indistinguishability) of the identical particles. Again, this happens only if the identical particles are separated and placed each in a given local environment.
\end{itemize}

In the following, on the basis of the above points, we shall give the techniques to determine the dynamics of the reduced density matrix of the two-particle system. These techniques are easily extendable to many-particle system and multiple environments, as we shall discuss later (see subsection \ref{subsec:gen}).

\subsection{Kraus representation for the reduced density matrix}

Assuming that system and environment are initially uncorrelated $\rho(0)=\rho_S(0)\otimes\rho_E(0)$, a typical condition fulfilled within the Born approximation \cite{nielsen2010quantum,breuer2002theory}, the evolved reduced density matrix of the system can be obtained by
\begin{equation}
        \rho_S(t)=\mathrm{Tr}_E\left[\hat{U}_L\otimes\hat{U}_R\ (\rho_S(0)\otimes\rho_E(0))\ 
        \hat{U}_L^\dagger\otimes\hat{U}_R^\dagger\right],
\end{equation}
where the trace is taken, as usual, over the environmental degrees of freedom \cite{breuer2002theory}. From this equation, one formally obtains the following operator-sum representation (or Kraus representation) for the reduced density matrix
\begin{equation}\label{Krausrep}
    \rho_S(t)=\sum_{\alpha,\beta}\mathcal{K}_{\alpha L}\cdot\mathcal{K}_{\beta R}\ \rho_S(0)\ 
    \mathcal{K}_{\alpha L}^{\dagger}\cdot\mathcal{K}_{\beta R}^{\dagger},
\end{equation}
where $\mathcal{K}_{\alpha X}=\bra{e_\alpha}\hat{U}_{X}\ket{e_0}$ ($X=L,R$) are Kraus operators defined by means of a complete set of orthonormal states $\{e_\alpha\}$ of the local environment $E_X$ \cite{kraus1983states} and $\mathcal{K}_{\alpha L}\cdot\mathcal{K}_{\beta R}$ indicates the product of the operators. 

We notice that the assumption of independent (separated) environments implies that the dynamical map of Eq.~(\ref{Krausrep}) is given by means of a factorization of operators related to each local environment. However, as expected in this case due to the general particle indistinguishability, each (local) Kraus operator is a two-particle operator which cannot be written in a tensor product form of single-particle operators, that is $\mathcal{K}_{\alpha X}\neq\mathcal{K}_{\alpha}^{(1)}\otimes\mathcal{K}_{\alpha}^{(2)}$. On the other hand, when the particles are separated, one in $E_L$ and one in $E_R$ and so spatially distinguishable, the local Kraus operators assume the tensor product form $\mathcal{K}_{\alpha L}=\mathcal{K}_{\alpha}^{(1)}\otimes\openone_2^{(2)}$ and $\mathcal{K}_{\beta R}=\openone_2^{(1)}\otimes\mathcal{K}_{\beta}^{(2)}$, where $\openone_2$ is the $2\times 2$ identity matrix: this means that each environmental operator only acts on an individual particle of the state vector. As can be immediately seen, when the Kraus operators take this tensor product form, the dynamical map of Eq.~(\ref{Krausrep}) reduces to the well-known operator-sum representation for nonidentical (distinguishable) particles 
$ \rho_S(t)=\sum_{\alpha,\beta}\mathcal{K}_{\alpha}^{(1)}\otimes
\mathcal{K}_{\beta}^{(2)}\ \rho_S(0)\ 
   \mathcal{K}_{\alpha}^{(1)\dagger}\otimes  \mathcal{K}_{\beta}^{(2)\dagger}$ \cite{breuer2002theory,nielsen2010quantum}. 
   
The Kraus representation of Eq.~(\ref{Krausrep}) for indistinguishable particles remains a formal solution for the system dynamics, which can be of difficult practical utility. Finding explicit expressions for the Kraus operators under general conditions is in fact a challenging task. The next step is therefore to look for a convenient solution by means of a master equation.

\subsection{Master equation for the reduced density matrix}

The most general dynamics of an open quantum system, under Born-Markov approximation, can be described by the Gorini-Kossakowski-Sudarshan-Lindblad equation \cite{lindblad1976generators, gorini1976completely,breuer2002theory}. We are interested in deriving this master equation for our system of two indistinguishable qubits, considering local memoryless bosonic environments at zero temperature. Each local noise channel, at the site $X$, is thus assumed to phenomenologically induce a typical single-particle decay rate $\gamma_{0X}$. Following the standard procedure, i.e. performing the Born-Markov approximation (weak coupling) $\rho_{SE}=\rho_S\otimes\rho_E$ and tracing out the environment degrees of freedom, we obtain the general structure of the master equation for the reduced density matrix of the two-qubit system in the interaction picture as follows
\begin{equation}
\label{master0}
        \Dot{\rho}_S =\sum_{X=L,R}\sum_{i,j=1,2}\gamma_{X}^{(i,j)}\left(\hat{\sigma}_X^{(i)}
        \rho_S\hat{\sigma}_X^{(j)\dagger}-
       \frac{1}{2}\{\hat{\sigma}_X^{(i)\dagger}\hat{\sigma}_X^{(j)},\rho_S\}\right),
\end{equation}
where $\hat{\sigma}^{(i)}$, $\hat{\sigma}^{(j)}$ are pseudospin single-particle jump operators, whose explicit form depends on the system-environment interaction, and  
\begin{equation}\label{EffectiveGamma}
\gamma_{X}^{(i,j)}=\gamma_{0X}|\bra{X}\psi_i\rangle\bra{\psi_j}X\rangle|, \quad (i,j=1,2)
\end{equation}
are effective decay rates associated to the noise action. We remark that these effective decay rates $\gamma_{X}^{(i,j)}$ are independent of particle statistics and crucially depend on particle spatial overlap, so manifesting the effects of spatial indistinguishability (even partial) of the identical qubits on the system evolution. 

When the qubits are spatially separated, Eq.~(\ref{master0}) reduces to the well-known master equation of two distinguishable particles each one embedded in its own local environment, as expected. In fact, in this case the qubits are distinguishable by their location, e.g., $\ket{\psi_1}=\ket{L}$ and $\ket{\psi_2}=\ket{R}$, so that the only nonzero effective decay rates are $\gamma_{L}^{(11)}=\gamma_{0L}$ and $\gamma_{R}^{(22)}=\gamma_{0R}$.

One can also immediately see the effect of maximal spatial indistinguishability for the two qubits. This is the situation when particles are present in the same region $X$ with same probability $P_X$ 
($|\bra{X}\psi_i\rangle| = |\bra{X}\psi_j\rangle|=\sqrt{P_X}$), so they are completely indistinguishable to the eyes of the local environments \cite{sunetalexp,franco2018indistinguishability}. In this case, the effective decay rates of Eq.~(\ref{EffectiveGamma}) are equal for a given location $X$, that is $\gamma_{X}^{(i,j)}=\gamma_{0X}P_X$ and the master equation of Eq.~(\ref{master0}) becomes
\begin{equation}
\label{master}
        \Dot{\rho}_S =\sum_{X=L,R}\gamma_{0X}P_X \left(\hat{\sigma}_X^\prime\rho_S\hat{\sigma}_X^{\prime\dagger}
        -\frac{1}{2}\{\hat{\sigma}_X^{\prime\dagger}\hat{\sigma}^\prime_X,\rho_S\}\right)
\end{equation}
where $\hat{\sigma}'_X=\hat{\sigma}_X^{(1)}+\hat{\sigma}_X^{(2)}$ is a collective single-particle pseudospin operator related to the noise action at location $X$. Notice that the above master equation is similar to the situation of noninteracting nonidentical (distinguishable) particles coupled to a common environment, which is known to give rise to collective effects \cite{gross1982superradiance,PhysRevA.99.052105}. However, we remark that here the collective interaction is only due to the spatial indistinguishability of the identical particles, which continuously rules the interplay between collective and individual effects of the localized environments. 

The master equation of Eq.~(\ref{master0}) represents the main tool to solve the dynamics of two identical qubits, with an arbitrary spatial overlap configuration, coupled to two local environments. The two-particle basis in which the master equation is solved will be conveniently chosen according to the specific system-environment interaction.
To complete the analysis, one just needs to recall the operational framework of sLOCC \cite{franco2018indistinguishability}, which is required to study the dynamics of quantum properties exploitable at separated sites, and a suitable degree of spatial indistinguishability \cite{nosrati2019control}. This shall be done in the following section for convenience.

\subsection{sLOCC operational framework and degree of spatial indistinguishability}

We briefly recall the sLOCC formalism \cite{franco2018indistinguishability} and the spatial indistinguishability measure \cite{nosrati2019control} needed to exploit, at time $t$, a state of two spatially overlapping identical particles distributed in two remote regions $L$ and $R$.

The state $\rho_S(t)$, determined by solving Eq.~(\ref{master0}), is finally projected by the operator $\Pi_{LR}^{(2)}=\sum_{s_1,s_2=\uparrow,\downarrow}\ket{Ls_1,Rs_2}\bra{Ls_1,Rs_2}$ onto the operational subspace, spanned by the basis $\mathcal{B}_\mathrm{LR}=\{\ket{\mathrm{L}\uparrow,\mathrm{R}\uparrow}, \ket{\mathrm{L}\uparrow,\mathrm{R}\downarrow}, \ket{\mathrm{L}\downarrow,\mathrm{R}\uparrow}, \ket{\mathrm{L}\downarrow,\mathrm{R}\downarrow}\}$, to get the distributed state \cite{nosrati2019control}
\begin{equation}\label{rhoLR}
\rho_{LR}(t)=\Pi_{LR}^{(2)}\rho_S(t)\Pi_{LR}^{(2)}/\mathrm{Tr}(\Pi_{LR}^{(2)}\rho_S(t)), 
\end{equation}
with sLOCC probability $P_{LR}(t)=\mathrm{Tr}(\Pi_{LR}^{(2)}\rho_S(t))$ (we provide in Appendix \ref{appe} detailed information about the sLOCC probability for the physical examples we shall treat in the following sections). Now the final state $\rho_{LR}(t)$ is activated to be used for any quantum information tasks. The action of the projection operator $\Pi_{LR}^{(2)}$ on the two-particle state is obtained by using the probability amplitude $
\scal{\phi'_1,\phi'_2}{\phi_1,\phi_2}=\scal{\phi'_1}{\phi_1}\scal{\phi'_2}{\phi_2}+\eta\scal{\phi'_1}{\phi_2}\scal{\phi'_2}{\phi_1}$ \cite{franco2016quantum,compagno2018dealing}, where $\eta = \pm 1$ with the upper or lower sign for bosons or fermions. Particle statistics is thus expected to play a role in the dynamics, especially in the behavior of the sLOCC probability. 
Once the state $\rho_{LR}(t)$ is obtained, with one particle in $L$ and another one $R$, the amount of useful entanglement of $\rho(t)$ within the sLOCC framework can be calculated by the concurrence \cite{nosrati2019control} 
\begin{equation}\label{CLR}
C_{LR}(\rho_S):=C(\rho_{LR})=\max \{0,\sqrt{\lambda_4}-\sqrt{\lambda_3}-\sqrt{\lambda_2}-\sqrt{\lambda_1}\},
\end{equation}
where $\lambda_i$ are the eigenvalues, in decreasing order, of the matrix $R=\rho_{LR}\Tilde{\rho}_{LR}$, being $\Tilde{\rho}_{LR}=\sigma^{L}_{y}\otimes\sigma_{y}^{R} 
\rho_{LR}^{*}\sigma_{y}^{L}\otimes\sigma_{y}^{R}$ with localized Pauli matrices $\sigma_{y}^{L}=\ket{{L}}\bra{{L}}\otimes\sigma_{y}$, $\sigma_{y}^\mathrm{R}=\ket{{R}}\bra{{R}}\otimes\sigma_{y}$. 

The degree of spatial indistinguishability of the two identical qubits, emerging from the outcomes of the joint sLOCC measurement $\Pi_{LR}^{(2)}$, is given by \cite{nosrati2019control}
\begin{align}
\begin{split}
\mathcal{I}=&-\dfrac{P_{L\psi_1}P_{R\psi_2}}{\mathcal{Z}} \log_2 \dfrac{P_{L\psi_1}P_{R\psi_2}}{\mathcal{Z}}\\
&-\dfrac{P_{{L}\psi_2}P_{{R}\psi_1}}{\mathcal{Z}}\log_2 \dfrac{P_{{L}\psi_2}P_{{R}\psi_1}}{\mathcal{Z}},
\end{split}
\label{Indistinguishability}
\end{align}
where $\mathcal{Z}=P_{{L\psi_1}}P_{{R}\psi_2}+P_{{L}\psi_2}P_{{R}\psi_1}$ and $P_{{X\psi_i}}=|\langle {X}|\psi_i\rangle|^2$ (${X}={L},{R}$ and $i=1,2$) is the probability of finding in $\ket{X}$ the particle coming from $\ket{\psi_i}$. One has $\mathcal{I}\in [0,1]$, being equal to 0 for separated (spatially distinguishable) particles (e.g., $P_{{L\psi_1}}=P_{{R\psi_2}}=1$) and equal to 1 for maximally indistinguishable particles ($P_{{X\psi_1}}=P_{{X\psi_2}}$).

Hereafter, we assume that the spatial wave function have the form
\begin{equation}\label{spatialWF}
\ket{\psi_1}=l\ket{\mathrm{L}}+r\ket{\mathrm{R}},\quad
\ket{\psi_2}=l'\ket{\mathrm{L}}+r'e^{i\theta}\ket{\mathrm{R}},
\end{equation} 
where $l$, $r$, $l'$, $r'$ are non-negative real numbers ($l^2+r^2=l'^2+r'^2=1$) and $\theta$ is a phase. Such a situation, illustrated in Fig.~\ref{fig1}, is particularly convenient for studying the role of spatial indistinguishability when one is interested in exploiting the state at the two distant operational locations $L$ and $R$. 
For this choice, we have $\mathcal{I}=0$ when ($l=r'=1$ or $l=r'=0$) and $\mathcal{I}=1$ (when $l=l'$). It is known that, switching the value of $\theta$, bosonic and fermionic behaviors can be interchanged for some initial state configurations \cite{nosrati2019control}. Of course, one has to pay attention to the specific values of the parameters defining the spatial wave functions of Eq.~(\ref{spatialWF}), in order that the system state at time $t$ obeys selection rules due to the particle statistics and the sLOCC probability remains nonzero. 

We now have all the elements to study the dynamics of entanglement of two spatially indistinguishable qubits. Before showing the study for three typical noise channels, we briefly discuss the generalization of the above formalism.


\subsection{Generalization to $N$ particles and $M$ separated environments}
\label{subsec:gen}

We notice that the dynamical formalism introduced above naturally leads to define the local coupling of $N$ noninteracting identical qubits to $M$ independent environments $E_{L_m}$ placed in separated locations $L_m$ ($m=1,\ldots,M$). In fact, one can easily extend Eqs.~(\ref{H1plusH2}) and (\ref{H1&H2}) to this more general case as
\begin{equation}\label{HINM}
    H_{I}:= \sum_{j=1}^N \sum_k\sum_{m=1}^M g_{kL_m}^{(j)}\hat{E}_{kL_m}
    \hat{\sigma}_{L_m}^{(j)}+h.c.,
\end{equation}
where $g_{kL_m}^{(j)}=g_{kL_m}\bra{L_m}\psi_j\rangle$ ($j=1,2,\ldots,N$) is the effective coupling constant between the particle with spatial wave function $\psi_j$ and the mode $k$ of the environment $E_{L_m}$. The operators $\hat{E}_{kL_m}$ are some environmental operators as a function of annihilation or creation operators of local field mode $k$, while $ \hat{\sigma}_{L_m}^{(j)}$ are single-particle pseudospin operators acting on the particle state at the $j$-th position in the global state vector. From Eq.~(\ref{HINM}), one immediately has 
\begin{equation}
    H_{I}= \sum_{j=1}^N H_I^{(j)} = \sum_{m}^M H_{I L_m},
\end{equation}
so that the evolved reduced density matrix $\rho_S(t)$ of the system of $N$ indistinguishable qubits can be formally derived by straightforward generalization of the master equation provided in Eq.~(\ref{master0}). 

We highlight that the evolution for a system made of $d$-level identical particles (qudits) can be easily derived by the above formalism, by simply changing the pseudospin operators $\hat{\sigma}$ appearing in the Hamiltonian of Eq.~(\ref{HINM}) with suitable $d$-level single-particle operators. Analogously, different types of local environments can be considered (such as spin-like or fermionic environments) by adopting the suitable environment operators in place of $\hat{E}_{kL_m}$. Finally, it is clear how the dynamical framework above is flexible and generalizable to specific system-environment interactions, provided that the identical subsystems are noninteracting and the environments are localized and separated.

\section{Phase damping noise} \label{PhaseD}

\begin{figure*}[t!] 
{\includegraphics[width=0.75\textwidth]{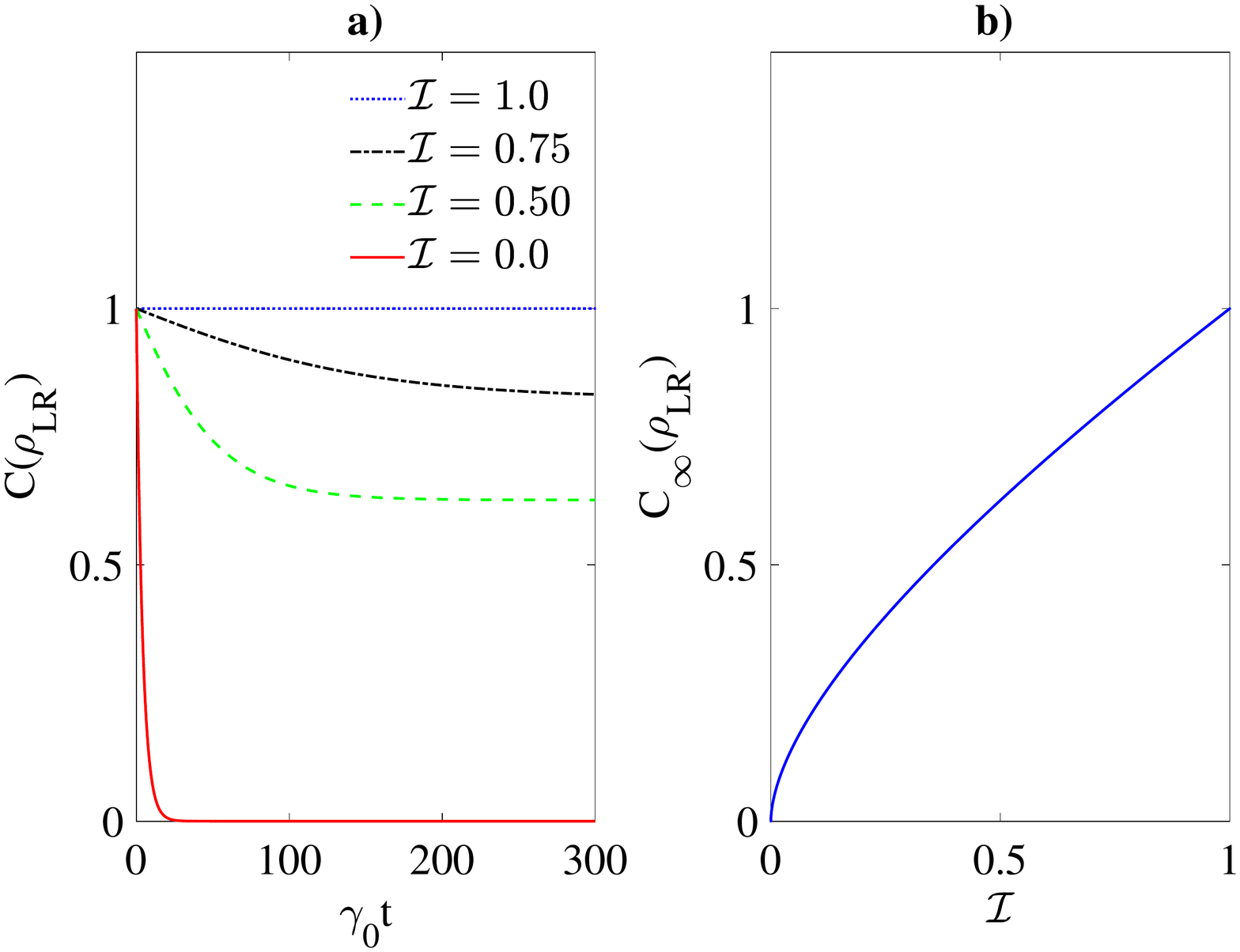}\vspace{-0.6cm}}
\caption{\textbf{Entanglement evolution under localized phase damping environments.} \textbf{a.} Concurrence dynamics $C(\rho_{LR}(t))$ as a function of dimensionless time $\gamma_0 t$ starting from the initial state $\ket{1_{-}}$ for different degrees of spatial indistinguishability $\mathcal{I}$,  fixing $l=r'$: blue dotted line $\mathcal{I}=1$ ($l=l'$), black dot-dashed line $\mathcal{I}=0.75$, green dashed line $\mathcal{I}=0.5$ and red solid line $\mathcal{I}=0$. \textbf{b.} Stationary entanglement $C_\infty(\rho_{LR})$ as a function of $\mathcal{I}$. All the plots are valid for both fermions (with 
$\theta=0$) and bosons (with $\theta=\pi$).}
\label{fig:phase-damping}
\end{figure*}

Phase damping as a noise process describes the loss of quantum coherence without loss of energy (nondissipative channel). For example, it is the result of random telegraph noise or phase noisy lasers \cite{zhou2010disentanglement, franco2012entanglement, bordone2012effect,bellomo2012noisylaser,caiSciRep,woldPRB} and characterizes typical nanodevices and superconducting qubits under low-frequency noise \cite{paladinoPRL,ithier2005decoherence, bylander2011noise,paladinoRMP,antonPRB,bellomoPRA}. 

We study the dynamics of two identical qubits interacting with two spatially separated phase damping environments. The environments are assumed to be Markovian and identical, so that $\gamma_{0X}=\gamma_0$ ($X=L,R$). From the general expressions of Eqs.~(\ref{H1plusH2}) and (\ref{H1&H2}), the total interaction Hamiltonian is $H_{I}=H_{I}^{(1)}+H_{I}^{(2)}$ with
\begin{equation}
        H_{I}^{(j)} =\sum_{X=L,R}\sum_k g_{kX}^{(j)}\hat{a}_{kX}\hat{\sigma}^{(j)}_{z} +h.c.,
\end{equation}
where $\hat{\sigma}^{(j)}_z$ is the usual Pauli pseudospin operator representing the pure dephasing interaction. 
Substituting $\hat{\sigma}_X=\hat{\sigma}_z$ in Eq.~(\ref{master0}), the master equation for the phase damping process is
\begin{equation}\label{PDMasterEq}
        \Dot{\rho}_S=\sum_{X=L,R}\sum_{i,j=1,2}\gamma_{X}^{(i,j)}\left(\sigma_{z}^{(i)}
        \rho_S\sigma_{z}^{(j)}-\frac{1}{2}\{\sigma_{z}^{(i)}\sigma_{z}^{(j)},\rho_S\}\right),
\end{equation}
where $\gamma_{X}^{(i,j)}=\gamma_{0}|\bra{X}\psi_i\rangle\bra{\psi_j}X\rangle|$ are the effective decay rates. 

For the case of two distinguishable qubits undergoing nondissipative channels (such as pure dephasing and depolarizing processes), a Bell-diagonal state structure is maintained during the evolution, the preferred basis being that of the four usual Bell states \cite{aaronsonPRA,silvaPRL}. We find here an analogous behavior. So, we study the dynamics in the basis of the four orthonormal Bell states $\mathcal{B}=\{\ket{{1}_\pm}, \ket{{2}_\pm}\}$ for spatially indistinguishable qubits, defined as \cite{nosrati2019control}
\begin{eqnarray}
\ket{{1}_{\pm}}&=&\dfrac{1}{\sqrt{2\mathcal{N}_{1_{\pm}}}}
(\ket{\psi_1\uparrow,\psi_2\downarrow}\pm\ket{\psi_1\downarrow,\psi_2\uparrow}),\nonumber\\
\ket{{2}_{\pm}}&=&\dfrac{1}{\sqrt{2\mathcal{N}_{2_{\pm}}}}
(\ket{\psi_1\uparrow,\psi_2\uparrow})\pm\ket{\psi_1 \downarrow,\psi_2\downarrow}),
\label{generalizedBS}
\end{eqnarray} 
where $\mathcal{N}_{1_-}=(1-\eta |\langle \psi_1|\psi_2\rangle|^2)$, $\mathcal{N}_{1_+}=\mathcal{N}_{2_{\pm}}=
(1+\eta |\langle \psi_1|\psi_2\rangle|^2)$. As a consequence, the evolved two-particle density matrix can be written in Bell-diagonal form as 
\begin{equation}\label{evolveddensitymatrix}
    \rho_S(t)=\sum_{u=\{1_\pm,2_\pm\}} {p}_{u}(t)\ket{u}\bra{u}
\end{equation}
where ${p}_{u}(t)$ is the probability that the system occupies the Bell state $\ket{u}$ at time $t$, with $\sum_{i}{p}_{u}(t)=1$. One expects that the time-dependent population coefficients $p_u(t)$ will be affected by the degree of spatial indistinguishability. Moreover, effects of the specific spatial wave functions reflect on the initial population coefficients $p_u(0)$, establishing selection rules allowing or forbidding some initial states due to the particle statistics. 

Using the master equation of Eq.~(\ref{PDMasterEq}) and for any $\psi_1$, $\psi_2$, we obtain the time-dependent populations $p_{u}(t)$ of the four Bell states 
\begin{equation}
    \begin{split} \label{Solutionofdephasing}
        p_{1\pm}(t)=& \frac{1}{2}\left(1+e^{-\gamma_-t/2}\right)p_{1\pm}(0)
        +\frac{1}{2}\left(1-e^{-\gamma_-t/2}\right)p_{1\mp}(0), \\
        p_{2\pm}(t)=& \frac{1}{2}\left(1+e^{-\gamma_+t/2}\right)p_{2\pm}(0)
        +\frac{1}{2}\left(1-e^{-\gamma_+t/2}\right)p_{2\mp}(0),
    \end{split}
\end{equation}
where the decay rates $\gamma_{\pm}$ are 
\begin{equation}\label{decay-rate}
        \gamma_{\pm}=\sum_{X=L,R}\sum_{i,j=1,2}(\pm 1)^{i+j}\gamma_{0}|\bra{X}\psi_i\rangle\bra{\psi_j}X\rangle|.
\end{equation}
Notice that, as expected, one-excitation states ($\ket{1_\pm}$) and two-excitation states ($\ket{2_\pm}$) remain in their own subspace during the dynamics.
The expressions above make it evident how specific choices of the spatial wave functions $\ket{\psi_1}$, $\ket{\psi_2}$ determine the two-particle state evolution under localized Markovian pure dephasing. This is a direct consequence of spatially indistinguishable particles. From Eq.~(\ref{Solutionofdephasing}) one can easily see that the system eventually tends to a steady state. 

We first analyze the dynamics of the two extreme cases: spatially distinguishable qubits ($\mathcal{I}=0$) and maximally indistinguishable qubits ($\mathcal{I}=1$). For spatially distinguishable qubits, when one qubit interacts with the environment $E_L$ and the other qubit with $E_R$ (e.g., $l=r'=1$), the decay rates of Eq.~(\ref{decay-rate}) are $\gamma_+=\gamma_-=2\gamma_0$ and the well-known dynamical behavior due to independent Markovian dephasing noises is retrieved.
When particles are perfectly spatially indistinguishable ($l=l'$), the dynamics drastically change. The decay rates from Eq.~(\ref{decay-rate}) become $\gamma_+=4\gamma_0$ and $\gamma_-=0$, so that the evolved Bell-state populations of Eq.~(\ref{Solutionofdephasing}) are 
\begin{equation}
    \begin{split} \label{I=1dephasing}
        p_{1\pm}(t)=& p_{1\pm}(0),\\
        p_{2\pm}(t)=& \frac{1}{2}\left(1+e^{-2\gamma_0t}\right)p_{2\pm}(0)
        +\frac{1}{2}\left(1-e^{-2\gamma_0 t}\right)p_{2\mp}(0). 
    \end{split}
\end{equation} 
The above equations clearly show an interesting consequence of maximum spatial indistinguishability, which decouple the states $\ket{1_\pm}$ from the phase damping process as noise-free subspaces, provided that the initial states are allowed by the particle statistics. Quantum properties contained in the states $\ket{1_\pm}$ are thus frozen and exploitable by sLOCC at any time.

Starting from the initial state $\ket{1_-}$, for fermions (bosons) with $\theta=0$ ($\theta=\pi$), and obtaining $\rho_{LR}(t)$ after sLOCC on $\rho_S(t)$, we calculate the entanglement amount $C_{LR}(\rho_S(t))=C(\rho_{LR}(t))$, which is plotted versus the dimensionless time $\gamma_0t$ in Fig.~\ref{fig:phase-damping}\textbf{a} for different values of $\mathcal{I}$, fixing $l=r'$. The sLOCC probability is $P_{LR}(t)\geq 0.5$ for any time (see Appendix~\ref{appe}; the choice $l=r'$ is just made to maximize the sLOCC probability in the case of maximal indistinguishability). As can be seen in Fig.~\ref{fig:phase-damping}\textbf{a}, entanglement sudden death (occurring for $\mathcal{I}=0$) is forbidden thanks to the spatial overlap of the identical qubits and the entanglement evolution achieves a stationary value which increases as spatial indistinguishability increases. The reason for this entangled steady state is that the randomization process of the relative phase between the states $\ket{L\uparrow,R\downarrow}$ and $\ket{L\downarrow,R\uparrow}$, due to the dephasing process, is mitigated or even prevented for $\mathcal{I}\neq 0$. The characteristic time of the evolution is $\tau_-=1/\gamma_-$ and spatial indistinguishability has the effect of inhibiting the decay rate $\gamma_-$. 
The asymptotic amount of entanglement can be analytically found as
\begin{equation}
    C_{\infty}(\rho_{LR})=\frac{2 (ll')^2}{l^4+l'^4},
\end{equation}
with associated sLOCC probability depending on the degree of spatial indistinguishability, namely: $P_{LR}=l^4+l'^4$ when $\mathcal{I}<1$, $P_\mathrm{LR}=1/2$ for fermions and $P_\mathrm{LR}=1$ for bosons when $\mathcal{I}=1$ (see Appendix~\ref{appe} for details). The stationary entanglement $C_{\infty}(\rho_{LR})$ is plotted as a function of $\mathcal{I}$ in Fig.~\ref{fig:phase-damping}\textbf{b}. It is remarkable that the amount of preserved quantum correlations directly depends on how much the two identical qubits are spatially indistinguishable in the two separated environmental sites. In particular, maximum entanglement is frozen ($C_{\infty}(\rho_{LR})=1$) for $\mathcal{I}=1$ ($l=l'$). Adjusting the value of $\mathcal{I}$ in the initial preparation stage allows one to exploit the remote entanglement at any time during the evolution, despite the action of the pure dephasing. Spatial indistinguishability is a shield for quantum correlations against local dephasing.

\section{Depolarizing noise} \label{depol}

\begin{figure*}[t!] 
{\includegraphics[width=0.75\textwidth]{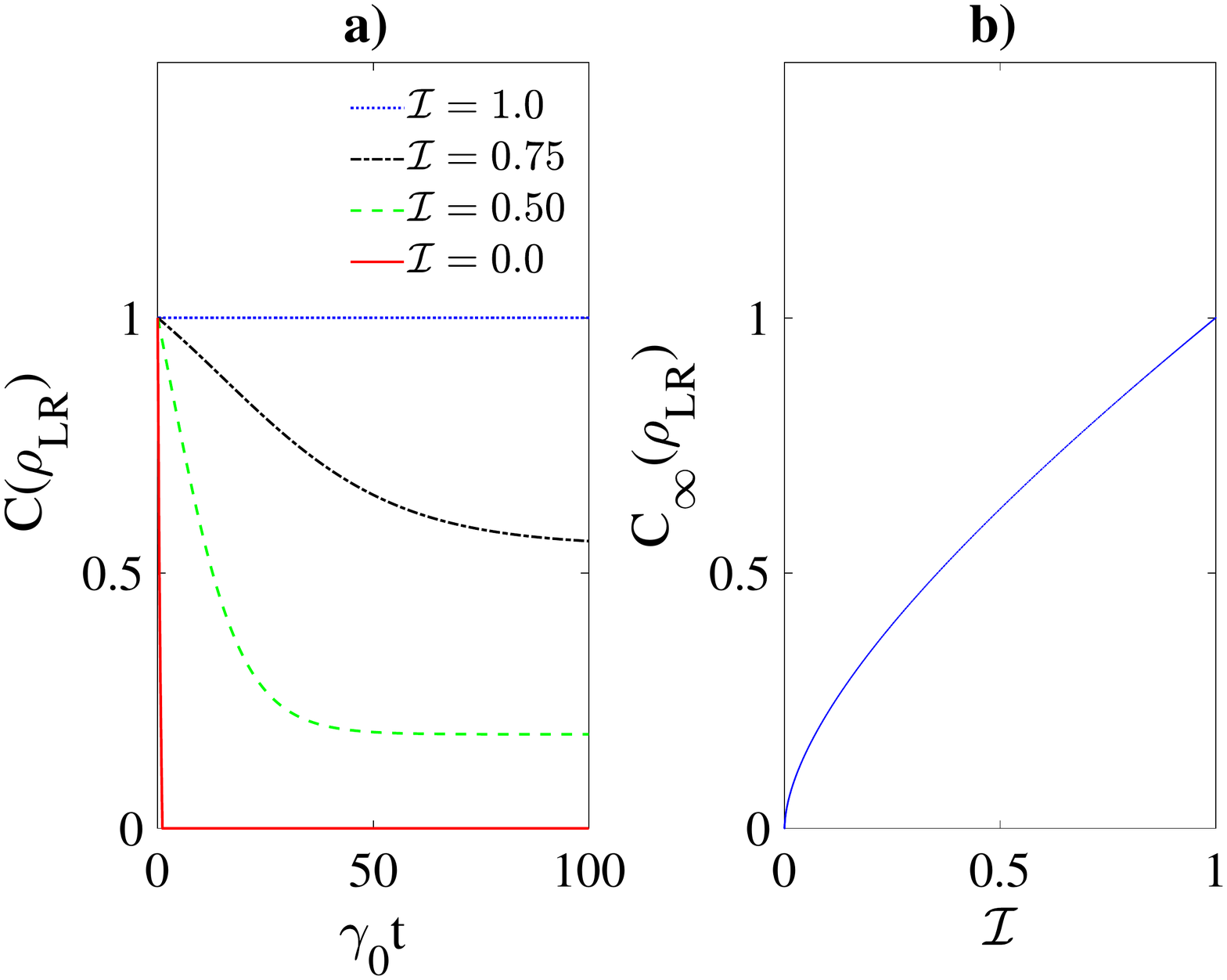}\vspace{-0.5 cm}}
\caption{\textbf{Entanglement evolution under a localized depolarizing environment.} \textbf{a.} Entanglement dynamics $C(\rho_{LR}(t))$ as a function of dimensionless time $\gamma_0 t$ starting from the initial state $\ket{1_{-}}$ for different degrees of spatial indistinguishability $\mathcal{I}$, fixing $l=r'$: blue dotted line $\mathcal{I}=1$ ($l=l'$), black dot-dashed line $\mathcal{I}=0.75$, green dashed line $\mathcal{I}=0.5$ and red solid line $\mathcal{I}=0$. \textbf{b.} Stationary entanglement $C_\infty(\rho_{LR})$ as a function of $\mathcal{I}$. All plots are valid for both fermions (with $\theta=0$) and bosons (with $\theta=\pi$).}
\label{fig:Depolarizing}
\end{figure*}

Depolarizing noise is an isotropic (symmetric) decoherence process of a qubit, such that the qubit state remains untouched with a given probability $p$ while an error (white noise) occurs with probability 
$1-p$ \cite{nielsen2010quantum,durPRA,Klimov2010}. This kind of noise can be the result of an isotropic interaction of a spin-$1/2$-like particle (qubit) with a bosonic or spin-like environment \cite{hamdouni2007time, romero2012simple, melikidze2004parity,bosePRA}. 
Experimentally, the depolarization process can be found in nuclear magnetic resonance setups \cite{xin2017quantum, ryan2009randomized} and Bose-Einstein condensates \cite{kasprzak2006bose, zipkes2010trapped}, where decoherence process is often caused by a residual fluctuating magnetic field. Depolarization processes for photons can also occur, caused by optical scattering when photons become randomly polarized \cite{puentes2005experimental,puentes2007entangled,shaham2011realizing}. 

We consider two identical qubits isotropically interacting with a single Markovian bosonic reservoir localized in region $L$, characterized by the decay rate $\gamma_{0L}=\gamma_0$ (there is no noise channel assumed at location $R$). 
From Eqs.~(\ref{H1plusH2}) and (\ref{H1&H2}), being $g_{kR}=0$, the total interaction Hamiltonian is $H_{I}=H_{I}^{(1)}+H_{I}^{(2)}$ with
\begin{equation}
        H_{I}^{(j)} =\sum_{n=1}^3 \sum_k g_{kL}^{(j)}\hat{\sigma}^{(j)}_n(\hat{a}_{kL}+\hat{a}_{kL}^\dagger),
\end{equation}
where $\hat{\sigma}^{(j)}_n$ ($n=1,2,3$) are the Pauli pseudospin operators. 
Substituting $\hat{\sigma}_X=\sum_{n=1}^3 \hat{\sigma}_n$ in Eq.~(\ref{master0}), the Markovian master equation corresponding to the depolarizing process is
\begin{equation}\label{masterDC}
        \Dot{\rho}_S=\sum_{n=1}^3\sum_{i,j}\gamma_{L}^{(i,j)}\left(\sigma_{n}^{(i)}\rho_S\sigma_{n}^{(j)}-\frac{1}{2}\{\sigma_{n}^{(i)}\sigma_{n}^{(j)},\rho_S\}\right),
\end{equation}
where the effective decay rate $\gamma_{L}^{(i,j)}$ is defined in Eq.~(\ref{EffectiveGamma}) with $X=L$.
The depolarizing noise is also nondissipative, like the phase damping noise, so the convenient basis for the dynamical description is the Bell-state basis of Eq.~(\ref{generalizedBS}). 
The solution of the master equation of Eq.~(\ref{masterDC}) gives an evolved density matrix of the form of Eq.~(\ref{evolveddensitymatrix}), with time-dependent population coefficients
\begin{equation}
    \begin{split} \label{Eq. DP}
       p_{1_-}(t)&=p_{1_-}(0)e^{-\gamma_{-}t}+\frac{1}{4}\left(1-e^{-\gamma_{-}t}\right), \\
       p_{u'}(t)&=p_{u'}(0)e^{-(3\gamma_{+}+\gamma_{-})t/4}+\frac{1}{4}(1-e^{-(3\gamma_{+}+\gamma_{-})t/4})
       \\&+\frac{1-4p_{1_-}(0)}{12}(e^{-\gamma_{-}t}-e^{-(3\gamma_{+}+\gamma_{-})t/4}), 
    \end{split}
\end{equation}
where ${u'}=\{\ket{1_+},\ket{2_\pm}\}$ and $\gamma_\pm$ are given in Eq.~(\ref{decay-rate}) with $X=L$ only, that is $\gamma_{\pm}=\sum_{i,j=1,2}(\pm 1)^{i+j}\gamma_{L}^{(i,j)}$.

In the case of distinguishable (separated) qubits ($\mathcal{I}=0$), the decay rates are $\gamma_+=\gamma_-=\gamma_0$, so that the state at time $t$ is determined by the population coefficients
\begin{equation}
    p_{u}(t)=p_{u}(0)e^{-\gamma_{0}t}+\frac{1}{4}\left(1-e^{-\gamma_{0}t}\right),
\end{equation}
which, from Eq.~(\ref{evolveddensitymatrix}), gives the well-known Werner state $\rho_W=p\rho_0+(1-p)\openone_4/4$ \cite{horodecki2009quantum,werner}, with $p=e^{-\gamma_{0}t}$, as expected. 
In fact, the Werner state describes the action of a depolarizing channel on an individual qubit of the pair, with the system which remains in its initial state $\rho_0$ with probability $p$ and degrades to the maximally mixed state $\openone_{4}/4$ with probability $1-p$. 
In contrast, for maximally spatially indistinguishable qubits ($\mathcal{I}=1$, $l=l'$), the decay rates become $\gamma_+=4l^2\gamma_0$ and $\gamma_-=0$. The most interesting consequence is that, from Eq.~(\ref{Eq. DP}), one has 
\begin{equation}\label{DP}
       p_{1_-}(t)=p_{1_-}(0),
\end{equation}
so that the initial state $\ket{1_-}$ ($p_{1_-}(0)=1$) is a noise-free state. Quantum features of this state can be utilized by sLOCC at any time. 

For the intermediate cases, we plot the entanglement dynamics individuated by $C(\rho_\mathrm{LR}(t))$ in Fig. \ref{fig:Depolarizing}\textbf{a} for different values of $\mathcal{I}$, fixing $l=r'$, starting from the initial state $\ket{1_-}$ for fermions (bosons) with $\theta=0$ ($\theta=\pi$). 
Analogously to the findings for the dephasing noise, the two-qubit entanglement is protected against the depolarizing noise thanks to nonzero spatial indistinguishability. Early-stage disentanglement, occurring for distinguishable qubits ($\mathcal{I}=0$) is prevented when $\mathcal{I}>0$, for which a stationary entangled state is achieved. The characteristic decay time of the evolution is $\tau_-=1/\gamma_-$, which increases as $\mathcal{I}$ increases. 
Eventually, the concurrence $C(\rho_\mathrm{LR}(t))$ converges to the asymptotic value
\begin{equation}\label{conc1}
C_{\infty}(\rho_\mathrm{LR})=\max\left\{0,\frac{3l^2l'^2}{2\left(l^4 +l'^4- l^2 l'^2\right)}-\frac{1}{2}\right\},
\end{equation}
which is plotted in Fig.~\ref{fig:Depolarizing}\textbf{b} as a function of $\mathcal{I}$.
For $\mathcal{I}=1$, one has $C_{\infty}(\rho_\mathrm{LR})=1$ with probability $P_{LR}=1/2$ for fermions and $P_{LR}=1$ for bosons (see Appendix~\ref{appe} for details on the sLOCC probabilities).
In contrast with what happens for distinguishable qubits, for which the steady-state $\rho(\infty)_{LR}=\openone_4/4$ is always a maximally mixed state, the spatial indistinguishability modifies the time-dependent population of each Bell state during the evolution so that the steady state can be entangled.  
The frozen entanglement is exclusively due to nonzero spatial indistinguishability, which tends to suppress the decay rate $\gamma_-$ and protects entanglement from depolarizing noise.

\section{Amplitude damping noise} \label{AM}

\begin{figure*}[t!] 
{\includegraphics[width=0.65\textwidth]{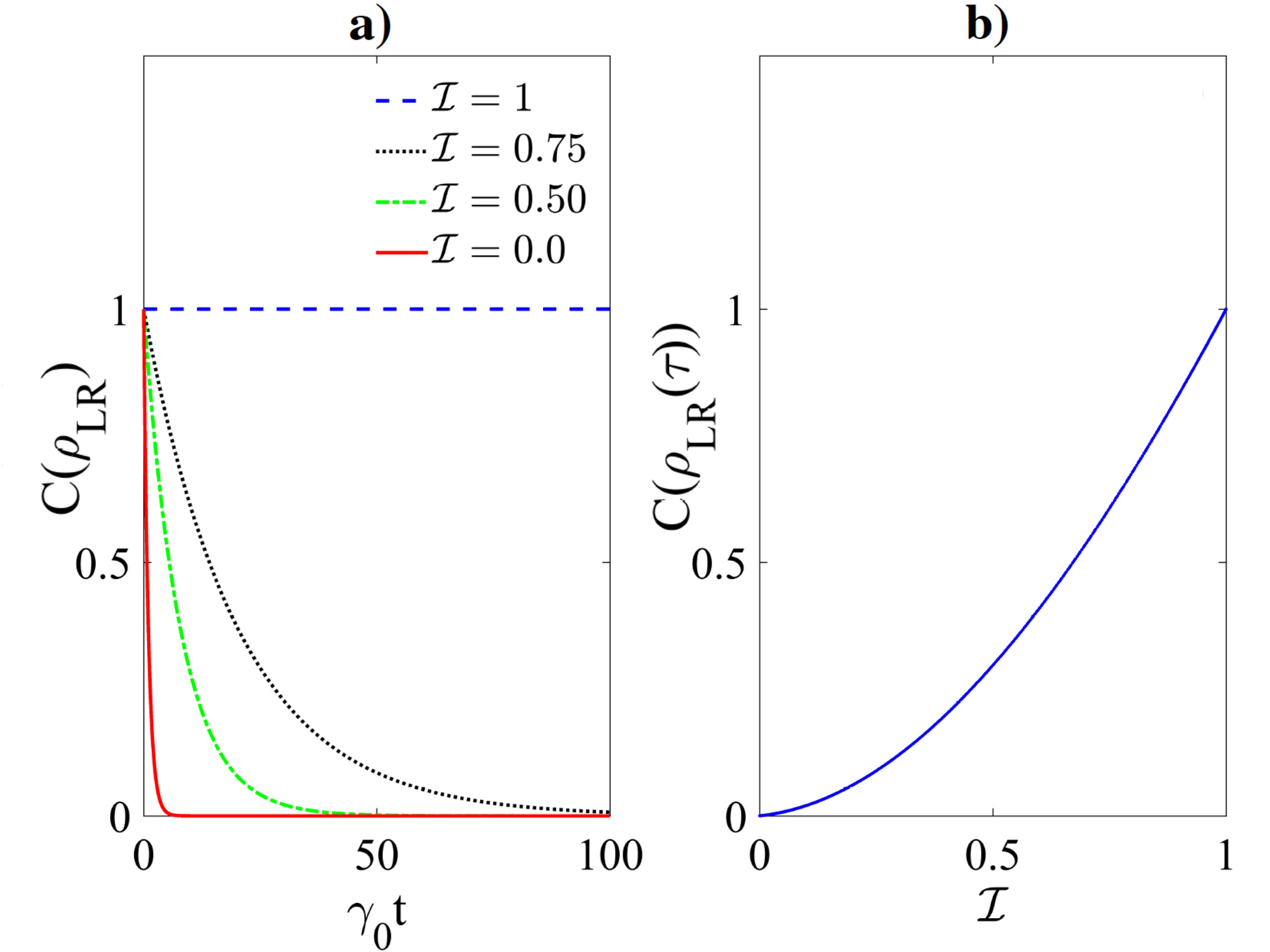}}
\caption{\textbf{Entanglement evolution under localized amplitude damping environments} \textbf{a.} Entanglement dynamics $C(\rho_{LR}(t))$ as a function of dimensionless time $\gamma_0 t$ starting from the initial state $\ket{1_{-}}$ for different degrees of spatial indistinguishability $\mathcal{I}$, fixing $l=r'$: blue dashed line $\mathcal{I}=1$ ($l=l'$), black dotted line $\mathcal{I}=0.75$, green dot-dashed line $\mathcal{I}=0.5$ and red solid line $\mathcal{I}=0$. \textbf{b.} Entanglement $C(\rho_{LR}(\tau))$ at the critical time $\tau$ versus $\mathcal{I}$. All plots are valid for both fermions and bosons with $\theta=\pi/2$.}
\label{fig:Amplitude damping}
\end{figure*}

The amplitude damping process is a well-known model for energy dissipation in open quantum systems. It describes a two-level system (a qubit) of transition frequency $\omega_0$ that interacts with a reservoir by excitation exchange due to spontaneous emission \cite{nielsen2010quantum}. This kind of dissipative channel is found in different contexts, such as atoms in cavities \cite{wineland2000,Schindler,Bruzewicz}, superconducting qubits in circuit QED \cite{blais2007,girvinNP}, spin chains with ferromagnetic Heisenberg interactions \cite{giovannettiPRA} or photon scattering from a single-mode optical fiber \cite{Satoh:99}, and is also well simulated by linear-optics devices \cite{Fisher_2012,6358698}.

We take here two identical qubits, with a given degree of spatial indistinguishability, interacting with two localized amplitude damping environments in $L$ and $R$ sites, respectively. The environments are assumed to be Markovian and identical, so that the (local) characteristic decay rates are $\gamma_{0X}=\gamma_0$ ($X=L,R$). From Eqs.~(\ref{H1plusH2}) and (\ref{H1&H2}), the total interaction Hamiltonian is $H_{I}=H_{I}^{(1)}+H_{I}^{(2)}$ with
\begin{equation}
        H_{I}^{(j)} =\sum_{X=L,R}\sum_k g_{kX}^{(j)}\hat{a}^\dagger_{kX}\hat{\sigma}^{(j)}_-+h.c., 
\end{equation}
where $\hat{\sigma}^{(j)}_\pm$ are single-particle pseudospin creation and annihilation operators. Substituting $\hat{\sigma}_X=\hat{\sigma}^{(j)}_-$ in Eq.~(\ref{master0}) we have the master equation for localized amplitude damping noises
\begin{equation}\label{masterAM}
        \Dot{\rho}=\sum_{X=L,R}\sum_{i,j}\gamma_{X}^{(i,j)}\left(\sigma_{-}^{(i)}\rho\sigma_{+}^{(j)}-\frac{1}{2}\{\sigma_{+}^{(i)}\sigma_{-}^{(j)},\rho\}\right).
\end{equation}
Considering the dissipative effects of this type of evolution, the general dynamics can be easily solved in the basis $\mathcal{B}_1=\{\ket{1_\pm}, \ket{2}, \ket{0} \}$, where $\ket{{1}_{\pm}}$ are the two Bell states defined in Eq.~(\ref{generalizedBS}) and
\begin{equation}
    \ket{2}=\frac{1}{\sqrt{\mathcal{N}_{1}}}\ket{\psi_1\uparrow,\psi_2\uparrow}, \quad    
    \ket{0}=\frac{1}{\sqrt{\mathcal{N}_{0}}}\ket{\psi_1\downarrow,\psi_2\downarrow},
\end{equation}
with $\mathcal{N}_{0}=\mathcal{N}_{1}=(1+\eta |\langle \psi_1|\psi_2\rangle|^2)$. Starting from a state which is a mixture of the four basis states above, the structure is maintained during the evolution: $\rho_S(t)=\sum_{u'} p_{u'}(t) \ket{u'}\bra{u'}$, where $\ket{u'}\in \mathcal{B}_1$. 
This choice leads to the following differential equations for the basis state populations
\begin{equation}\label{DEAM}
    \begin{split}
       \dot{p}_{1_-}&=(1-\xi)\gamma_{0}\left(p_2-p_{1_-}\right)  ,\\
        \dot{p}_{1_+}&=(1+\xi)\gamma_{0}\left(p_2-p_{1_+}\right),\\
      \dot{p}_{2}&=-2\gamma_0p_2 ,
    \end{split}
\end{equation}
with $p_0(t)=1-(p_{1_+}(t)+p_{1_-}(t)+p_{2}(t))$ and where $\xi=ll'+rr'$ is a spatial overlap parameter (see Appendix~\ref{app:AM} for details on the general solutions). 

For spatially separated qubits ($\mathcal{I}=\xi=0$), one retrieves the known dissipative two-qubit dynamics, eventually decaying to the ground state $\ket{0}$. On the other hand, for maximally indistinguishable qubits ($\mathcal{I}=\xi=1$), from Eq.~(\ref{DEAM}) we obtain
\begin{equation}
        p_{1_-}(t)=p_{1_-}(0),
\end{equation}
which clearly shows that state $\ket{1_-}$ is unaffected from the detrimental environment. Perfect spatial indistinguishability prevents the decay to the ground state of the two-qubit system and maintains the Bell state $\ket{1_-}$ frozen (noise-free).

The evolution of two-qubit entanglement, identified by $C(\rho_\mathrm{LR}(t))$, is displayed in Fig.~\ref{fig:Amplitude damping}\textbf{a} for different values of $\mathcal{I}$, starting from the initial state $\ket{1_-}$ for fermions and bosons with $\theta=\pi/2$. As can be seen, spatial indistinguishability significantly extends the timescale of entanglement decay in the system compared to the one for distinguishable particles (red solid line, entanglement asymptotically goes to zero \cite{lofrancoreview,Yu598}). 
In Fig.~\ref{fig:Amplitude damping}\textbf{b} we plot the concurrence $C(\rho_{LR}(\tau))$, calculated at a fixed critical time $\tau$ when the distinguishable-qubit entanglement is very small ($\sim10^{-5}$), as a function of $\mathcal{I}$, which highlights the direct relationship between preserved entanglement and spatial indistinguishability. When $\mathcal{I}<1$ the system eventually decays to the unentangled ground state $\ket{0}_{LR}=\ket{L\downarrow,R\downarrow}$ after sLOCC.
We remark that the probability to obtain the frozen exploitable maximum entanglement $C(\rho_\mathrm{LR}(t)=1$, contained in the state $\ket{1_-}$ when $\mathcal{I}=1$, is $P_{LR}=1/3$ for fermions and $P_{LR}=1$ for bosons (see Appendix~\ref{appe} for details on the sLOCC probabilities). 
Collective effects enabled by spatial indistinguishability can thus very efficiently preserve two-qubit entanglement from decoherence also under dissipative noise.

\section{Conclusions} \label{Conclusion}

In conclusion, we have introduced a formalism that allows us in principle to obtain the dynamics of a system of noninteracting identical particles, with an arbitrary degree of spatial indistinguishability $\mathcal{I}$, coupled to separated localized environments. The value of $\mathcal{I}$ is related to the amount of spatial overlap of the particle wave functions in the operational sites where the noises operate. The distributed resource state of the system at a given time is then activated by spatially localized operations and classical communications (sLOCC). The procedure is valid for any form of system-environment interaction and for any particle statistics (fermions and bosons). 
As a general aspect, we have found that environmental collective effects on the many-particle system are inevitable as long as particles are spatially indistinguishable. In fact, the dynamics of identical particles is not only determined by the localized interaction between system and environment but also directly by the value of $\mathcal{I}$. Spatial indistinguishability is the property of the system acting as a continuous knob which governs the interplay between collective and individual effects of the localized environments. 

The introduced dynamical framework has been then applied to the typical case of two identical qubits weakly interacting with local Markovian (memoryless) environments. In particular, we have considered  paradigmatic quantum noises such as phase damping, depolarizing and amplitude damping. As a result valid for all the noise channels, we have found that spatial indistinguishability plays a sort of dynamical decoupling role of some states of the system from the environment, in that it inhibits the effective decay rate: the larger $\mathcal{I}$, the smaller the decay rate. For maximal spatial indistinguishability ($\mathcal{I}=1$), a maximally entangled state can be perfectly preserved against noise, leading to frozen entanglement. Ultimately, spatial indistinguishability is a property capable to efficiently shield exploitable entangled states at distant sites against noise.

In agreement with previous results \cite{nosrati2019control}, these findings are very promising to fault-tolerant quantum information tasks, since controllable indistinguishability is an intrinsic resource of a composite quantum network made of identical subsystems. Experimental verification of our predictions is thus one of the main outlook of this work. Various experimental setups can be indeed brought up for implementing protection by spatial indistinguishability, from photonic and circuit QED platforms to Bose-Einstein condensates. 

Our results motivate analyses of non-Markovian system-environment interactions and generalizations to composite systems of larger size (many-qudit systems). They open the way to further studies investigating the effects of indistinguishability for other types of local environments, e.g. spin-like surroundings and classical noises, different classes of initial states and various figures of merit of quantumness.

\section*{Acknowledgements}
F.N. and R.L.F. acknowledge Vittorio Giovannetti and Antonella De Pasquale for discussions during the visit at the Scuola Normale Superiore in Pisa (Italy). F.N. would also like to thank Ilenia Tinnirello for supporting this research.

\appendix
\section{sLOCC probability} \label{appe}

\begin{figure*}[t!] 
{\includegraphics[width=0.95\textwidth]{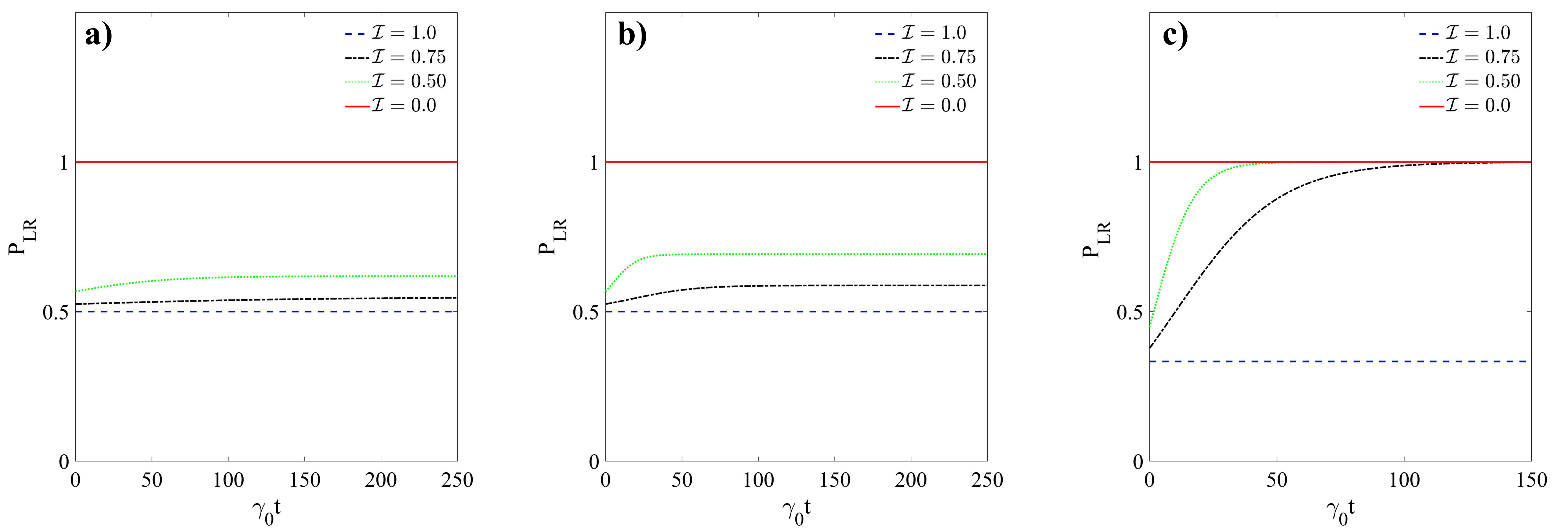}}
\caption{\textbf{sLOCC probability for fermions.} $P_{LR}$ as a function of dimensionless time $\gamma_0 t$, starting from the initial state $\ket{1_{-}}$, fixing $l=r'$, under localized: \textbf{a)} phase damping noise, \textbf{b)} depolarizing noise, \textbf{c)} amplitude damping noise. The plots are given for different degrees of spatial indistinguishability: $\mathcal{I}=1$ (blue dashed line), $\mathcal{I}=0.75$ (black dot-dashed line), $\mathcal{I}=0.5$ (green dotted line) and $\mathcal{I}=0$ (red solid line).} 
\label{fig:pr}
\end{figure*}

\begin{figure*}[t!] 
{\includegraphics[width=0.95\textwidth]{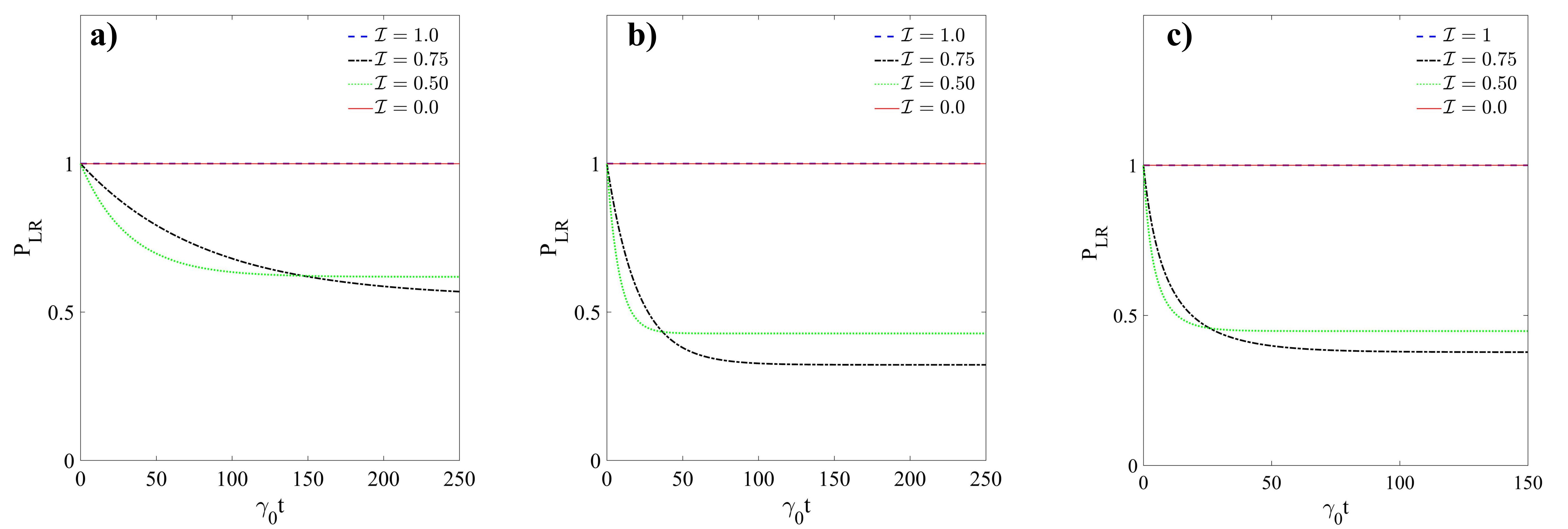}}
\caption{\textbf{sLOCC probability for bosons.} $P_{LR}$ as a function of dimensionless time $\gamma_0 t$, starting from the initial state $\ket{1_{-}}$, fixing $l=r'$, under localized: \textbf{a)} phase damping noise, \textbf{b)} depolarizing noise, \textbf{c)} amplitude damping noise. The plots are given for different degrees of spatial indistinguishability: $\mathcal{I}=1$ (blue dashed line), $\mathcal{I}=0.75$ (black dot-dashed line), $\mathcal{I}=0.5$ (green dotted line) and $\mathcal{I}=0$ (red solid line).} 
\label{fig:prBos}
\end{figure*}

In this appendix we provide the expressions of the the time dependent sLOCC probability for the three localized noise channels considered in the manuscript. 

In general, the sLOCC measurement giving the distributed resource state $\rho_{LR}(t)$ of Eq.~(\ref{rhoLR}) comes with a cost, since it corresponds to the post-selection procedure to find one particle (qubit) in the region $L$ and the other one in the region $R$. 
This post-selection probability is given by the sLOCC probability   
\begin{equation}
    P_{LR}(t)=\mathrm{Tr}(\Pi_{LR}^{(2)}\rho_S(t),
\end{equation}
where $\rho_S(t)$ is the two-qubit density matrix solution of the master equation of Eq.~(\ref{master0}) for the given noise channel. As a result, the exploitable entanglement of $\rho_{LR}(t)$ is conditional in general. So, it is important that the value of $P_{LR}(t)$ remains high enough during the evolution to be of experimental relevance. Here we explicitly show that this is the case.

We consider the state $\ket{1_-}$ of Eq.~(\ref{generalizedBS}) as initial state since it shows noise-free characteristics. In the main text, we have already reported that, for maximum spatial indistinguishability  $\mathcal{I}=1$ with $l=r'$ in the spatial wave functions of Eq.~(\ref{spatialWF}) (meaning $l=r=l'=r'=1/\sqrt{2}$), $P_{LR}(t)=1$ for bosons at any time for all the three types of local noise. This means that for bosonic qubits the maximum entanglement protection is even deterministic. The choice $l=r'$ is made such as to maximize the sLOCC probability when $\mathcal{I}=1$.

\textbf{Phase damping noise.} For this kind of local system-environment interaction, the general expression for the sLOCC probability is
\begin{equation}
    P_{LR}(t)=\frac{ |l'r|^2+ |l r'|^2 -2\eta ll' r r' e^{-\gamma_{-}t}\cos{\theta}}{1-\eta e^{-\gamma_{-}t}\left(|l'l|^2+ |r r'|^2 +2ll'r r'\cos{\theta}\right)},
    \label{Pr_Phase Damping}
\end{equation}
where $\gamma_-$ is defined in Eq.~(\ref{decay-rate}). It is interesting to notice that, when 
$\mathcal{I}<1$, that is $\gamma_->0$, the sLOCC probability at the stationary state ($t\rightarrow\infty$) is simply $P_{LR}=|l'r|^2+ |l r'|^2$. For maximal indistinguishability $\mathcal{I}=1$ ($l=l'$ and $r=r'$), for which $\gamma_{-}=0$, from Eq.~(\ref{Pr_Phase Damping}) one has $P_\mathrm{LR}(t)=1$ for bosons with $\theta=\pi$ and $P_{LR}(t)=2 l^2 \left(1-l^2\right)$ for fermions with $\theta=0$ (the latter becomes $1/2$ when $l=r'$). In Figs.~\ref{fig:pr}\textbf{a} and \ref{fig:prBos}\textbf{a} we plot $P_{LR}(t)$ for the phase damping process, fixing $l=r'$, associated to fermions and bosons, respectively, considering different degrees of spatial indistinguishability.

\textbf{Depolarizing noise.} For a local depolarizing system-environment interaction, the general expression for the sLOCC probability is
\begin{equation} \label{Pr_Depolarizing}
P_{LR}(t)=\frac{2\left(l^2r'^2+l'^2r^2+\eta lr'rl'\cos\theta\ (1-3e^{-\gamma_- t})\right)}{2+\eta(1-3e^{-\gamma_-t}) \left(l^2l'^2+r^2r'^2+2ll'r r'\cos\theta\right)}.
\end{equation}
For maximal indistinguishability $\mathcal{I}=1$ ($l=l'$ and $r=r'$), for which $\gamma_{-}=0$, from Eq.~(\ref{Pr_Depolarizing}) with $l=r'$ we again find, as before, that $P_\mathrm{LR}(t)=1$ for bosons with $\theta=\pi$ and $P_{LR}(t)=1/2$ for fermions with $\theta=0$. In Figs.~\ref{fig:pr}\textbf{b} and \ref{fig:prBos}\textbf{b} we plot $P_{LR}(t)$ for the depolarizing process, fixing $l=r'$, associated to fermions and bosons, respectively, for different values of $\mathcal{I}$.

\textbf{Amplitude damping noise.} For localized amplitude damping system-environment interaction, the general expression for the sLOCC probability is
\begin{equation}\label{Pr_AM}
P_{LR}(t)=\frac{l^2r'^2+l'^2r^2+2\eta lr'rl'\cos\theta\ (1-2e^{-\gamma_0(1-\xi) t})}{1+\eta(1-2e^{-\gamma_0(1-\xi)t}) \left(l^2l'^2+r^2r'^2+2ll'r r'\cos\theta\right)},
\end{equation}
where $\xi=ll'+rr'$ is the spatial overlap parameter. 
For maximal indistinguishability $\mathcal{I}=1$ ($l=l'$ and $r=r'$), for which $\xi=1$, from Eq.~(\ref{Pr_AM}) with $\theta=\pi/2$, we have that $P_{LR}(t)=1$ for bosons and $P_{LR}(t)=\frac{l^2 \left(1-l^2\right)}{1-l^2 \left(1-l^2\right)}$ for fermions (the latter becomes $P_{LR}(t)=1/3$ for $l=r'$).
When $\mathcal{I}_{LR}<1$ ($\xi<1$), the sLOCC probability for the stationary ground state ($t\rightarrow\infty$) with $\theta=\pi/2$ and $l=r'$ is $P_{LR}=\frac{l^4+l'^4}{1+2\eta l^2l'^2}$.
We display in Figs.~\ref{fig:pr}\textbf{c} and \ref{fig:prBos}\textbf{c} the time behavior of $P_{LR}(t)$ for the amplitude damping process, fixing $l=r'$, corresponding to fermions and bosons, respectively, for different values of $\mathcal{I}$.

The general conclusion of this detailed analysis is that the entanglement resource, shielded by indistinguishability from the noise, is efficiently exploited by sLOCC, since the sLOCC probabilities remains of experimental value during the dynamics. In particular, bosonic qubits admit a deterministic entanglement protection. We finally point out that for both the two extreme cases, $\mathcal{I}=1,0$, the sLOCC probability is time-independent.

\section{Solution of the amplitude damping channel} \label{app:AM}

From the master equation for the local amplitude damping channels acting on two spatially overlapping identical qubits, given in Eq.~(\ref{masterAM}), we have obtained the differential equations of Eq.~(\ref{DEAM}). These differential equations can be conveniently solved separately for the cases when $\mathcal{I}<1$ and for maximal spatial indistinguishability $\mathcal{I}=1$. 

The solution for the case when the spatial indistinguishability is not maximum, $\mathcal{I}<1$, so that the spatial overlap parameter $0\leq\xi<1$, 
\begin{equation} \label{t-pop1}
    \begin{split}
        p_{2}(t)&=p_{2}(0)e^{-2\gamma_{0}t},\\
        p_{1_+}(t)&=p_{1_+}(0)e^{-\left(1+\xi\right)\gamma_{0}t}+
        \left(\frac{1+\xi}{1-\xi}\right)p_{2}(0)\left(e^{-\left(1+\xi\right)\gamma_{0}t}-e^{-2\gamma_{0}t}\right), \\
        p_{1_-}(t)&=p_{1_-}(0)e^{-\left(1-\xi\right)\gamma_{0}t}+
        \left(\frac{1-\xi}{1+\xi}\right)p_{2}(0)\left(e^{-\left(1-\xi\right)\gamma_{0}t}-e^{-2\gamma_{0}t}\right),\\
        p_0(t)&=1- p_{1_+}(t) - p_{1_-}(t) - p_{2}(t),
    \end{split}
\end{equation}
where $\xi=ll'+rr'$. The above equations clearly show that the system state eventually decays to the ground state $\ket{0}=\ket{\psi_1\downarrow,\psi_2 \downarrow}$, where $\ket{\psi_1}$, $\ket{\psi_2}$ are defined in Eq.~(\ref{spatialWF}). In the special case when the two identical qubits are spatially separated, one in the site $L$ and one in $R$, one has $l=r'=1$ or $l=r'=0$, leading to $\xi=0$ (distinguishable particles $\mathcal{I}=0$). In this case, the time-dependent populations given in Eq.~(\ref{t-pop1}) reduce to the usual decaying functions for the one-excitation Bell states 
$\ket{1_\pm}=(\ket{L\uparrow,R \downarrow}\pm\ket{L\downarrow,R \uparrow})/\sqrt{2}$ and for the two-excitation state $\ket{2}=\ket{L\uparrow,R \uparrow}$. 

When the qubits are maximally indistinguishable ($\mathcal{I}_{LR}=1$ and $\xi=1$), the time-dependent populations are
\begin{equation}
    \begin{split}
      p_{1_-}(t)&=p_{1_-}(0)\\
        p_{1_+}(t)&=p_{1_+}(0)e^{-2\gamma_{0}t}+2p_{1}(0)\gamma_{0}te^{-2\gamma_{0}t} \\
      p_{2}(t)&=p_{2}(0)e^{-2\gamma_{0}t},\\
      p_0(t)&=1- p_{1_+}(t) - p_{1_-}(t) - p_{2}(t),
    \end{split}
\end{equation}
from which the system dynamics of an initial state being a mixture of the basis states $\mathcal{B}_1=\{\ket{1_\pm}, \ket{2}, \ket{0} \}$ can be completely determined. 

Of course, different classes of initial states can be considered which will lead to different solutions. This is left to detailed investigations elsewhere.


\begin{thebibliography}{124}%
\makeatletter
\providecommand \@ifxundefined [1]{%
 \@ifx{#1\undefined}
}%
\providecommand \@ifnum [1]{%
 \ifnum #1\expandafter \@firstoftwo
 \else \expandafter \@secondoftwo
 \fi
}%
\providecommand \@ifx [1]{%
 \ifx #1\expandafter \@firstoftwo
 \else \expandafter \@secondoftwo
 \fi
}%
\providecommand \natexlab [1]{#1}%
\providecommand \enquote  [1]{``#1''}%
\providecommand \bibnamefont  [1]{#1}%
\providecommand \bibfnamefont [1]{#1}%
\providecommand \citenamefont [1]{#1}%
\providecommand \href@noop [0]{\@secondoftwo}%
\providecommand \href [0]{\begingroup \@sanitize@url \@href}%
\providecommand \@href[1]{\@@startlink{#1}\@@href}%
\providecommand \@@href[1]{\endgroup#1\@@endlink}%
\providecommand \@sanitize@url [0]{\catcode `\\12\catcode `\$12\catcode
  `\&12\catcode `\#12\catcode `\^12\catcode `\_12\catcode `\%12\relax}%
\providecommand \@@startlink[1]{}%
\providecommand \@@endlink[0]{}%
\providecommand \url  [0]{\begingroup\@sanitize@url \@url }%
\providecommand \@url [1]{\endgroup\@href {#1}{\urlprefix }}%
\providecommand \urlprefix  [0]{URL }%
\providecommand \Eprint [0]{\href }%
\providecommand \doibase [0]{http://dx.doi.org/}%
\providecommand \selectlanguage [0]{\@gobble}%
\providecommand \bibinfo  [0]{\@secondoftwo}%
\providecommand \bibfield  [0]{\@secondoftwo}%
\providecommand \translation [1]{[#1]}%
\providecommand \BibitemOpen [0]{}%
\providecommand \bibitemStop [0]{}%
\providecommand \bibitemNoStop [0]{.\EOS\space}%
\providecommand \EOS [0]{\spacefactor3000\relax}%
\providecommand \BibitemShut  [1]{\csname bibitem#1\endcsname}%
\let\auto@bib@innerbib\@empty
\bibitem [{\citenamefont {Reusch}\ \emph {et~al.}(2015)\citenamefont {Reusch},
  \citenamefont {Sperling},\ and\ \citenamefont {Vogel}}]{vogel2015PRA}%
  \BibitemOpen
  \bibfield  {author} {\bibinfo {author} {\bibfnamefont {A.}~\bibnamefont
  {Reusch}}, \bibinfo {author} {\bibfnamefont {J.}~\bibnamefont {Sperling}}, \
  and\ \bibinfo {author} {\bibfnamefont {W.}~\bibnamefont {Vogel}},\
  }\href@noop {} {\bibfield  {journal} {\bibinfo  {journal} {Phys. Rev. A}\
  }\textbf {\bibinfo {volume} {91}},\ \bibinfo {pages} {042324} (\bibinfo
  {year} {2015})}\BibitemShut {NoStop}%
\bibitem [{\citenamefont {Shi}(2003)}]{Shi2003PRA}%
  \BibitemOpen
  \bibfield  {author} {\bibinfo {author} {\bibfnamefont {Y.}~\bibnamefont
  {Shi}},\ }\href@noop {} {\bibfield  {journal} {\bibinfo  {journal} {Phys.
  Rev. A}\ }\textbf {\bibinfo {volume} {67}},\ \bibinfo {pages} {024301}
  (\bibinfo {year} {2003})}\BibitemShut {NoStop}%
\bibitem [{\citenamefont {Li}\ \emph {et~al.}(2001)\citenamefont {Li},
  \citenamefont {Zeng}, \citenamefont {Liu},\ and\ \citenamefont
  {Long}}]{Li2001PRA}%
  \BibitemOpen
  \bibfield  {author} {\bibinfo {author} {\bibfnamefont {Y.-S.}\ \bibnamefont
  {Li}}, \bibinfo {author} {\bibfnamefont {B.}~\bibnamefont {Zeng}}, \bibinfo
  {author} {\bibfnamefont {X.-S.}\ \bibnamefont {Liu}}, \ and\ \bibinfo
  {author} {\bibfnamefont {G.-L.}\ \bibnamefont {Long}},\ }\href@noop {}
  {\bibfield  {journal} {\bibinfo  {journal} {Phys. Rev. A}\ }\textbf {\bibinfo
  {volume} {64}},\ \bibinfo {pages} {054302} (\bibinfo {year}
  {2001})}\BibitemShut {NoStop}%
\bibitem [{\citenamefont {Paskauskas}\ and\ \citenamefont
  {You}(2001)}]{Paskauskas2001PRA}%
  \BibitemOpen
  \bibfield  {author} {\bibinfo {author} {\bibfnamefont {R.}~\bibnamefont
  {Paskauskas}}\ and\ \bibinfo {author} {\bibfnamefont {L.}~\bibnamefont
  {You}},\ }\href@noop {} {\bibfield  {journal} {\bibinfo  {journal} {Phys.
  Rev. A}\ }\textbf {\bibinfo {volume} {64}},\ \bibinfo {pages} {042310}
  (\bibinfo {year} {2001})}\BibitemShut {NoStop}%
\bibitem [{\citenamefont {Schliemann}\ \emph {et~al.}(2001)\citenamefont
  {Schliemann}, \citenamefont {Cirac}, \citenamefont {Kus}, \citenamefont
  {Lewenstein},\ and\ \citenamefont {Loss}}]{cirac2001PRA}%
  \BibitemOpen
  \bibfield  {author} {\bibinfo {author} {\bibfnamefont {J.}~\bibnamefont
  {Schliemann}}, \bibinfo {author} {\bibfnamefont {J.~I.}\ \bibnamefont
  {Cirac}}, \bibinfo {author} {\bibfnamefont {M.}~\bibnamefont {Kus}}, \bibinfo
  {author} {\bibfnamefont {M.}~\bibnamefont {Lewenstein}}, \ and\ \bibinfo
  {author} {\bibfnamefont {D.}~\bibnamefont {Loss}},\ }\href@noop {} {\bibfield
   {journal} {\bibinfo  {journal} {Phys. Rev. A}\ }\textbf {\bibinfo {volume}
  {64}},\ \bibinfo {pages} {022303} (\bibinfo {year} {2001})}\BibitemShut
  {NoStop}%
\bibitem [{\citenamefont {Zanardi}(2002)}]{zanardiPRA}%
  \BibitemOpen
  \bibfield  {author} {\bibinfo {author} {\bibfnamefont {P.}~\bibnamefont
  {Zanardi}},\ }\href@noop {} {\bibfield  {journal} {\bibinfo  {journal} {Phys.
  Rev. A}\ }\textbf {\bibinfo {volume} {65}},\ \bibinfo {pages} {042101}
  (\bibinfo {year} {2002})}\BibitemShut {NoStop}%
\bibitem [{\citenamefont {Tichy}\ \emph {et~al.}(2013)\citenamefont {Tichy},
  \citenamefont {{de Melo}}, \citenamefont {Kus}, \citenamefont {Mintert},\
  and\ \citenamefont {Buchleitner}}]{tichyFort}%
  \BibitemOpen
  \bibfield  {author} {\bibinfo {author} {\bibfnamefont {M.~C.}\ \bibnamefont
  {Tichy}}, \bibinfo {author} {\bibfnamefont {F.}~\bibnamefont {{de Melo}}},
  \bibinfo {author} {\bibfnamefont {M.}~\bibnamefont {Kus}}, \bibinfo {author}
  {\bibfnamefont {F.}~\bibnamefont {Mintert}}, \ and\ \bibinfo {author}
  {\bibfnamefont {A.}~\bibnamefont {Buchleitner}},\ }\href@noop {} {\bibfield
  {journal} {\bibinfo  {journal} {Fortschr. Phys.}\ }\textbf {\bibinfo {volume}
  {61}},\ \bibinfo {pages} {225} (\bibinfo {year} {2013})}\BibitemShut
  {NoStop}%
\bibitem [{\citenamefont {Tichy}\ \emph {et~al.}(2011)\citenamefont {Tichy},
  \citenamefont {Mintert},\ and\ \citenamefont
  {Buchleitner}}]{tichy2011essential}%
  \BibitemOpen
  \bibfield  {author} {\bibinfo {author} {\bibfnamefont {M.~C.}\ \bibnamefont
  {Tichy}}, \bibinfo {author} {\bibfnamefont {F.}~\bibnamefont {Mintert}}, \
  and\ \bibinfo {author} {\bibfnamefont {A.}~\bibnamefont {Buchleitner}},\
  }\href@noop {} {\bibfield  {journal} {\bibinfo  {journal} {J. Phys. B: At.
  Mol. Opt. Phys.}\ }\textbf {\bibinfo {volume} {44}},\ \bibinfo {pages}
  {192001} (\bibinfo {year} {2011})}\BibitemShut {NoStop}%
\bibitem [{\citenamefont {Bose}\ and\ \citenamefont {Home}(2013)}]{bose2013}%
  \BibitemOpen
  \bibfield  {author} {\bibinfo {author} {\bibfnamefont {S.}~\bibnamefont
  {Bose}}\ and\ \bibinfo {author} {\bibfnamefont {D.}~\bibnamefont {Home}},\
  }\href@noop {} {\bibfield  {journal} {\bibinfo  {journal} {Phys. Rev. Lett.}\
  }\textbf {\bibinfo {volume} {110}},\ \bibinfo {pages} {140404} (\bibinfo
  {year} {2013})}\BibitemShut {NoStop}%
\bibitem [{\citenamefont {Sasaki}\ \emph {et~al.}(2011)\citenamefont {Sasaki},
  \citenamefont {Ichikawa},\ and\ \citenamefont {Tsutsui}}]{sasaki2011PRA}%
  \BibitemOpen
  \bibfield  {author} {\bibinfo {author} {\bibfnamefont {T.}~\bibnamefont
  {Sasaki}}, \bibinfo {author} {\bibfnamefont {T.}~\bibnamefont {Ichikawa}}, \
  and\ \bibinfo {author} {\bibfnamefont {I.}~\bibnamefont {Tsutsui}},\
  }\href@noop {} {\bibfield  {journal} {\bibinfo  {journal} {Phys. Rev. A}\
  }\textbf {\bibinfo {volume} {83}},\ \bibinfo {pages} {012113} (\bibinfo
  {year} {2011})}\BibitemShut {NoStop}%
\bibitem [{\citenamefont {Ghirardi}\ \emph {et~al.}(2002)\citenamefont
  {Ghirardi}, \citenamefont {Marinatto},\ and\ \citenamefont
  {Weber}}]{ghirardi2002}%
  \BibitemOpen
  \bibfield  {author} {\bibinfo {author} {\bibfnamefont {G.~C.}\ \bibnamefont
  {Ghirardi}}, \bibinfo {author} {\bibfnamefont {L.}~\bibnamefont {Marinatto}},
  \ and\ \bibinfo {author} {\bibfnamefont {T.}~\bibnamefont {Weber}},\
  }\href@noop {} {\bibfield  {journal} {\bibinfo  {journal} {J. Stat. Phys.}\
  }\textbf {\bibinfo {volume} {108}},\ \bibinfo {pages} {49} (\bibinfo {year}
  {2002})}\BibitemShut {NoStop}%
\bibitem [{\citenamefont {Balachandran}\ \emph {et~al.}(2013)\citenamefont
  {Balachandran}, \citenamefont {Govindarajan}, \citenamefont {de~Queiroz},\
  and\ \citenamefont {Reyes-Lega}}]{balachandranPRL}%
  \BibitemOpen
  \bibfield  {author} {\bibinfo {author} {\bibfnamefont {A.}~\bibnamefont
  {Balachandran}}, \bibinfo {author} {\bibfnamefont {T.}~\bibnamefont
  {Govindarajan}}, \bibinfo {author} {\bibfnamefont {A.~R.}\ \bibnamefont
  {de~Queiroz}}, \ and\ \bibinfo {author} {\bibfnamefont {A.}~\bibnamefont
  {Reyes-Lega}},\ }\href@noop {} {\bibfield  {journal} {\bibinfo  {journal}
  {Phys. Rev. Lett.}\ }\textbf {\bibinfo {volume} {110}},\ \bibinfo {pages}
  {080503} (\bibinfo {year} {2013})}\BibitemShut {NoStop}%
\bibitem [{\citenamefont {Killoran}\ \emph {et~al.}(2014)\citenamefont
  {Killoran}, \citenamefont {Cramer},\ and\ \citenamefont
  {Plenio}}]{plenio2014PRL}%
  \BibitemOpen
  \bibfield  {author} {\bibinfo {author} {\bibfnamefont {N.}~\bibnamefont
  {Killoran}}, \bibinfo {author} {\bibfnamefont {M.}~\bibnamefont {Cramer}}, \
  and\ \bibinfo {author} {\bibfnamefont {M.~B.}\ \bibnamefont {Plenio}},\
  }\href@noop {} {\bibfield  {journal} {\bibinfo  {journal} {Phys. Rev. Lett.}\
  }\textbf {\bibinfo {volume} {112}},\ \bibinfo {pages} {150501} (\bibinfo
  {year} {2014})}\BibitemShut {NoStop}%
\bibitem [{\citenamefont {Benatti}\ \emph
  {et~al.}(2014{\natexlab{a}})\citenamefont {Benatti}, \citenamefont
  {Floreanini},\ and\ \citenamefont {Titimbo}}]{benatti2014review}%
  \BibitemOpen
  \bibfield  {author} {\bibinfo {author} {\bibfnamefont {F.}~\bibnamefont
  {Benatti}}, \bibinfo {author} {\bibfnamefont {R.}~\bibnamefont {Floreanini}},
  \ and\ \bibinfo {author} {\bibfnamefont {K.}~\bibnamefont {Titimbo}},\
  }\href@noop {} {\bibfield  {journal} {\bibinfo  {journal} {Open Syst. Inf.
  Dyn.}\ }\textbf {\bibinfo {volume} {21}},\ \bibinfo {pages} {1440003}
  (\bibinfo {year} {2014}{\natexlab{a}})}\BibitemShut {NoStop}%
\bibitem [{\citenamefont {Benatti}\ \emph {et~al.}(2017)\citenamefont
  {Benatti}, \citenamefont {Floreanini}, \citenamefont {Franchini},\ and\
  \citenamefont {Marzolino}}]{benattiOSID2017}%
  \BibitemOpen
  \bibfield  {author} {\bibinfo {author} {\bibfnamefont {F.}~\bibnamefont
  {Benatti}}, \bibinfo {author} {\bibfnamefont {R.}~\bibnamefont {Floreanini}},
  \bibinfo {author} {\bibfnamefont {F.}~\bibnamefont {Franchini}}, \ and\
  \bibinfo {author} {\bibfnamefont {U.}~\bibnamefont {Marzolino}},\ }\href@noop
  {} {\bibfield  {journal} {\bibinfo  {journal} {Open Sys. Inform. Dyn.}\
  }\textbf {\bibinfo {volume} {24}},\ \bibinfo {pages} {1740004} (\bibinfo
  {year} {2017})}\BibitemShut {NoStop}%
\bibitem [{\citenamefont {{Lo Franco}}\ and\ \citenamefont
  {Compagno}(2016)}]{franco2016quantum}%
  \BibitemOpen
  \bibfield  {author} {\bibinfo {author} {\bibfnamefont {R.}~\bibnamefont {{Lo
  Franco}}}\ and\ \bibinfo {author} {\bibfnamefont {G.}~\bibnamefont
  {Compagno}},\ }\href@noop {} {\bibfield  {journal} {\bibinfo  {journal} {Sci.
  Rep.}\ }\textbf {\bibinfo {volume} {6}},\ \bibinfo {pages} {20603} (\bibinfo
  {year} {2016})}\BibitemShut {NoStop}%
\bibitem [{\citenamefont {Compagno}\ \emph {et~al.}(2018)\citenamefont
  {Compagno}, \citenamefont {Castellini},\ and\ \citenamefont {{Lo
  Franco}}}]{compagno2018dealing}%
  \BibitemOpen
  \bibfield  {author} {\bibinfo {author} {\bibfnamefont {G.}~\bibnamefont
  {Compagno}}, \bibinfo {author} {\bibfnamefont {A.}~\bibnamefont
  {Castellini}}, \ and\ \bibinfo {author} {\bibfnamefont {R.}~\bibnamefont {{Lo
  Franco}}},\ }\href@noop {} {\bibfield  {journal} {\bibinfo  {journal} {Phil.
  Trans. R. Soc. A}\ }\textbf {\bibinfo {volume} {376}},\ \bibinfo {pages}
  {20170317} (\bibinfo {year} {2018})}\BibitemShut {NoStop}%
\bibitem [{\citenamefont {Louren\ifmmode~\mbox{\c{c}}\else \c{c}\fi{}o}\ \emph
  {et~al.}(2019)\citenamefont {Louren\ifmmode~\mbox{\c{c}}\else \c{c}\fi{}o},
  \citenamefont {Debarba},\ and\ \citenamefont {Duzzioni}}]{duzzioniPRA}%
  \BibitemOpen
  \bibfield  {author} {\bibinfo {author} {\bibfnamefont {A.~C.}\ \bibnamefont
  {Louren\ifmmode~\mbox{\c{c}}\else \c{c}\fi{}o}}, \bibinfo {author}
  {\bibfnamefont {T.}~\bibnamefont {Debarba}}, \ and\ \bibinfo {author}
  {\bibfnamefont {E.~I.}\ \bibnamefont {Duzzioni}},\ }\href@noop {} {\bibfield
  {journal} {\bibinfo  {journal} {Phys. Rev. A}\ }\textbf {\bibinfo {volume}
  {99}},\ \bibinfo {pages} {012341} (\bibinfo {year} {2019})}\BibitemShut
  {NoStop}%
\bibitem [{\citenamefont {Buscemi}\ and\ \citenamefont
  {Bordone}(2011)}]{bordone2011}%
  \BibitemOpen
  \bibfield  {author} {\bibinfo {author} {\bibfnamefont {F.}~\bibnamefont
  {Buscemi}}\ and\ \bibinfo {author} {\bibfnamefont {P.}~\bibnamefont
  {Bordone}},\ }\href@noop {} {\bibfield  {journal} {\bibinfo  {journal} {Phys.
  Rev. A}\ }\textbf {\bibinfo {volume} {84}},\ \bibinfo {pages} {022303}
  (\bibinfo {year} {2011})}\BibitemShut {NoStop}%
\bibitem [{\citenamefont {Morris}\ \emph {et~al.}(2019)\citenamefont {Morris},
  \citenamefont {Yadin}, \citenamefont {Fadel}, \citenamefont {Zibold},
  \citenamefont {Treutlein},\ and\ \citenamefont {Adesso}}]{morris2019}%
  \BibitemOpen
  \bibfield  {author} {\bibinfo {author} {\bibfnamefont {B.}~\bibnamefont
  {Morris}}, \bibinfo {author} {\bibfnamefont {B.}~\bibnamefont {Yadin}},
  \bibinfo {author} {\bibfnamefont {M.}~\bibnamefont {Fadel}}, \bibinfo
  {author} {\bibfnamefont {T.}~\bibnamefont {Zibold}}, \bibinfo {author}
  {\bibfnamefont {P.}~\bibnamefont {Treutlein}}, \ and\ \bibinfo {author}
  {\bibfnamefont {G.}~\bibnamefont {Adesso}},\ }\href@noop {} {\enquote
  {\bibinfo {title} {Entanglement between identical particles is a useful and
  consistent resource},}\ } (\bibinfo {year} {2019}),\ \Eprint
  {http://arxiv.org/abs/1908.11735} {arXiv:1908.11735 [quant-ph]} \BibitemShut
  {NoStop}%
\bibitem [{\citenamefont {Benatti}\ \emph {et~al.}(2020)\citenamefont
  {Benatti}, \citenamefont {Floreanini}, \citenamefont {Franchini},\ and\
  \citenamefont {Marzolino}}]{benatti2020entanglement}%
  \BibitemOpen
  \bibfield  {author} {\bibinfo {author} {\bibfnamefont {F.}~\bibnamefont
  {Benatti}}, \bibinfo {author} {\bibfnamefont {R.}~\bibnamefont {Floreanini}},
  \bibinfo {author} {\bibfnamefont {F.}~\bibnamefont {Franchini}}, \ and\
  \bibinfo {author} {\bibfnamefont {U.}~\bibnamefont {Marzolino}},\ }\href@noop
  {} {\enquote {\bibinfo {title} {Entanglement in indistinguishable particle
  systems},}\ } (\bibinfo {year} {2020}),\ \Eprint
  {http://arxiv.org/abs/2007.06253} {arXiv:2007.06253 [quant-ph]} \BibitemShut
  {NoStop}%
\bibitem [{\citenamefont {Cohen-Tannoudji}\ \emph {et~al.}(2006)\citenamefont
  {Cohen-Tannoudji}, \citenamefont {Diu}, \citenamefont {Laloe},\ and\
  \citenamefont {Dui}}]{cohen2006quantum}%
  \BibitemOpen
  \bibfield  {author} {\bibinfo {author} {\bibfnamefont {C.}~\bibnamefont
  {Cohen-Tannoudji}}, \bibinfo {author} {\bibfnamefont {B.}~\bibnamefont
  {Diu}}, \bibinfo {author} {\bibfnamefont {F.}~\bibnamefont {Laloe}}, \ and\
  \bibinfo {author} {\bibfnamefont {B.}~\bibnamefont {Dui}},\ }\href@noop {}
  {\emph {\bibinfo {title} {Quantum Mechanics}}}\ (\bibinfo  {publisher}
  {Wiley-Interscience},\ \bibinfo {year} {2006})\BibitemShut {NoStop}%
\bibitem [{\citenamefont {Nosrati}\ \emph {et~al.}(2020)\citenamefont
  {Nosrati}, \citenamefont {Castellini}, \citenamefont {Compagno},\ and\
  \citenamefont {{Lo Franco}}}]{nosrati2019control}%
  \BibitemOpen
  \bibfield  {author} {\bibinfo {author} {\bibfnamefont {F.}~\bibnamefont
  {Nosrati}}, \bibinfo {author} {\bibfnamefont {A.}~\bibnamefont {Castellini}},
  \bibinfo {author} {\bibfnamefont {G.}~\bibnamefont {Compagno}}, \ and\
  \bibinfo {author} {\bibfnamefont {R.}~\bibnamefont {{Lo Franco}}},\
  }\href@noop {} {\bibfield  {journal} {\bibinfo  {journal} {npj Quantum
  Information}\ }\textbf {\bibinfo {volume} {6}},\ \bibinfo {pages} {39}
  (\bibinfo {year} {2020})}\BibitemShut {NoStop}%
\bibitem [{\citenamefont {Horodecki}\ \emph {et~al.}(2009)\citenamefont
  {Horodecki}, \citenamefont {Horodecki}, \citenamefont {Horodecki},\ and\
  \citenamefont {Horodecki}}]{horodecki2009quantum}%
  \BibitemOpen
  \bibfield  {author} {\bibinfo {author} {\bibfnamefont {R.}~\bibnamefont
  {Horodecki}}, \bibinfo {author} {\bibfnamefont {P.}~\bibnamefont
  {Horodecki}}, \bibinfo {author} {\bibfnamefont {M.}~\bibnamefont
  {Horodecki}}, \ and\ \bibinfo {author} {\bibfnamefont {K.}~\bibnamefont
  {Horodecki}},\ }\href@noop {} {\bibfield  {journal} {\bibinfo  {journal}
  {Rev. Mod. Phys.}\ }\textbf {\bibinfo {volume} {81}},\ \bibinfo {pages} {865}
  (\bibinfo {year} {2009})}\BibitemShut {NoStop}%
\bibitem [{\citenamefont {Chitambar}\ \emph {et~al.}(2014)\citenamefont
  {Chitambar}, \citenamefont {Leung}, \citenamefont {Man{\v{c}}inska},
  \citenamefont {Ozols},\ and\ \citenamefont
  {Winter}}]{chitambar2014everything}%
  \BibitemOpen
  \bibfield  {author} {\bibinfo {author} {\bibfnamefont {E.}~\bibnamefont
  {Chitambar}}, \bibinfo {author} {\bibfnamefont {D.}~\bibnamefont {Leung}},
  \bibinfo {author} {\bibfnamefont {L.}~\bibnamefont {Man{\v{c}}inska}},
  \bibinfo {author} {\bibfnamefont {M.}~\bibnamefont {Ozols}}, \ and\ \bibinfo
  {author} {\bibfnamefont {A.}~\bibnamefont {Winter}},\ }\href@noop {}
  {\bibfield  {journal} {\bibinfo  {journal} {Comm. Math. Phys.}\ }\textbf
  {\bibinfo {volume} {328}},\ \bibinfo {pages} {303} (\bibinfo {year}
  {2014})}\BibitemShut {NoStop}%
\bibitem [{\citenamefont {Nielsen}\ and\ \citenamefont
  {Chuang}(2010)}]{nielsen2010quantum}%
  \BibitemOpen
  \bibfield  {author} {\bibinfo {author} {\bibfnamefont {M.~A.}\ \bibnamefont
  {Nielsen}}\ and\ \bibinfo {author} {\bibfnamefont {I.~L.}\ \bibnamefont
  {Chuang}},\ }\href@noop {} {\emph {\bibinfo {title} {Quantum Computation and
  Quantum Information}}}\ (\bibinfo  {publisher} {Cambridge university press},\
  \bibinfo {year} {2010})\BibitemShut {NoStop}%
\bibitem [{\citenamefont {Breuer}\ \emph {et~al.}(2002)\citenamefont {Breuer},
  \citenamefont {Petruccione} \emph {et~al.}}]{breuer2002theory}%
  \BibitemOpen
  \bibfield  {author} {\bibinfo {author} {\bibfnamefont {H.-P.}\ \bibnamefont
  {Breuer}}, \bibinfo {author} {\bibfnamefont {F.}~\bibnamefont {Petruccione}},
   \emph {et~al.},\ }\href@noop {} {\emph {\bibinfo {title} {The theory of open
  quantum systems}}}\ (\bibinfo  {publisher} {Oxford University Press on
  Demand},\ \bibinfo {year} {2002})\BibitemShut {NoStop}%
\bibitem [{\citenamefont {Aolita}\ \emph {et~al.}(2015)\citenamefont {Aolita},
  \citenamefont {{de Melo}},\ and\ \citenamefont {Davidovich}}]{aolitareview}%
  \BibitemOpen
  \bibfield  {author} {\bibinfo {author} {\bibfnamefont {L.}~\bibnamefont
  {Aolita}}, \bibinfo {author} {\bibfnamefont {F.}~\bibnamefont {{de Melo}}}, \
  and\ \bibinfo {author} {\bibfnamefont {L.}~\bibnamefont {Davidovich}},\
  }\href@noop {} {\bibfield  {journal} {\bibinfo  {journal} {Rep. Prog. Phys.}\
  }\textbf {\bibinfo {volume} {78}},\ \bibinfo {pages} {042001} (\bibinfo
  {year} {2015})}\BibitemShut {NoStop}%
\bibitem [{\citenamefont {{Lo Franco}}\ and\ \citenamefont
  {Compagno}(2018)}]{franco2018indistinguishability}%
  \BibitemOpen
  \bibfield  {author} {\bibinfo {author} {\bibfnamefont {R.}~\bibnamefont {{Lo
  Franco}}}\ and\ \bibinfo {author} {\bibfnamefont {G.}~\bibnamefont
  {Compagno}},\ }\href@noop {} {\bibfield  {journal} {\bibinfo  {journal}
  {Phys. Rev. Lett.}\ }\textbf {\bibinfo {volume} {120}},\ \bibinfo {pages}
  {240403} (\bibinfo {year} {2018})}\BibitemShut {NoStop}%
\bibitem [{\citenamefont {Sun}\ \emph {et~al.}(2020)\citenamefont {Sun},
  \citenamefont {Wang}, \citenamefont {Liu}, \citenamefont {Xu}, \citenamefont
  {Xu}, \citenamefont {Li}, \citenamefont {Guo}, \citenamefont {Castellini},
  \citenamefont {Nosrati}, \citenamefont {Compagno},\ and\ \citenamefont {{Lo
  Franco}}}]{sunetalexp}%
  \BibitemOpen
  \bibfield  {author} {\bibinfo {author} {\bibfnamefont {K.}~\bibnamefont
  {Sun}}, \bibinfo {author} {\bibfnamefont {Y.}~\bibnamefont {Wang}}, \bibinfo
  {author} {\bibfnamefont {Z.-H.}\ \bibnamefont {Liu}}, \bibinfo {author}
  {\bibfnamefont {X.-Y.}\ \bibnamefont {Xu}}, \bibinfo {author} {\bibfnamefont
  {J.-S.}\ \bibnamefont {Xu}}, \bibinfo {author} {\bibfnamefont {C.-F.}\
  \bibnamefont {Li}}, \bibinfo {author} {\bibfnamefont {G.-C.}\ \bibnamefont
  {Guo}}, \bibinfo {author} {\bibfnamefont {A.}~\bibnamefont {Castellini}},
  \bibinfo {author} {\bibfnamefont {F.}~\bibnamefont {Nosrati}}, \bibinfo
  {author} {\bibfnamefont {G.}~\bibnamefont {Compagno}}, \ and\ \bibinfo
  {author} {\bibfnamefont {R.}~\bibnamefont {{Lo Franco}}},\ }\href@noop {}
  {\enquote {\bibinfo {title} {Experimental control of remote spatial
  indistinguishability of photons to realize entanglement and teleportation},}\
  } (\bibinfo {year} {2020}),\ \Eprint {http://arxiv.org/abs/2003.10659}
  {arXiv:2003.10659 [quant-ph]} \BibitemShut {NoStop}%
\bibitem [{\citenamefont {Barros}\ \emph {et~al.}(2019)\citenamefont {Barros},
  \citenamefont {Chin}, \citenamefont {Pramanik}, \citenamefont {Lim},
  \citenamefont {Cho}, \citenamefont {Huh},\ and\ \citenamefont
  {Kim}}]{barros2019entangling}%
  \BibitemOpen
  \bibfield  {author} {\bibinfo {author} {\bibfnamefont {M.~R.}\ \bibnamefont
  {Barros}}, \bibinfo {author} {\bibfnamefont {S.}~\bibnamefont {Chin}},
  \bibinfo {author} {\bibfnamefont {T.}~\bibnamefont {Pramanik}}, \bibinfo
  {author} {\bibfnamefont {H.-T.}\ \bibnamefont {Lim}}, \bibinfo {author}
  {\bibfnamefont {Y.-W.}\ \bibnamefont {Cho}}, \bibinfo {author} {\bibfnamefont
  {J.}~\bibnamefont {Huh}}, \ and\ \bibinfo {author} {\bibfnamefont {Y.-S.}\
  \bibnamefont {Kim}},\ }\href@noop {} {\enquote {\bibinfo {title} {Entangling
  bosons through particle indistinguishability and spatial overlap},}\ }
  (\bibinfo {year} {2019}),\ \Eprint {http://arxiv.org/abs/1912.04208}
  {arXiv:1912.04208 [quant-ph]} \BibitemShut {NoStop}%
\bibitem [{\citenamefont {Chin}\ and\ \citenamefont
  {Huh}(2019{\natexlab{a}})}]{chin2019entanglement}%
  \BibitemOpen
  \bibfield  {author} {\bibinfo {author} {\bibfnamefont {S.}~\bibnamefont
  {Chin}}\ and\ \bibinfo {author} {\bibfnamefont {J.}~\bibnamefont {Huh}},\
  }\href@noop {} {\bibfield  {journal} {\bibinfo  {journal} {Phys. Rev. A}\
  }\textbf {\bibinfo {volume} {99}},\ \bibinfo {pages} {052345} (\bibinfo
  {year} {2019}{\natexlab{a}})}\BibitemShut {NoStop}%
\bibitem [{\citenamefont {Louren{\c{c}}o}\ \emph {et~al.}(2019)\citenamefont
  {Louren{\c{c}}o}, \citenamefont {Debarba},\ and\ \citenamefont
  {Duzzioni}}]{lourencco2019entanglement}%
  \BibitemOpen
  \bibfield  {author} {\bibinfo {author} {\bibfnamefont {A.~C.}\ \bibnamefont
  {Louren{\c{c}}o}}, \bibinfo {author} {\bibfnamefont {T.}~\bibnamefont
  {Debarba}}, \ and\ \bibinfo {author} {\bibfnamefont {E.~I.}\ \bibnamefont
  {Duzzioni}},\ }\href@noop {} {\bibfield  {journal} {\bibinfo  {journal}
  {Phys. Rev. A}\ }\textbf {\bibinfo {volume} {99}},\ \bibinfo {pages} {012341}
  (\bibinfo {year} {2019})}\BibitemShut {NoStop}%
\bibitem [{\citenamefont {Chin}\ and\ \citenamefont
  {Huh}(2019{\natexlab{b}})}]{chin2019reduced}%
  \BibitemOpen
  \bibfield  {author} {\bibinfo {author} {\bibfnamefont {S.}~\bibnamefont
  {Chin}}\ and\ \bibinfo {author} {\bibfnamefont {J.}~\bibnamefont {Huh}},\
  }\href@noop {} {\enquote {\bibinfo {title} {Reduced density matrix of
  nonlocal identical particles},}\ } (\bibinfo {year} {2019}{\natexlab{b}}),\
  \Eprint {http://arxiv.org/abs/1906.00542} {arXiv:1906.00542 [quant-ph]}
  \BibitemShut {NoStop}%
\bibitem [{\citenamefont {Qureshi}\ and\ \citenamefont
  {Rizwan}(2017)}]{Quanta66}%
  \BibitemOpen
  \bibfield  {author} {\bibinfo {author} {\bibfnamefont {T.}~\bibnamefont
  {Qureshi}}\ and\ \bibinfo {author} {\bibfnamefont {U.}~\bibnamefont
  {Rizwan}},\ }\href@noop {} {\bibfield  {journal} {\bibinfo  {journal}
  {Quanta}\ }\textbf {\bibinfo {volume} {6}},\ \bibinfo {pages} {61} (\bibinfo
  {year} {2017})}\BibitemShut {NoStop}%
\bibitem [{\citenamefont {Mani}\ \emph {et~al.}(2020)\citenamefont {Mani},
  \citenamefont {Ramadas},\ and\ \citenamefont {Sreedhar}}]{mani2020quantum}%
  \BibitemOpen
  \bibfield  {author} {\bibinfo {author} {\bibfnamefont {H.}~\bibnamefont
  {Mani}}, \bibinfo {author} {\bibfnamefont {N.}~\bibnamefont {Ramadas}}, \
  and\ \bibinfo {author} {\bibfnamefont {V.}~\bibnamefont {Sreedhar}},\
  }\href@noop {} {\bibfield  {journal} {\bibinfo  {journal} {Phys. Rev. A}\
  }\textbf {\bibinfo {volume} {101}},\ \bibinfo {pages} {022314} (\bibinfo
  {year} {2020})}\BibitemShut {NoStop}%
\bibitem [{\citenamefont {Omar}\ \emph {et~al.}(2002)\citenamefont {Omar},
  \citenamefont {Paunkovi\ifmmode~\acute{c}\else \'{c}\fi{}}, \citenamefont
  {Bose},\ and\ \citenamefont {Vedral}}]{PhysRevA.65.062305}%
  \BibitemOpen
  \bibfield  {author} {\bibinfo {author} {\bibfnamefont {Y.}~\bibnamefont
  {Omar}}, \bibinfo {author} {\bibfnamefont {N.}~\bibnamefont
  {Paunkovi\ifmmode~\acute{c}\else \'{c}\fi{}}}, \bibinfo {author}
  {\bibfnamefont {S.}~\bibnamefont {Bose}}, \ and\ \bibinfo {author}
  {\bibfnamefont {V.}~\bibnamefont {Vedral}},\ }\href@noop {} {\bibfield
  {journal} {\bibinfo  {journal} {Phys. Rev. A}\ }\textbf {\bibinfo {volume}
  {65}},\ \bibinfo {pages} {062305} (\bibinfo {year} {2002})}\BibitemShut
  {NoStop}%
\bibitem [{\citenamefont {Bose}\ \emph {et~al.}(2003)\citenamefont {Bose},
  \citenamefont {Ekert}, \citenamefont {Omar}, \citenamefont
  {Paunkovi\ifmmode~\acute{c}\else \'{c}\fi{}},\ and\ \citenamefont
  {Vedral}}]{PhysRevA.68.052309}%
  \BibitemOpen
  \bibfield  {author} {\bibinfo {author} {\bibfnamefont {S.}~\bibnamefont
  {Bose}}, \bibinfo {author} {\bibfnamefont {A.}~\bibnamefont {Ekert}},
  \bibinfo {author} {\bibfnamefont {Y.}~\bibnamefont {Omar}}, \bibinfo {author}
  {\bibfnamefont {N.}~\bibnamefont {Paunkovi\ifmmode~\acute{c}\else
  \'{c}\fi{}}}, \ and\ \bibinfo {author} {\bibfnamefont {V.}~\bibnamefont
  {Vedral}},\ }\href@noop {} {\bibfield  {journal} {\bibinfo  {journal} {Phys.
  Rev. A}\ }\textbf {\bibinfo {volume} {68}},\ \bibinfo {pages} {052309}
  (\bibinfo {year} {2003})}\BibitemShut {NoStop}%
\bibitem [{\citenamefont {Castellini}\ \emph
  {et~al.}(2019{\natexlab{a}})\citenamefont {Castellini}, \citenamefont
  {Bellomo}, \citenamefont {Compagno},\ and\ \citenamefont {{Lo
  Franco}}}]{castellini2019activating}%
  \BibitemOpen
  \bibfield  {author} {\bibinfo {author} {\bibfnamefont {A.}~\bibnamefont
  {Castellini}}, \bibinfo {author} {\bibfnamefont {B.}~\bibnamefont {Bellomo}},
  \bibinfo {author} {\bibfnamefont {G.}~\bibnamefont {Compagno}}, \ and\
  \bibinfo {author} {\bibfnamefont {R.}~\bibnamefont {{Lo Franco}}},\
  }\href@noop {} {\bibfield  {journal} {\bibinfo  {journal} {Phys. Rev. A}\
  }\textbf {\bibinfo {volume} {99}},\ \bibinfo {pages} {062322} (\bibinfo
  {year} {2019}{\natexlab{a}})}\BibitemShut {NoStop}%
\bibitem [{\citenamefont {Benatti}\ \emph
  {et~al.}(2014{\natexlab{b}})\citenamefont {Benatti}, \citenamefont
  {Alipour},\ and\ \citenamefont {Rezakhani}}]{benatti2014dissipative}%
  \BibitemOpen
  \bibfield  {author} {\bibinfo {author} {\bibfnamefont {F.}~\bibnamefont
  {Benatti}}, \bibinfo {author} {\bibfnamefont {S.}~\bibnamefont {Alipour}}, \
  and\ \bibinfo {author} {\bibfnamefont {A.}~\bibnamefont {Rezakhani}},\
  }\href@noop {} {\bibfield  {journal} {\bibinfo  {journal} {New J. Phys.}\
  }\textbf {\bibinfo {volume} {16}},\ \bibinfo {pages} {015023} (\bibinfo
  {year} {2014}{\natexlab{b}})}\BibitemShut {NoStop}%
\bibitem [{\citenamefont {Castellini}\ \emph
  {et~al.}(2019{\natexlab{b}})\citenamefont {Castellini}, \citenamefont {{Lo
  Franco}}, \citenamefont {Lami}, \citenamefont {Winter}, \citenamefont
  {Adesso},\ and\ \citenamefont {Compagno}}]{PhysRevA.100.012308}%
  \BibitemOpen
  \bibfield  {author} {\bibinfo {author} {\bibfnamefont {A.}~\bibnamefont
  {Castellini}}, \bibinfo {author} {\bibfnamefont {R.}~\bibnamefont {{Lo
  Franco}}}, \bibinfo {author} {\bibfnamefont {L.}~\bibnamefont {Lami}},
  \bibinfo {author} {\bibfnamefont {A.}~\bibnamefont {Winter}}, \bibinfo
  {author} {\bibfnamefont {G.}~\bibnamefont {Adesso}}, \ and\ \bibinfo {author}
  {\bibfnamefont {G.}~\bibnamefont {Compagno}},\ }\href@noop {} {\bibfield
  {journal} {\bibinfo  {journal} {Phys. Rev. A}\ }\textbf {\bibinfo {volume}
  {100}},\ \bibinfo {pages} {012308} (\bibinfo {year}
  {2019}{\natexlab{b}})}\BibitemShut {NoStop}%
\bibitem [{\citenamefont {Argentieri}\ \emph {et~al.}(2011)\citenamefont
  {Argentieri}, \citenamefont {Benatti}, \citenamefont {Floreanini},\ and\
  \citenamefont {Marzolino}}]{argentieri2011}%
  \BibitemOpen
  \bibfield  {author} {\bibinfo {author} {\bibfnamefont {G.}~\bibnamefont
  {Argentieri}}, \bibinfo {author} {\bibfnamefont {F.}~\bibnamefont {Benatti}},
  \bibinfo {author} {\bibfnamefont {R.}~\bibnamefont {Floreanini}}, \ and\
  \bibinfo {author} {\bibfnamefont {U.}~\bibnamefont {Marzolino}},\ }\href@noop
  {} {\bibfield  {journal} {\bibinfo  {journal} {Int. J. Quant. Inf.}\ }\textbf
  {\bibinfo {volume} {9}},\ \bibinfo {pages} {1745} (\bibinfo {year}
  {2011})}\BibitemShut {NoStop}%
\bibitem [{\citenamefont {Marzolino}(2013)}]{Marzolino2013}%
  \BibitemOpen
  \bibfield  {author} {\bibinfo {author} {\bibfnamefont {U.}~\bibnamefont
  {Marzolino}},\ }\href@noop {} {\bibfield  {journal} {\bibinfo  {journal}
  {{EPL} (Europhysics Letters)}\ }\textbf {\bibinfo {volume} {104}},\ \bibinfo
  {pages} {40004} (\bibinfo {year} {2013})}\BibitemShut {NoStop}%
\bibitem [{\citenamefont {Beggi}\ \emph {et~al.}(2016)\citenamefont {Beggi},
  \citenamefont {Buscemi},\ and\ \citenamefont {Bordone}}]{bordone2016}%
  \BibitemOpen
  \bibfield  {author} {\bibinfo {author} {\bibfnamefont {A.}~\bibnamefont
  {Beggi}}, \bibinfo {author} {\bibfnamefont {F.}~\bibnamefont {Buscemi}}, \
  and\ \bibinfo {author} {\bibfnamefont {P.}~\bibnamefont {Bordone}},\
  }\href@noop {} {\bibfield  {journal} {\bibinfo  {journal} {Quantum Inf.
  Process.}\ }\textbf {\bibinfo {volume} {15}},\ \bibinfo {pages} {3711}
  (\bibinfo {year} {2016})}\BibitemShut {NoStop}%
\bibitem [{\citenamefont {Zurek}(2003)}]{Zurek2003}%
  \BibitemOpen
  \bibfield  {author} {\bibinfo {author} {\bibfnamefont {W.~H.}\ \bibnamefont
  {Zurek}},\ }\href@noop {} {\bibfield  {journal} {\bibinfo  {journal} {Rev.
  Mod. Phys.}\ }\textbf {\bibinfo {volume} {75}},\ \bibinfo {pages} {715}
  (\bibinfo {year} {2003})}\BibitemShut {NoStop}%
\bibitem [{\citenamefont {Peres}(1984)}]{peres1984stability}%
  \BibitemOpen
  \bibfield  {author} {\bibinfo {author} {\bibfnamefont {A.}~\bibnamefont
  {Peres}},\ }\href@noop {} {\bibfield  {journal} {\bibinfo  {journal} {Phys.
  Rev. A}\ }\textbf {\bibinfo {volume} {30}},\ \bibinfo {pages} {1610}
  (\bibinfo {year} {1984})}\BibitemShut {NoStop}%
\bibitem [{\citenamefont {Jalabert}\ and\ \citenamefont
  {Pastawski}(2001)}]{PhysRevLett.86.2490}%
  \BibitemOpen
  \bibfield  {author} {\bibinfo {author} {\bibfnamefont {R.~A.}\ \bibnamefont
  {Jalabert}}\ and\ \bibinfo {author} {\bibfnamefont {H.~M.}\ \bibnamefont
  {Pastawski}},\ }\href@noop {} {\bibfield  {journal} {\bibinfo  {journal}
  {Phys. Rev. Lett.}\ }\textbf {\bibinfo {volume} {86}},\ \bibinfo {pages}
  {2490} (\bibinfo {year} {2001})}\BibitemShut {NoStop}%
\bibitem [{\citenamefont {Yu}\ and\ \citenamefont
  {Eberly}(2004)}]{yu2004finite}%
  \BibitemOpen
  \bibfield  {author} {\bibinfo {author} {\bibfnamefont {T.}~\bibnamefont
  {Yu}}\ and\ \bibinfo {author} {\bibfnamefont {J.}~\bibnamefont {Eberly}},\
  }\href@noop {} {\bibfield  {journal} {\bibinfo  {journal} {Phys. Rev. Lett.}\
  }\textbf {\bibinfo {volume} {93}},\ \bibinfo {pages} {140404} (\bibinfo
  {year} {2004})}\BibitemShut {NoStop}%
\bibitem [{\citenamefont {Yu}\ and\ \citenamefont {Eberly}(2009)}]{Yu598}%
  \BibitemOpen
  \bibfield  {author} {\bibinfo {author} {\bibfnamefont {T.}~\bibnamefont
  {Yu}}\ and\ \bibinfo {author} {\bibfnamefont {J.~H.}\ \bibnamefont
  {Eberly}},\ }\href@noop {} {\bibfield  {journal} {\bibinfo  {journal}
  {Science}\ }\textbf {\bibinfo {volume} {323}},\ \bibinfo {pages} {598}
  (\bibinfo {year} {2009})}\BibitemShut {NoStop}%
\bibitem [{\citenamefont {Almeida}\ \emph {et~al.}(2007)\citenamefont
  {Almeida}, \citenamefont {de~Melo}, \citenamefont {Hor-Meyll}, \citenamefont
  {Salles}, \citenamefont {Walborn}, \citenamefont {Ribeiro},\ and\
  \citenamefont {Davidovich}}]{almeida2007environment}%
  \BibitemOpen
  \bibfield  {author} {\bibinfo {author} {\bibfnamefont {M.~P.}\ \bibnamefont
  {Almeida}}, \bibinfo {author} {\bibfnamefont {F.}~\bibnamefont {de~Melo}},
  \bibinfo {author} {\bibfnamefont {M.}~\bibnamefont {Hor-Meyll}}, \bibinfo
  {author} {\bibfnamefont {A.}~\bibnamefont {Salles}}, \bibinfo {author}
  {\bibfnamefont {S.}~\bibnamefont {Walborn}}, \bibinfo {author} {\bibfnamefont
  {P.~S.}\ \bibnamefont {Ribeiro}}, \ and\ \bibinfo {author} {\bibfnamefont
  {L.}~\bibnamefont {Davidovich}},\ }\href@noop {} {\bibfield  {journal}
  {\bibinfo  {journal} {Science}\ }\textbf {\bibinfo {volume} {316}},\ \bibinfo
  {pages} {579} (\bibinfo {year} {2007})}\BibitemShut {NoStop}%
\bibitem [{\citenamefont {Laurat}\ \emph {et~al.}(2007)\citenamefont {Laurat},
  \citenamefont {Choi}, \citenamefont {Deng}, \citenamefont {Chou},\ and\
  \citenamefont {Kimble}}]{laurat2007heralded}%
  \BibitemOpen
  \bibfield  {author} {\bibinfo {author} {\bibfnamefont {J.}~\bibnamefont
  {Laurat}}, \bibinfo {author} {\bibfnamefont {K.}~\bibnamefont {Choi}},
  \bibinfo {author} {\bibfnamefont {H.}~\bibnamefont {Deng}}, \bibinfo {author}
  {\bibfnamefont {C.}~\bibnamefont {Chou}}, \ and\ \bibinfo {author}
  {\bibfnamefont {H.}~\bibnamefont {Kimble}},\ }\href@noop {} {\bibfield
  {journal} {\bibinfo  {journal} {Phys. Rev. Lett.}\ }\textbf {\bibinfo
  {volume} {99}},\ \bibinfo {pages} {180504} (\bibinfo {year}
  {2007})}\BibitemShut {NoStop}%
\bibitem [{\citenamefont {Bellomo}\ \emph {et~al.}(2007)\citenamefont
  {Bellomo}, \citenamefont {{Lo Franco}},\ and\ \citenamefont
  {Compagno}}]{bellomo2007non}%
  \BibitemOpen
  \bibfield  {author} {\bibinfo {author} {\bibfnamefont {B.}~\bibnamefont
  {Bellomo}}, \bibinfo {author} {\bibfnamefont {R.}~\bibnamefont {{Lo
  Franco}}}, \ and\ \bibinfo {author} {\bibfnamefont {G.}~\bibnamefont
  {Compagno}},\ }\href@noop {} {\bibfield  {journal} {\bibinfo  {journal}
  {Phys. Rev. Lett.}\ }\textbf {\bibinfo {volume} {99}},\ \bibinfo {pages}
  {160502} (\bibinfo {year} {2007})}\BibitemShut {NoStop}%
\bibitem [{\citenamefont {Bellomo}\ \emph
  {et~al.}(2008{\natexlab{a}})\citenamefont {Bellomo}, \citenamefont {{Lo
  Franco}},\ and\ \citenamefont {Compagno}}]{bellomo2008entanglement}%
  \BibitemOpen
  \bibfield  {author} {\bibinfo {author} {\bibfnamefont {B.}~\bibnamefont
  {Bellomo}}, \bibinfo {author} {\bibfnamefont {R.}~\bibnamefont {{Lo
  Franco}}}, \ and\ \bibinfo {author} {\bibfnamefont {G.}~\bibnamefont
  {Compagno}},\ }\href@noop {} {\bibfield  {journal} {\bibinfo  {journal}
  {Phys. Rev. A}\ }\textbf {\bibinfo {volume} {77}},\ \bibinfo {pages} {032342}
  (\bibinfo {year} {2008}{\natexlab{a}})}\BibitemShut {NoStop}%
\bibitem [{\citenamefont {{Lo Franco}}\ \emph {et~al.}(2013)\citenamefont {{Lo
  Franco}}, \citenamefont {Bellomo}, \citenamefont {Maniscalco},\ and\
  \citenamefont {Compagno}}]{lofrancoreview}%
  \BibitemOpen
  \bibfield  {author} {\bibinfo {author} {\bibfnamefont {R.}~\bibnamefont {{Lo
  Franco}}}, \bibinfo {author} {\bibfnamefont {B.}~\bibnamefont {Bellomo}},
  \bibinfo {author} {\bibfnamefont {S.}~\bibnamefont {Maniscalco}}, \ and\
  \bibinfo {author} {\bibfnamefont {G.}~\bibnamefont {Compagno}},\ }\href@noop
  {} {\bibfield  {journal} {\bibinfo  {journal} {Int. J. Mod. Phys. B}\
  }\textbf {\bibinfo {volume} {27}},\ \bibinfo {pages} {1345053} (\bibinfo
  {year} {2013})}\BibitemShut {NoStop}%
\bibitem [{\citenamefont {{Lo Franco}}\ and\ \citenamefont
  {Compagno}(2017)}]{lofrancoClassical}%
  \BibitemOpen
  \bibfield  {author} {\bibinfo {author} {\bibfnamefont {R.}~\bibnamefont {{Lo
  Franco}}}\ and\ \bibinfo {author} {\bibfnamefont {G.}~\bibnamefont
  {Compagno}},\ }\enquote {\bibinfo {title} {Overview on the phenomenon of
  two-qubit entanglement revivals in classical environments},}\ in\ \href@noop
  {} {\emph {\bibinfo {booktitle} {Lectures on General Quantum Correlations and
  their Applications. Quantum Science and Technology}}},\ \bibinfo {editor}
  {edited by\ \bibinfo {editor} {\bibfnamefont {F.}~\bibnamefont {Fanchini}},
  \bibinfo {editor} {\bibfnamefont {D.}~\bibnamefont {{Soares-Pinto}}}, \ and\
  \bibinfo {editor} {\bibfnamefont {G.}~\bibnamefont {Adesso}}}\ (\bibinfo
  {publisher} {Springer, Cham},\ \bibinfo {year} {2017})\ pp.\ \bibinfo {pages}
  {367--391}\BibitemShut {NoStop}%
\bibitem [{\citenamefont {{D'Arrigo}}\ \emph {et~al.}(2014)\citenamefont
  {{D'Arrigo}}, \citenamefont {{Lo Franco}}, \citenamefont {Benenti},
  \citenamefont {Paladino},\ and\ \citenamefont {Falci}}]{darrigo2012AOP}%
  \BibitemOpen
  \bibfield  {author} {\bibinfo {author} {\bibfnamefont {A.}~\bibnamefont
  {{D'Arrigo}}}, \bibinfo {author} {\bibfnamefont {R.}~\bibnamefont {{Lo
  Franco}}}, \bibinfo {author} {\bibfnamefont {G.}~\bibnamefont {Benenti}},
  \bibinfo {author} {\bibfnamefont {E.}~\bibnamefont {Paladino}}, \ and\
  \bibinfo {author} {\bibfnamefont {G.}~\bibnamefont {Falci}},\ }\href@noop {}
  {\bibfield  {journal} {\bibinfo  {journal} {Ann. Phys.}\ }\textbf {\bibinfo
  {volume} {350}},\ \bibinfo {pages} {211} (\bibinfo {year}
  {2014})}\BibitemShut {NoStop}%
\bibitem [{\citenamefont {Orieux}\ \emph
  {et~al.}(2015{\natexlab{a}})\citenamefont {Orieux}, \citenamefont {Ferranti},
  \citenamefont {D'Arrigo}, \citenamefont {{Lo Franco}}, \citenamefont
  {Benenti}, \citenamefont {Paladino}, \citenamefont {Falci}, \citenamefont
  {Sciarrino},\ and\ \citenamefont {Mataloni}}]{adeline2014}%
  \BibitemOpen
  \bibfield  {author} {\bibinfo {author} {\bibfnamefont {A.}~\bibnamefont
  {Orieux}}, \bibinfo {author} {\bibfnamefont {G.}~\bibnamefont {Ferranti}},
  \bibinfo {author} {\bibfnamefont {A.}~\bibnamefont {D'Arrigo}}, \bibinfo
  {author} {\bibfnamefont {R.}~\bibnamefont {{Lo Franco}}}, \bibinfo {author}
  {\bibfnamefont {G.}~\bibnamefont {Benenti}}, \bibinfo {author} {\bibfnamefont
  {E.}~\bibnamefont {Paladino}}, \bibinfo {author} {\bibfnamefont
  {G.}~\bibnamefont {Falci}}, \bibinfo {author} {\bibfnamefont
  {F.}~\bibnamefont {Sciarrino}}, \ and\ \bibinfo {author} {\bibfnamefont
  {P.}~\bibnamefont {Mataloni}},\ }\href@noop {} {\bibfield  {journal}
  {\bibinfo  {journal} {Sci. Rep.}\ }\textbf {\bibinfo {volume} {5}},\ \bibinfo
  {pages} {8575} (\bibinfo {year} {2015}{\natexlab{a}})}\BibitemShut {NoStop}%
\bibitem [{\citenamefont {{Lo Franco}}\ \emph
  {et~al.}(2012{\natexlab{a}})\citenamefont {{Lo Franco}}, \citenamefont
  {Bellomo}, \citenamefont {Andersson},\ and\ \citenamefont
  {Compagno}}]{lofranco2012PRA}%
  \BibitemOpen
  \bibfield  {author} {\bibinfo {author} {\bibfnamefont {R.}~\bibnamefont {{Lo
  Franco}}}, \bibinfo {author} {\bibfnamefont {B.}~\bibnamefont {Bellomo}},
  \bibinfo {author} {\bibfnamefont {E.}~\bibnamefont {Andersson}}, \ and\
  \bibinfo {author} {\bibfnamefont {G.}~\bibnamefont {Compagno}},\ }\href@noop
  {} {\bibfield  {journal} {\bibinfo  {journal} {Phys. Rev. A}\ }\textbf
  {\bibinfo {volume} {85}},\ \bibinfo {pages} {032318} (\bibinfo {year}
  {2012}{\natexlab{a}})}\BibitemShut {NoStop}%
\bibitem [{\citenamefont {Xu}\ \emph {et~al.}(2013)\citenamefont {Xu},
  \citenamefont {S.}, \citenamefont {Li}, \citenamefont {Xu}, \citenamefont
  {Guo}, \citenamefont {Andersson}, \citenamefont {{Lo Franco}},\ and\
  \citenamefont {Compagno}}]{LoFrancoNatCom}%
  \BibitemOpen
  \bibfield  {author} {\bibinfo {author} {\bibfnamefont {J.-S.}\ \bibnamefont
  {Xu}}, \bibinfo {author} {\bibfnamefont {K.}~\bibnamefont {S.}}, \bibinfo
  {author} {\bibfnamefont {C.-F.}\ \bibnamefont {Li}}, \bibinfo {author}
  {\bibfnamefont {X.-Y.}\ \bibnamefont {Xu}}, \bibinfo {author} {\bibfnamefont
  {G.-C.}\ \bibnamefont {Guo}}, \bibinfo {author} {\bibfnamefont
  {E.}~\bibnamefont {Andersson}}, \bibinfo {author} {\bibfnamefont
  {R.}~\bibnamefont {{Lo Franco}}}, \ and\ \bibinfo {author} {\bibfnamefont
  {G.}~\bibnamefont {Compagno}},\ }\href@noop {} {\bibfield  {journal}
  {\bibinfo  {journal} {Nat. Comm.}\ }\textbf {\bibinfo {volume} {4}},\
  \bibinfo {pages} {2851} (\bibinfo {year} {2013})}\BibitemShut {NoStop}%
\bibitem [{\citenamefont {Trapani}\ \emph {et~al.}(2015)\citenamefont
  {Trapani}, \citenamefont {Bina}, \citenamefont {Maniscalco},\ and\
  \citenamefont {Paris}}]{trapaniPRA}%
  \BibitemOpen
  \bibfield  {author} {\bibinfo {author} {\bibfnamefont {J.}~\bibnamefont
  {Trapani}}, \bibinfo {author} {\bibfnamefont {M.}~\bibnamefont {Bina}},
  \bibinfo {author} {\bibfnamefont {S.}~\bibnamefont {Maniscalco}}, \ and\
  \bibinfo {author} {\bibfnamefont {M.~G.~A.}\ \bibnamefont {Paris}},\
  }\href@noop {} {\bibfield  {journal} {\bibinfo  {journal} {Phys. Rev. A}\
  }\textbf {\bibinfo {volume} {91}},\ \bibinfo {pages} {022113} (\bibinfo
  {year} {2015})}\BibitemShut {NoStop}%
\bibitem [{\citenamefont {de~Vega}\ and\ \citenamefont
  {Alonso}(2017)}]{devegaRMP}%
  \BibitemOpen
  \bibfield  {author} {\bibinfo {author} {\bibfnamefont {I.}~\bibnamefont
  {de~Vega}}\ and\ \bibinfo {author} {\bibfnamefont {D.}~\bibnamefont
  {Alonso}},\ }\href@noop {} {\bibfield  {journal} {\bibinfo  {journal} {Rev.
  Mod. Phys.}\ }\textbf {\bibinfo {volume} {89}},\ \bibinfo {pages} {015001}
  (\bibinfo {year} {2017})}\BibitemShut {NoStop}%
\bibitem [{\citenamefont {Xu}\ \emph {et~al.}(2010)\citenamefont {Xu},
  \citenamefont {Li}, \citenamefont {Gong}, \citenamefont {Zou}, \citenamefont
  {Shi}, \citenamefont {Chen},\ and\ \citenamefont {Guo}}]{xuPRL}%
  \BibitemOpen
  \bibfield  {author} {\bibinfo {author} {\bibfnamefont {J.-S.}\ \bibnamefont
  {Xu}}, \bibinfo {author} {\bibfnamefont {C.-F.}\ \bibnamefont {Li}}, \bibinfo
  {author} {\bibfnamefont {M.}~\bibnamefont {Gong}}, \bibinfo {author}
  {\bibfnamefont {X.-B.}\ \bibnamefont {Zou}}, \bibinfo {author} {\bibfnamefont
  {C.-H.}\ \bibnamefont {Shi}}, \bibinfo {author} {\bibfnamefont
  {G.}~\bibnamefont {Chen}}, \ and\ \bibinfo {author} {\bibfnamefont {G.-C.}\
  \bibnamefont {Guo}},\ }\href@noop {} {\bibfield  {journal} {\bibinfo
  {journal} {Phys. Rev. Lett.}\ }\textbf {\bibinfo {volume} {104}},\ \bibinfo
  {pages} {100502} (\bibinfo {year} {2010})}\BibitemShut {NoStop}%
\bibitem [{\citenamefont {Maniscalco}\ \emph {et~al.}(2008)\citenamefont
  {Maniscalco}, \citenamefont {Francica}, \citenamefont {Zaffino},
  \citenamefont {Gullo},\ and\ \citenamefont
  {Plastina}}]{maniscalco2008protecting}%
  \BibitemOpen
  \bibfield  {author} {\bibinfo {author} {\bibfnamefont {S.}~\bibnamefont
  {Maniscalco}}, \bibinfo {author} {\bibfnamefont {F.}~\bibnamefont
  {Francica}}, \bibinfo {author} {\bibfnamefont {R.~L.}\ \bibnamefont
  {Zaffino}}, \bibinfo {author} {\bibfnamefont {N.~L.}\ \bibnamefont {Gullo}},
  \ and\ \bibinfo {author} {\bibfnamefont {F.}~\bibnamefont {Plastina}},\
  }\href@noop {} {\bibfield  {journal} {\bibinfo  {journal} {Phys. Rev. Lett.}\
  }\textbf {\bibinfo {volume} {100}},\ \bibinfo {pages} {090503} (\bibinfo
  {year} {2008})}\BibitemShut {NoStop}%
\bibitem [{\citenamefont {Mazzola}\ \emph {et~al.}(2009)\citenamefont
  {Mazzola}, \citenamefont {Maniscalco}, \citenamefont {Piilo}, \citenamefont
  {Suominen},\ and\ \citenamefont {Garraway}}]{mazzola2009sudden}%
  \BibitemOpen
  \bibfield  {author} {\bibinfo {author} {\bibfnamefont {L.}~\bibnamefont
  {Mazzola}}, \bibinfo {author} {\bibfnamefont {S.}~\bibnamefont {Maniscalco}},
  \bibinfo {author} {\bibfnamefont {J.}~\bibnamefont {Piilo}}, \bibinfo
  {author} {\bibfnamefont {K.-A.}\ \bibnamefont {Suominen}}, \ and\ \bibinfo
  {author} {\bibfnamefont {B.~M.}\ \bibnamefont {Garraway}},\ }\href@noop {}
  {\bibfield  {journal} {\bibinfo  {journal} {Phys. Rev. A}\ }\textbf {\bibinfo
  {volume} {79}},\ \bibinfo {pages} {042302} (\bibinfo {year}
  {2009})}\BibitemShut {NoStop}%
\bibitem [{\citenamefont {Bellomo}\ \emph
  {et~al.}(2008{\natexlab{b}})\citenamefont {Bellomo}, \citenamefont {{Lo
  Franco}}, \citenamefont {Maniscalco},\ and\ \citenamefont
  {Compagno}}]{bellomo2008entanglementtr}%
  \BibitemOpen
  \bibfield  {author} {\bibinfo {author} {\bibfnamefont {B.}~\bibnamefont
  {Bellomo}}, \bibinfo {author} {\bibfnamefont {R.}~\bibnamefont {{Lo
  Franco}}}, \bibinfo {author} {\bibfnamefont {S.}~\bibnamefont {Maniscalco}},
  \ and\ \bibinfo {author} {\bibfnamefont {G.}~\bibnamefont {Compagno}},\
  }\href@noop {} {\bibfield  {journal} {\bibinfo  {journal} {Phys. Rev. A}\
  }\textbf {\bibinfo {volume} {78}},\ \bibinfo {pages} {060302} (\bibinfo
  {year} {2008}{\natexlab{b}})}\BibitemShut {NoStop}%
\bibitem [{\citenamefont {Man}\ \emph {et~al.}(2015)\citenamefont {Man},
  \citenamefont {Xia},\ and\ \citenamefont {{Lo Franco}}}]{Man2015a}%
  \BibitemOpen
  \bibfield  {author} {\bibinfo {author} {\bibfnamefont {Z.-X.}\ \bibnamefont
  {Man}}, \bibinfo {author} {\bibfnamefont {Y.-J.}\ \bibnamefont {Xia}}, \ and\
  \bibinfo {author} {\bibfnamefont {R.}~\bibnamefont {{Lo Franco}}},\
  }\href@noop {} {\bibfield  {journal} {\bibinfo  {journal} {Sci. Rep.}\
  }\textbf {\bibinfo {volume} {5}},\ \bibinfo {pages} {13843} (\bibinfo {year}
  {2015})}\BibitemShut {NoStop}%
\bibitem [{\citenamefont {Bennett}\ \emph {et~al.}(1996)\citenamefont
  {Bennett}, \citenamefont {Brassard}, \citenamefont {Popescu}, \citenamefont
  {Schumacher}, \citenamefont {Smolin},\ and\ \citenamefont
  {Wootters}}]{Bennett1996}%
  \BibitemOpen
  \bibfield  {author} {\bibinfo {author} {\bibfnamefont {C.~H.}\ \bibnamefont
  {Bennett}}, \bibinfo {author} {\bibfnamefont {G.}~\bibnamefont {Brassard}},
  \bibinfo {author} {\bibfnamefont {S.}~\bibnamefont {Popescu}}, \bibinfo
  {author} {\bibfnamefont {B.}~\bibnamefont {Schumacher}}, \bibinfo {author}
  {\bibfnamefont {J.~A.}\ \bibnamefont {Smolin}}, \ and\ \bibinfo {author}
  {\bibfnamefont {W.~K.}\ \bibnamefont {Wootters}},\ }\href@noop {} {\bibfield
  {journal} {\bibinfo  {journal} {Phys. Rev. Lett.}\ }\textbf {\bibinfo
  {volume} {76}},\ \bibinfo {pages} {722} (\bibinfo {year} {1996})}\BibitemShut
  {NoStop}%
\bibitem [{\citenamefont {Zanardi}\ and\ \citenamefont
  {Rasetti}(1997)}]{zanardi1997noiseless}%
  \BibitemOpen
  \bibfield  {author} {\bibinfo {author} {\bibfnamefont {P.}~\bibnamefont
  {Zanardi}}\ and\ \bibinfo {author} {\bibfnamefont {M.}~\bibnamefont
  {Rasetti}},\ }\href@noop {} {\bibfield  {journal} {\bibinfo  {journal} {Phys.
  Rev. Lett.}\ }\textbf {\bibinfo {volume} {79}},\ \bibinfo {pages} {3306}
  (\bibinfo {year} {1997})}\BibitemShut {NoStop}%
\bibitem [{\citenamefont {Lidar}\ \emph {et~al.}(1998)\citenamefont {Lidar},
  \citenamefont {Chuang},\ and\ \citenamefont {Whaley}}]{lidar1998decoherence}%
  \BibitemOpen
  \bibfield  {author} {\bibinfo {author} {\bibfnamefont {D.~A.}\ \bibnamefont
  {Lidar}}, \bibinfo {author} {\bibfnamefont {I.~L.}\ \bibnamefont {Chuang}}, \
  and\ \bibinfo {author} {\bibfnamefont {K.~B.}\ \bibnamefont {Whaley}},\
  }\href@noop {} {\bibfield  {journal} {\bibinfo  {journal} {Phys. Rev. Lett.}\
  }\textbf {\bibinfo {volume} {81}},\ \bibinfo {pages} {2594} (\bibinfo {year}
  {1998})}\BibitemShut {NoStop}%
\bibitem [{\citenamefont {Viola}\ and\ \citenamefont
  {Lloyd}(1998)}]{Viola1998}%
  \BibitemOpen
  \bibfield  {author} {\bibinfo {author} {\bibfnamefont {L.}~\bibnamefont
  {Viola}}\ and\ \bibinfo {author} {\bibfnamefont {S.}~\bibnamefont {Lloyd}},\
  }\href@noop {} {\bibfield  {journal} {\bibinfo  {journal} {Phys. Rev. A}\
  }\textbf {\bibinfo {volume} {58}},\ \bibinfo {pages} {2733} (\bibinfo {year}
  {1998})}\BibitemShut {NoStop}%
\bibitem [{\citenamefont {Viola}\ and\ \citenamefont
  {Knill}(2005)}]{viola2005random}%
  \BibitemOpen
  \bibfield  {author} {\bibinfo {author} {\bibfnamefont {L.}~\bibnamefont
  {Viola}}\ and\ \bibinfo {author} {\bibfnamefont {E.}~\bibnamefont {Knill}},\
  }\href@noop {} {\bibfield  {journal} {\bibinfo  {journal} {Phys. Rev. Lett.}\
  }\textbf {\bibinfo {volume} {94}},\ \bibinfo {pages} {060502} (\bibinfo
  {year} {2005})}\BibitemShut {NoStop}%
\bibitem [{\citenamefont {{Lo Franco}}\ \emph {et~al.}(2014)\citenamefont {{Lo
  Franco}}, \citenamefont {D'Arrigo}, \citenamefont {Falci}, \citenamefont
  {Compagno},\ and\ \citenamefont {Paladino}}]{franco2014preserving}%
  \BibitemOpen
  \bibfield  {author} {\bibinfo {author} {\bibfnamefont {R.}~\bibnamefont {{Lo
  Franco}}}, \bibinfo {author} {\bibfnamefont {A.}~\bibnamefont {D'Arrigo}},
  \bibinfo {author} {\bibfnamefont {G.}~\bibnamefont {Falci}}, \bibinfo
  {author} {\bibfnamefont {G.}~\bibnamefont {Compagno}}, \ and\ \bibinfo
  {author} {\bibfnamefont {E.}~\bibnamefont {Paladino}},\ }\href@noop {}
  {\bibfield  {journal} {\bibinfo  {journal} {Phys. Rev. B}\ }\textbf {\bibinfo
  {volume} {90}},\ \bibinfo {pages} {054304} (\bibinfo {year}
  {2014})}\BibitemShut {NoStop}%
\bibitem [{\citenamefont {Orieux}\ \emph
  {et~al.}(2015{\natexlab{b}})\citenamefont {Orieux}, \citenamefont {D'Arrigo},
  \citenamefont {Ferranti}, \citenamefont {{Lo Franco}}, \citenamefont
  {Benenti}, \citenamefont {Paladino}, \citenamefont {Falci}, \citenamefont
  {Sciarrino},\ and\ \citenamefont {Mataloni}}]{orieux2015experimental}%
  \BibitemOpen
  \bibfield  {author} {\bibinfo {author} {\bibfnamefont {A.}~\bibnamefont
  {Orieux}}, \bibinfo {author} {\bibfnamefont {A.}~\bibnamefont {D'Arrigo}},
  \bibinfo {author} {\bibfnamefont {G.}~\bibnamefont {Ferranti}}, \bibinfo
  {author} {\bibfnamefont {R.}~\bibnamefont {{Lo Franco}}}, \bibinfo {author}
  {\bibfnamefont {G.}~\bibnamefont {Benenti}}, \bibinfo {author} {\bibfnamefont
  {E.}~\bibnamefont {Paladino}}, \bibinfo {author} {\bibfnamefont
  {G.}~\bibnamefont {Falci}}, \bibinfo {author} {\bibfnamefont
  {F.}~\bibnamefont {Sciarrino}}, \ and\ \bibinfo {author} {\bibfnamefont
  {P.}~\bibnamefont {Mataloni}},\ }\href@noop {} {\bibfield  {journal}
  {\bibinfo  {journal} {Sci. Rep.}\ }\textbf {\bibinfo {volume} {5}},\ \bibinfo
  {pages} {1} (\bibinfo {year} {2015}{\natexlab{b}})}\BibitemShut {NoStop}%
\bibitem [{\citenamefont {Damodarakurup}\ \emph {et~al.}(2009)\citenamefont
  {Damodarakurup}, \citenamefont {Lucamarini}, \citenamefont {Di~Giuseppe},
  \citenamefont {Vitali},\ and\ \citenamefont
  {Tombesi}}]{damodarakurup2009experimental}%
  \BibitemOpen
  \bibfield  {author} {\bibinfo {author} {\bibfnamefont {S.}~\bibnamefont
  {Damodarakurup}}, \bibinfo {author} {\bibfnamefont {M.}~\bibnamefont
  {Lucamarini}}, \bibinfo {author} {\bibfnamefont {G.}~\bibnamefont
  {Di~Giuseppe}}, \bibinfo {author} {\bibfnamefont {D.}~\bibnamefont {Vitali}},
  \ and\ \bibinfo {author} {\bibfnamefont {P.}~\bibnamefont {Tombesi}},\
  }\href@noop {} {\bibfield  {journal} {\bibinfo  {journal} {Phys. Rev. Lett.}\
  }\textbf {\bibinfo {volume} {103}},\ \bibinfo {pages} {040502} (\bibinfo
  {year} {2009})}\BibitemShut {NoStop}%
\bibitem [{\citenamefont {Cuevas}\ \emph {et~al.}(2017)\citenamefont {Cuevas},
  \citenamefont {Mari}, \citenamefont {De~Pasquale}, \citenamefont {Orieux},
  \citenamefont {Massaro}, \citenamefont {Sciarrino}, \citenamefont
  {Mataloni},\ and\ \citenamefont {Giovannetti}}]{cuevas2017cut}%
  \BibitemOpen
  \bibfield  {author} {\bibinfo {author} {\bibfnamefont {{\'A}.}~\bibnamefont
  {Cuevas}}, \bibinfo {author} {\bibfnamefont {A.}~\bibnamefont {Mari}},
  \bibinfo {author} {\bibfnamefont {A.}~\bibnamefont {De~Pasquale}}, \bibinfo
  {author} {\bibfnamefont {A.}~\bibnamefont {Orieux}}, \bibinfo {author}
  {\bibfnamefont {M.}~\bibnamefont {Massaro}}, \bibinfo {author} {\bibfnamefont
  {F.}~\bibnamefont {Sciarrino}}, \bibinfo {author} {\bibfnamefont
  {P.}~\bibnamefont {Mataloni}}, \ and\ \bibinfo {author} {\bibfnamefont
  {V.}~\bibnamefont {Giovannetti}},\ }\href@noop {} {\bibfield  {journal}
  {\bibinfo  {journal} {Phys. Rev. A}\ }\textbf {\bibinfo {volume} {96}},\
  \bibinfo {pages} {012314} (\bibinfo {year} {2017})}\BibitemShut {NoStop}%
\bibitem [{\citenamefont {Mortezapour}\ and\ \citenamefont {{Lo
  Franco}}(2018)}]{mortezapour2018protecting}%
  \BibitemOpen
  \bibfield  {author} {\bibinfo {author} {\bibfnamefont {A.}~\bibnamefont
  {Mortezapour}}\ and\ \bibinfo {author} {\bibfnamefont {R.}~\bibnamefont {{Lo
  Franco}}},\ }\href@noop {} {\bibfield  {journal} {\bibinfo  {journal} {Sci.
  Rep.}\ }\textbf {\bibinfo {volume} {8}},\ \bibinfo {pages} {14304} (\bibinfo
  {year} {2018})}\BibitemShut {NoStop}%
\bibitem [{\citenamefont {Preskill}(1998)}]{preskill1998reliable}%
  \BibitemOpen
  \bibfield  {author} {\bibinfo {author} {\bibfnamefont {J.}~\bibnamefont
  {Preskill}},\ }\href@noop {} {\bibfield  {journal} {\bibinfo  {journal}
  {Proc. R. Soc. London Series A}\ }\textbf {\bibinfo {volume} {454}},\
  \bibinfo {pages} {385} (\bibinfo {year} {1998})}\BibitemShut {NoStop}%
\bibitem [{\citenamefont {Knill}(2005)}]{knill2005quantum}%
  \BibitemOpen
  \bibfield  {author} {\bibinfo {author} {\bibfnamefont {E.}~\bibnamefont
  {Knill}},\ }\href@noop {} {\bibfield  {journal} {\bibinfo  {journal}
  {Nature}\ }\textbf {\bibinfo {volume} {434}},\ \bibinfo {pages} {39}
  (\bibinfo {year} {2005})}\BibitemShut {NoStop}%
\bibitem [{\citenamefont {Shor}(1995)}]{shor1995scheme}%
  \BibitemOpen
  \bibfield  {author} {\bibinfo {author} {\bibfnamefont {P.~W.}\ \bibnamefont
  {Shor}},\ }\href@noop {} {\bibfield  {journal} {\bibinfo  {journal} {Phys.
  Rev. A}\ }\textbf {\bibinfo {volume} {52}},\ \bibinfo {pages} {R2493}
  (\bibinfo {year} {1995})}\BibitemShut {NoStop}%
\bibitem [{\citenamefont {Kitaev}(2003)}]{kitaev2003fault}%
  \BibitemOpen
  \bibfield  {author} {\bibinfo {author} {\bibfnamefont {A.~Y.}\ \bibnamefont
  {Kitaev}},\ }\href@noop {} {\bibfield  {journal} {\bibinfo  {journal} {Ann.
  Phys.}\ }\textbf {\bibinfo {volume} {303}},\ \bibinfo {pages} {2} (\bibinfo
  {year} {2003})}\BibitemShut {NoStop}%
\bibitem [{\citenamefont {Freedman}\ \emph {et~al.}(2003)\citenamefont
  {Freedman}, \citenamefont {Kitaev}, \citenamefont {Larsen},\ and\
  \citenamefont {Wang}}]{freedman2003topological}%
  \BibitemOpen
  \bibfield  {author} {\bibinfo {author} {\bibfnamefont {M.}~\bibnamefont
  {Freedman}}, \bibinfo {author} {\bibfnamefont {A.}~\bibnamefont {Kitaev}},
  \bibinfo {author} {\bibfnamefont {M.}~\bibnamefont {Larsen}}, \ and\ \bibinfo
  {author} {\bibfnamefont {Z.}~\bibnamefont {Wang}},\ }\href@noop {} {\bibfield
   {journal} {\bibinfo  {journal} {Bull. Am. Math. Soc.}\ }\textbf {\bibinfo
  {volume} {40}},\ \bibinfo {pages} {31} (\bibinfo {year} {2003})}\BibitemShut
  {NoStop}%
\bibitem [{\citenamefont {Suter}\ and\ \citenamefont
  {{\'A}lvarez}(2016)}]{suter2016colloquium}%
  \BibitemOpen
  \bibfield  {author} {\bibinfo {author} {\bibfnamefont {D.}~\bibnamefont
  {Suter}}\ and\ \bibinfo {author} {\bibfnamefont {G.~A.}\ \bibnamefont
  {{\'A}lvarez}},\ }\href@noop {} {\bibfield  {journal} {\bibinfo  {journal}
  {Rev. Mod. Phys.}\ }\textbf {\bibinfo {volume} {88}},\ \bibinfo {pages}
  {041001} (\bibinfo {year} {2016})}\BibitemShut {NoStop}%
\bibitem [{\citenamefont {Kraus}(1983)}]{kraus1983states}%
  \BibitemOpen
  \bibfield  {author} {\bibinfo {author} {\bibfnamefont {K.}~\bibnamefont
  {Kraus}},\ }\href@noop {} {\emph {\bibinfo {title} {States, effects and
  operations: fundamental notions of quantum theory}}}\ (\bibinfo  {publisher}
  {Springer},\ \bibinfo {year} {1983})\BibitemShut {NoStop}%
\bibitem [{\citenamefont {Lindblad}(1976)}]{lindblad1976generators}%
  \BibitemOpen
  \bibfield  {author} {\bibinfo {author} {\bibfnamefont {G.}~\bibnamefont
  {Lindblad}},\ }\href@noop {} {\bibfield  {journal} {\bibinfo  {journal}
  {Comm. Math. Phys.}\ }\textbf {\bibinfo {volume} {48}},\ \bibinfo {pages}
  {119} (\bibinfo {year} {1976})}\BibitemShut {NoStop}%
\bibitem [{\citenamefont {Gorini}\ \emph {et~al.}(1976)\citenamefont {Gorini},
  \citenamefont {Kossakowski},\ and\ \citenamefont
  {Sudarshan}}]{gorini1976completely}%
  \BibitemOpen
  \bibfield  {author} {\bibinfo {author} {\bibfnamefont {V.}~\bibnamefont
  {Gorini}}, \bibinfo {author} {\bibfnamefont {A.}~\bibnamefont {Kossakowski}},
  \ and\ \bibinfo {author} {\bibfnamefont {E.~C.~G.}\ \bibnamefont
  {Sudarshan}},\ }\href@noop {} {\bibfield  {journal} {\bibinfo  {journal} {J.
  Math. Phys.}\ }\textbf {\bibinfo {volume} {17}},\ \bibinfo {pages} {821}
  (\bibinfo {year} {1976})}\BibitemShut {NoStop}%
\bibitem [{\citenamefont {Gross}\ and\ \citenamefont
  {Haroche}(1982)}]{gross1982superradiance}%
  \BibitemOpen
  \bibfield  {author} {\bibinfo {author} {\bibfnamefont {M.}~\bibnamefont
  {Gross}}\ and\ \bibinfo {author} {\bibfnamefont {S.}~\bibnamefont
  {Haroche}},\ }\href@noop {} {\bibfield  {journal} {\bibinfo  {journal} {Phys.
  Rep.}\ }\textbf {\bibinfo {volume} {93}},\ \bibinfo {pages} {301} (\bibinfo
  {year} {1982})}\BibitemShut {NoStop}%
\bibitem [{\citenamefont {Latune}\ \emph {et~al.}(2019)\citenamefont {Latune},
  \citenamefont {Sinayskiy},\ and\ \citenamefont
  {Petruccione}}]{PhysRevA.99.052105}%
  \BibitemOpen
  \bibfield  {author} {\bibinfo {author} {\bibfnamefont {C.~L.}\ \bibnamefont
  {Latune}}, \bibinfo {author} {\bibfnamefont {I.}~\bibnamefont {Sinayskiy}}, \
  and\ \bibinfo {author} {\bibfnamefont {F.}~\bibnamefont {Petruccione}},\
  }\href@noop {} {\bibfield  {journal} {\bibinfo  {journal} {Phys. Rev. A}\
  }\textbf {\bibinfo {volume} {99}},\ \bibinfo {pages} {052105} (\bibinfo
  {year} {2019})}\BibitemShut {NoStop}%
\bibitem [{\citenamefont {Zhou}\ \emph {et~al.}(2010)\citenamefont {Zhou},
  \citenamefont {Lang},\ and\ \citenamefont {Joynt}}]{zhou2010disentanglement}%
  \BibitemOpen
  \bibfield  {author} {\bibinfo {author} {\bibfnamefont {D.}~\bibnamefont
  {Zhou}}, \bibinfo {author} {\bibfnamefont {A.}~\bibnamefont {Lang}}, \ and\
  \bibinfo {author} {\bibfnamefont {R.}~\bibnamefont {Joynt}},\ }\href@noop {}
  {\bibfield  {journal} {\bibinfo  {journal} {Quantum Information Processing}\
  }\textbf {\bibinfo {volume} {9}},\ \bibinfo {pages} {727} (\bibinfo {year}
  {2010})}\BibitemShut {NoStop}%
\bibitem [{\citenamefont {{Lo Franco}}\ \emph
  {et~al.}(2012{\natexlab{b}})\citenamefont {{Lo Franco}}, \citenamefont
  {D'Arrigo}, \citenamefont {Falci}, \citenamefont {Compagno},\ and\
  \citenamefont {Paladino}}]{franco2012entanglement}%
  \BibitemOpen
  \bibfield  {author} {\bibinfo {author} {\bibfnamefont {R.}~\bibnamefont {{Lo
  Franco}}}, \bibinfo {author} {\bibfnamefont {A.}~\bibnamefont {D'Arrigo}},
  \bibinfo {author} {\bibfnamefont {G.}~\bibnamefont {Falci}}, \bibinfo
  {author} {\bibfnamefont {G.}~\bibnamefont {Compagno}}, \ and\ \bibinfo
  {author} {\bibfnamefont {E.}~\bibnamefont {Paladino}},\ }\href@noop {}
  {\bibfield  {journal} {\bibinfo  {journal} {Phys. Scr.}\ }\textbf {\bibinfo
  {volume} {2012}},\ \bibinfo {pages} {014019} (\bibinfo {year}
  {2012}{\natexlab{b}})}\BibitemShut {NoStop}%
\bibitem [{\citenamefont {Bordone}\ \emph {et~al.}(2012)\citenamefont
  {Bordone}, \citenamefont {Buscemi},\ and\ \citenamefont
  {Benedetti}}]{bordone2012effect}%
  \BibitemOpen
  \bibfield  {author} {\bibinfo {author} {\bibfnamefont {P.}~\bibnamefont
  {Bordone}}, \bibinfo {author} {\bibfnamefont {F.}~\bibnamefont {Buscemi}}, \
  and\ \bibinfo {author} {\bibfnamefont {C.}~\bibnamefont {Benedetti}},\
  }\href@noop {} {\bibfield  {journal} {\bibinfo  {journal} {Fluctuation and
  Noise Lett.}\ }\textbf {\bibinfo {volume} {11}},\ \bibinfo {pages} {1242003}
  (\bibinfo {year} {2012})}\BibitemShut {NoStop}%
\bibitem [{\citenamefont {Bellomo}\ \emph {et~al.}(2012)\citenamefont
  {Bellomo}, \citenamefont {{Lo Franco}}, \citenamefont {Andersson},
  \citenamefont {Cresser},\ and\ \citenamefont
  {Compagno}}]{bellomo2012noisylaser}%
  \BibitemOpen
  \bibfield  {author} {\bibinfo {author} {\bibfnamefont {B.}~\bibnamefont
  {Bellomo}}, \bibinfo {author} {\bibfnamefont {R.}~\bibnamefont {{Lo
  Franco}}}, \bibinfo {author} {\bibfnamefont {E.}~\bibnamefont {Andersson}},
  \bibinfo {author} {\bibfnamefont {J.~D.}\ \bibnamefont {Cresser}}, \ and\
  \bibinfo {author} {\bibfnamefont {G.}~\bibnamefont {Compagno}},\ }\href@noop
  {} {\bibfield  {journal} {\bibinfo  {journal} {Phys. Scr.}\ }\textbf
  {\bibinfo {volume} {T147}},\ \bibinfo {pages} {014004} (\bibinfo {year}
  {2012})}\BibitemShut {NoStop}%
\bibitem [{\citenamefont {Cai}(2020)}]{caiSciRep}%
  \BibitemOpen
  \bibfield  {author} {\bibinfo {author} {\bibfnamefont {X.}~\bibnamefont
  {Cai}},\ }\href@noop {} {\bibfield  {journal} {\bibinfo  {journal} {Sci.
  Rep.}\ }\textbf {\bibinfo {volume} {10}},\ \bibinfo {pages} {88} (\bibinfo
  {year} {2020})}\BibitemShut {NoStop}%
\bibitem [{\citenamefont {Wold}\ \emph {et~al.}(2012)\citenamefont {Wold},
  \citenamefont {Brox}, \citenamefont {Galperin},\ and\ \citenamefont
  {Bergli}}]{woldPRB}%
  \BibitemOpen
  \bibfield  {author} {\bibinfo {author} {\bibfnamefont {H.~J.}\ \bibnamefont
  {Wold}}, \bibinfo {author} {\bibfnamefont {H.}~\bibnamefont {Brox}}, \bibinfo
  {author} {\bibfnamefont {Y.~M.}\ \bibnamefont {Galperin}}, \ and\ \bibinfo
  {author} {\bibfnamefont {J.}~\bibnamefont {Bergli}},\ }\href@noop {}
  {\bibfield  {journal} {\bibinfo  {journal} {Phys. Rev. B}\ }\textbf {\bibinfo
  {volume} {86}},\ \bibinfo {pages} {205404} (\bibinfo {year}
  {2012})}\BibitemShut {NoStop}%
\bibitem [{\citenamefont {Paladino}\ \emph {et~al.}(2002)\citenamefont
  {Paladino}, \citenamefont {Faoro}, \citenamefont {Falci},\ and\ \citenamefont
  {Fazio}}]{paladinoPRL}%
  \BibitemOpen
  \bibfield  {author} {\bibinfo {author} {\bibfnamefont {E.}~\bibnamefont
  {Paladino}}, \bibinfo {author} {\bibfnamefont {L.}~\bibnamefont {Faoro}},
  \bibinfo {author} {\bibfnamefont {G.}~\bibnamefont {Falci}}, \ and\ \bibinfo
  {author} {\bibfnamefont {R.}~\bibnamefont {Fazio}},\ }\href@noop {}
  {\bibfield  {journal} {\bibinfo  {journal} {Phys. Rev. Lett.}\ }\textbf
  {\bibinfo {volume} {88}},\ \bibinfo {pages} {228304} (\bibinfo {year}
  {2002})}\BibitemShut {NoStop}%
\bibitem [{\citenamefont {Ithier}\ \emph {et~al.}(2005)\citenamefont {Ithier},
  \citenamefont {Collin}, \citenamefont {Joyez}, \citenamefont {Meeson},
  \citenamefont {Vion}, \citenamefont {Esteve}, \citenamefont {Chiarello},
  \citenamefont {Shnirman}, \citenamefont {Makhlin}, \citenamefont {Schriefl}
  \emph {et~al.}}]{ithier2005decoherence}%
  \BibitemOpen
  \bibfield  {author} {\bibinfo {author} {\bibfnamefont {G.}~\bibnamefont
  {Ithier}}, \bibinfo {author} {\bibfnamefont {E.}~\bibnamefont {Collin}},
  \bibinfo {author} {\bibfnamefont {P.}~\bibnamefont {Joyez}}, \bibinfo
  {author} {\bibfnamefont {P.}~\bibnamefont {Meeson}}, \bibinfo {author}
  {\bibfnamefont {D.}~\bibnamefont {Vion}}, \bibinfo {author} {\bibfnamefont
  {D.}~\bibnamefont {Esteve}}, \bibinfo {author} {\bibfnamefont
  {F.}~\bibnamefont {Chiarello}}, \bibinfo {author} {\bibfnamefont
  {A.}~\bibnamefont {Shnirman}}, \bibinfo {author} {\bibfnamefont
  {Y.}~\bibnamefont {Makhlin}}, \bibinfo {author} {\bibfnamefont
  {J.}~\bibnamefont {Schriefl}},  \emph {et~al.},\ }\href@noop {} {\bibfield
  {journal} {\bibinfo  {journal} {Phys. Rev. B}\ }\textbf {\bibinfo {volume}
  {72}},\ \bibinfo {pages} {134519} (\bibinfo {year} {2005})}\BibitemShut
  {NoStop}%
\bibitem [{\citenamefont {Bylander}\ \emph {et~al.}(2011)\citenamefont
  {Bylander}, \citenamefont {Gustavsson}, \citenamefont {Yan}, \citenamefont
  {Yoshihara}, \citenamefont {Harrabi}, \citenamefont {Fitch}, \citenamefont
  {Cory}, \citenamefont {Nakamura}, \citenamefont {Tsai},\ and\ \citenamefont
  {Oliver}}]{bylander2011noise}%
  \BibitemOpen
  \bibfield  {author} {\bibinfo {author} {\bibfnamefont {J.}~\bibnamefont
  {Bylander}}, \bibinfo {author} {\bibfnamefont {S.}~\bibnamefont
  {Gustavsson}}, \bibinfo {author} {\bibfnamefont {F.}~\bibnamefont {Yan}},
  \bibinfo {author} {\bibfnamefont {F.}~\bibnamefont {Yoshihara}}, \bibinfo
  {author} {\bibfnamefont {K.}~\bibnamefont {Harrabi}}, \bibinfo {author}
  {\bibfnamefont {G.}~\bibnamefont {Fitch}}, \bibinfo {author} {\bibfnamefont
  {D.~G.}\ \bibnamefont {Cory}}, \bibinfo {author} {\bibfnamefont
  {Y.}~\bibnamefont {Nakamura}}, \bibinfo {author} {\bibfnamefont {J.-S.}\
  \bibnamefont {Tsai}}, \ and\ \bibinfo {author} {\bibfnamefont {W.~D.}\
  \bibnamefont {Oliver}},\ }\href@noop {} {\bibfield  {journal} {\bibinfo
  {journal} {Nature Physics}\ }\textbf {\bibinfo {volume} {7}},\ \bibinfo
  {pages} {565} (\bibinfo {year} {2011})}\BibitemShut {NoStop}%
\bibitem [{\citenamefont {Paladino}\ \emph {et~al.}(2014)\citenamefont
  {Paladino}, \citenamefont {Galperin}, \citenamefont {Falci},\ and\
  \citenamefont {Altshuler}}]{paladinoRMP}%
  \BibitemOpen
  \bibfield  {author} {\bibinfo {author} {\bibfnamefont {E.}~\bibnamefont
  {Paladino}}, \bibinfo {author} {\bibfnamefont {Y.~M.}\ \bibnamefont
  {Galperin}}, \bibinfo {author} {\bibfnamefont {G.}~\bibnamefont {Falci}}, \
  and\ \bibinfo {author} {\bibfnamefont {B.~L.}\ \bibnamefont {Altshuler}},\
  }\href@noop {} {\bibfield  {journal} {\bibinfo  {journal} {Rev. Mod. Phys.}\
  }\textbf {\bibinfo {volume} {86}},\ \bibinfo {pages} {361} (\bibinfo {year}
  {2014})}\BibitemShut {NoStop}%
\bibitem [{\citenamefont {Anton}\ \emph {et~al.}(2012)\citenamefont {Anton},
  \citenamefont {M\"uller}, \citenamefont {Birenbaum}, \citenamefont
  {O'Kelley}, \citenamefont {Fefferman}, \citenamefont {Golubev}, \citenamefont
  {Hilton}, \citenamefont {Cho}, \citenamefont {Irwin}, \citenamefont
  {Wellstood}, \citenamefont {Sch\"on}, \citenamefont {Shnirman},\ and\
  \citenamefont {Clarke}}]{antonPRB}%
  \BibitemOpen
  \bibfield  {author} {\bibinfo {author} {\bibfnamefont {S.~M.}\ \bibnamefont
  {Anton}}, \bibinfo {author} {\bibfnamefont {C.}~\bibnamefont {M\"uller}},
  \bibinfo {author} {\bibfnamefont {J.~S.}\ \bibnamefont {Birenbaum}}, \bibinfo
  {author} {\bibfnamefont {S.~R.}\ \bibnamefont {O'Kelley}}, \bibinfo {author}
  {\bibfnamefont {A.~D.}\ \bibnamefont {Fefferman}}, \bibinfo {author}
  {\bibfnamefont {D.~S.}\ \bibnamefont {Golubev}}, \bibinfo {author}
  {\bibfnamefont {G.~C.}\ \bibnamefont {Hilton}}, \bibinfo {author}
  {\bibfnamefont {H.-M.}\ \bibnamefont {Cho}}, \bibinfo {author} {\bibfnamefont
  {K.~D.}\ \bibnamefont {Irwin}}, \bibinfo {author} {\bibfnamefont {F.~C.}\
  \bibnamefont {Wellstood}}, \bibinfo {author} {\bibfnamefont {G.}~\bibnamefont
  {Sch\"on}}, \bibinfo {author} {\bibfnamefont {A.}~\bibnamefont {Shnirman}}, \
  and\ \bibinfo {author} {\bibfnamefont {J.}~\bibnamefont {Clarke}},\
  }\href@noop {} {\bibfield  {journal} {\bibinfo  {journal} {Phys. Rev. B}\
  }\textbf {\bibinfo {volume} {85}},\ \bibinfo {pages} {224505} (\bibinfo
  {year} {2012})}\BibitemShut {NoStop}%
\bibitem [{\citenamefont {Bellomo}\ \emph {et~al.}(2010)\citenamefont
  {Bellomo}, \citenamefont {Compagno}, \citenamefont {D'Arrigo}, \citenamefont
  {Falci}, \citenamefont {Lo~Franco},\ and\ \citenamefont
  {Paladino}}]{bellomoPRA}%
  \BibitemOpen
  \bibfield  {author} {\bibinfo {author} {\bibfnamefont {B.}~\bibnamefont
  {Bellomo}}, \bibinfo {author} {\bibfnamefont {G.}~\bibnamefont {Compagno}},
  \bibinfo {author} {\bibfnamefont {A.}~\bibnamefont {D'Arrigo}}, \bibinfo
  {author} {\bibfnamefont {G.}~\bibnamefont {Falci}}, \bibinfo {author}
  {\bibfnamefont {R.}~\bibnamefont {Lo~Franco}}, \ and\ \bibinfo {author}
  {\bibfnamefont {E.}~\bibnamefont {Paladino}},\ }\href@noop {} {\bibfield
  {journal} {\bibinfo  {journal} {Phys. Rev. A}\ }\textbf {\bibinfo {volume}
  {81}},\ \bibinfo {pages} {062309} (\bibinfo {year} {2010})}\BibitemShut
  {NoStop}%
\bibitem [{\citenamefont {Aaronson}\ \emph {et~al.}(2013)\citenamefont
  {Aaronson}, \citenamefont {Lo~Franco},\ and\ \citenamefont
  {Adesso}}]{aaronsonPRA}%
  \BibitemOpen
  \bibfield  {author} {\bibinfo {author} {\bibfnamefont {B.}~\bibnamefont
  {Aaronson}}, \bibinfo {author} {\bibfnamefont {R.}~\bibnamefont {Lo~Franco}},
  \ and\ \bibinfo {author} {\bibfnamefont {G.}~\bibnamefont {Adesso}},\
  }\href@noop {} {\bibfield  {journal} {\bibinfo  {journal} {Phys. Rev. A}\
  }\textbf {\bibinfo {volume} {88}},\ \bibinfo {pages} {012120} (\bibinfo
  {year} {2013})}\BibitemShut {NoStop}%
\bibitem [{\citenamefont {Silva}\ \emph {et~al.}(2016)\citenamefont {Silva},
  \citenamefont {Souza}, \citenamefont {Bromley}, \citenamefont {Cianciaruso},
  \citenamefont {Marx}, \citenamefont {Sarthour}, \citenamefont {Oliveira},
  \citenamefont {Lo~Franco}, \citenamefont {Glaser}, \citenamefont {deAzevedo},
  \citenamefont {Soares-Pinto},\ and\ \citenamefont {Adesso}}]{silvaPRL}%
  \BibitemOpen
  \bibfield  {author} {\bibinfo {author} {\bibfnamefont {I.~A.}\ \bibnamefont
  {Silva}}, \bibinfo {author} {\bibfnamefont {A.~M.}\ \bibnamefont {Souza}},
  \bibinfo {author} {\bibfnamefont {T.~R.}\ \bibnamefont {Bromley}}, \bibinfo
  {author} {\bibfnamefont {M.}~\bibnamefont {Cianciaruso}}, \bibinfo {author}
  {\bibfnamefont {R.}~\bibnamefont {Marx}}, \bibinfo {author} {\bibfnamefont
  {R.~S.}\ \bibnamefont {Sarthour}}, \bibinfo {author} {\bibfnamefont {I.~S.}\
  \bibnamefont {Oliveira}}, \bibinfo {author} {\bibfnamefont {R.}~\bibnamefont
  {Lo~Franco}}, \bibinfo {author} {\bibfnamefont {S.~J.}\ \bibnamefont
  {Glaser}}, \bibinfo {author} {\bibfnamefont {E.~R.}\ \bibnamefont
  {deAzevedo}}, \bibinfo {author} {\bibfnamefont {D.~O.}\ \bibnamefont
  {Soares-Pinto}}, \ and\ \bibinfo {author} {\bibfnamefont {G.}~\bibnamefont
  {Adesso}},\ }\href@noop {} {\bibfield  {journal} {\bibinfo  {journal} {Phys.
  Rev. Lett.}\ }\textbf {\bibinfo {volume} {117}},\ \bibinfo {pages} {160402}
  (\bibinfo {year} {2016})}\BibitemShut {NoStop}%
\bibitem [{\citenamefont {D\"ur}\ \emph {et~al.}(2005)\citenamefont {D\"ur},
  \citenamefont {Hein}, \citenamefont {Cirac},\ and\ \citenamefont
  {Briegel}}]{durPRA}%
  \BibitemOpen
  \bibfield  {author} {\bibinfo {author} {\bibfnamefont {W.}~\bibnamefont
  {D\"ur}}, \bibinfo {author} {\bibfnamefont {M.}~\bibnamefont {Hein}},
  \bibinfo {author} {\bibfnamefont {J.~I.}\ \bibnamefont {Cirac}}, \ and\
  \bibinfo {author} {\bibfnamefont {H.-J.}\ \bibnamefont {Briegel}},\
  }\href@noop {} {\bibfield  {journal} {\bibinfo  {journal} {Phys. Rev. A}\
  }\textbf {\bibinfo {volume} {72}},\ \bibinfo {pages} {052326} (\bibinfo
  {year} {2005})}\BibitemShut {NoStop}%
\bibitem [{\citenamefont {Klimov}\ and\ \citenamefont
  {S{\'{a}}nchez-Soto}(2010)}]{Klimov2010}%
  \BibitemOpen
  \bibfield  {author} {\bibinfo {author} {\bibfnamefont {A.~B.}\ \bibnamefont
  {Klimov}}\ and\ \bibinfo {author} {\bibfnamefont {L.~L.}\ \bibnamefont
  {S{\'{a}}nchez-Soto}},\ }\href@noop {} {\bibfield  {journal} {\bibinfo
  {journal} {Phys. Scr.}\ }\textbf {\bibinfo {volume} {T140}},\ \bibinfo
  {pages} {014009} (\bibinfo {year} {2010})}\BibitemShut {NoStop}%
\bibitem [{\citenamefont {Hamdouni}\ and\ \citenamefont
  {Petruccione}(2007)}]{hamdouni2007time}%
  \BibitemOpen
  \bibfield  {author} {\bibinfo {author} {\bibfnamefont {Y.}~\bibnamefont
  {Hamdouni}}\ and\ \bibinfo {author} {\bibfnamefont {F.}~\bibnamefont
  {Petruccione}},\ }\href@noop {} {\bibfield  {journal} {\bibinfo  {journal}
  {Phys. Rev. B}\ }\textbf {\bibinfo {volume} {76}},\ \bibinfo {pages} {174306}
  (\bibinfo {year} {2007})}\BibitemShut {NoStop}%
\bibitem [{\citenamefont {Romero}\ and\ \citenamefont
  {Franco}(2012)}]{romero2012simple}%
  \BibitemOpen
  \bibfield  {author} {\bibinfo {author} {\bibfnamefont {K.~F.}\ \bibnamefont
  {Romero}}\ and\ \bibinfo {author} {\bibfnamefont {R.~L.}\ \bibnamefont
  {Franco}},\ }\href@noop {} {\bibfield  {journal} {\bibinfo  {journal} {Phys.
  Scr.}\ }\textbf {\bibinfo {volume} {86}},\ \bibinfo {pages} {065004}
  (\bibinfo {year} {2012})}\BibitemShut {NoStop}%
\bibitem [{\citenamefont {Melikidze}\ \emph {et~al.}(2004)\citenamefont
  {Melikidze}, \citenamefont {Dobrovitski}, \citenamefont {De~Raedt},
  \citenamefont {Katsnelson},\ and\ \citenamefont
  {Harmon}}]{melikidze2004parity}%
  \BibitemOpen
  \bibfield  {author} {\bibinfo {author} {\bibfnamefont {A.}~\bibnamefont
  {Melikidze}}, \bibinfo {author} {\bibfnamefont {V.}~\bibnamefont
  {Dobrovitski}}, \bibinfo {author} {\bibfnamefont {H.}~\bibnamefont
  {De~Raedt}}, \bibinfo {author} {\bibfnamefont {M.}~\bibnamefont
  {Katsnelson}}, \ and\ \bibinfo {author} {\bibfnamefont {B.}~\bibnamefont
  {Harmon}},\ }\href@noop {} {\bibfield  {journal} {\bibinfo  {journal} {Phys.
  Rev. B}\ }\textbf {\bibinfo {volume} {70}},\ \bibinfo {pages} {014435}
  (\bibinfo {year} {2004})}\BibitemShut {NoStop}%
\bibitem [{\citenamefont {Hutton}\ and\ \citenamefont {Bose}(2004)}]{bosePRA}%
  \BibitemOpen
  \bibfield  {author} {\bibinfo {author} {\bibfnamefont {A.}~\bibnamefont
  {Hutton}}\ and\ \bibinfo {author} {\bibfnamefont {S.}~\bibnamefont {Bose}},\
  }\href@noop {} {\bibfield  {journal} {\bibinfo  {journal} {Phys. Rev. A}\
  }\textbf {\bibinfo {volume} {69}},\ \bibinfo {pages} {042312} (\bibinfo
  {year} {2004})}\BibitemShut {NoStop}%
\bibitem [{\citenamefont {Xin}\ \emph {et~al.}(2017)\citenamefont {Xin},
  \citenamefont {Wei}, \citenamefont {Pedernales}, \citenamefont {Solano},\
  and\ \citenamefont {Long}}]{xin2017quantum}%
  \BibitemOpen
  \bibfield  {author} {\bibinfo {author} {\bibfnamefont {T.}~\bibnamefont
  {Xin}}, \bibinfo {author} {\bibfnamefont {S.-J.}\ \bibnamefont {Wei}},
  \bibinfo {author} {\bibfnamefont {J.~S.}\ \bibnamefont {Pedernales}},
  \bibinfo {author} {\bibfnamefont {E.}~\bibnamefont {Solano}}, \ and\ \bibinfo
  {author} {\bibfnamefont {G.-L.}\ \bibnamefont {Long}},\ }\href@noop {}
  {\bibfield  {journal} {\bibinfo  {journal} {Phys. Rev. A}\ }\textbf {\bibinfo
  {volume} {96}},\ \bibinfo {pages} {062303} (\bibinfo {year}
  {2017})}\BibitemShut {NoStop}%
\bibitem [{\citenamefont {Ryan}\ \emph {et~al.}(2009)\citenamefont {Ryan},
  \citenamefont {Laforest},\ and\ \citenamefont
  {Laflamme}}]{ryan2009randomized}%
  \BibitemOpen
  \bibfield  {author} {\bibinfo {author} {\bibfnamefont {C.}~\bibnamefont
  {Ryan}}, \bibinfo {author} {\bibfnamefont {M.}~\bibnamefont {Laforest}}, \
  and\ \bibinfo {author} {\bibfnamefont {R.}~\bibnamefont {Laflamme}},\
  }\href@noop {} {\bibfield  {journal} {\bibinfo  {journal} {New J. Phys.}\
  }\textbf {\bibinfo {volume} {11}},\ \bibinfo {pages} {013034} (\bibinfo
  {year} {2009})}\BibitemShut {NoStop}%
\bibitem [{\citenamefont {Kasprzak}\ \emph {et~al.}(2006)\citenamefont
  {Kasprzak}, \citenamefont {Richard}, \citenamefont {Kundermann},
  \citenamefont {Baas}, \citenamefont {Jeambrun}, \citenamefont {Keeling},
  \citenamefont {Marchetti}, \citenamefont {Szyma{\'n}ska}, \citenamefont
  {Andr{\'e}}, \citenamefont {Staehli} \emph {et~al.}}]{kasprzak2006bose}%
  \BibitemOpen
  \bibfield  {author} {\bibinfo {author} {\bibfnamefont {J.}~\bibnamefont
  {Kasprzak}}, \bibinfo {author} {\bibfnamefont {M.}~\bibnamefont {Richard}},
  \bibinfo {author} {\bibfnamefont {S.}~\bibnamefont {Kundermann}}, \bibinfo
  {author} {\bibfnamefont {A.}~\bibnamefont {Baas}}, \bibinfo {author}
  {\bibfnamefont {P.}~\bibnamefont {Jeambrun}}, \bibinfo {author}
  {\bibfnamefont {J.}~\bibnamefont {Keeling}}, \bibinfo {author} {\bibfnamefont
  {F.}~\bibnamefont {Marchetti}}, \bibinfo {author} {\bibfnamefont
  {M.}~\bibnamefont {Szyma{\'n}ska}}, \bibinfo {author} {\bibfnamefont
  {R.}~\bibnamefont {Andr{\'e}}}, \bibinfo {author} {\bibfnamefont
  {J.}~\bibnamefont {Staehli}},  \emph {et~al.},\ }\href@noop {} {\bibfield
  {journal} {\bibinfo  {journal} {Nature}\ }\textbf {\bibinfo {volume} {443}},\
  \bibinfo {pages} {409} (\bibinfo {year} {2006})}\BibitemShut {NoStop}%
\bibitem [{\citenamefont {Zipkes}\ \emph {et~al.}(2010)\citenamefont {Zipkes},
  \citenamefont {Palzer}, \citenamefont {Sias},\ and\ \citenamefont
  {K{\"o}hl}}]{zipkes2010trapped}%
  \BibitemOpen
  \bibfield  {author} {\bibinfo {author} {\bibfnamefont {C.}~\bibnamefont
  {Zipkes}}, \bibinfo {author} {\bibfnamefont {S.}~\bibnamefont {Palzer}},
  \bibinfo {author} {\bibfnamefont {C.}~\bibnamefont {Sias}}, \ and\ \bibinfo
  {author} {\bibfnamefont {M.}~\bibnamefont {K{\"o}hl}},\ }\href@noop {}
  {\bibfield  {journal} {\bibinfo  {journal} {Nature}\ }\textbf {\bibinfo
  {volume} {464}},\ \bibinfo {pages} {388} (\bibinfo {year}
  {2010})}\BibitemShut {NoStop}%
\bibitem [{\citenamefont {Puentes}\ \emph {et~al.}(2005)\citenamefont
  {Puentes}, \citenamefont {Voigt}, \citenamefont {Aiello},\ and\ \citenamefont
  {Woerdman}}]{puentes2005experimental}%
  \BibitemOpen
  \bibfield  {author} {\bibinfo {author} {\bibfnamefont {G.}~\bibnamefont
  {Puentes}}, \bibinfo {author} {\bibfnamefont {D.}~\bibnamefont {Voigt}},
  \bibinfo {author} {\bibfnamefont {A.}~\bibnamefont {Aiello}}, \ and\ \bibinfo
  {author} {\bibfnamefont {J.}~\bibnamefont {Woerdman}},\ }\href@noop {}
  {\bibfield  {journal} {\bibinfo  {journal} {Optics Lett.}\ }\textbf {\bibinfo
  {volume} {30}},\ \bibinfo {pages} {3216} (\bibinfo {year}
  {2005})}\BibitemShut {NoStop}%
\bibitem [{\citenamefont {Puentes}\ \emph {et~al.}(2007)\citenamefont
  {Puentes}, \citenamefont {Aiello}, \citenamefont {Voigt},\ and\ \citenamefont
  {Woerdman}}]{puentes2007entangled}%
  \BibitemOpen
  \bibfield  {author} {\bibinfo {author} {\bibfnamefont {G.}~\bibnamefont
  {Puentes}}, \bibinfo {author} {\bibfnamefont {A.}~\bibnamefont {Aiello}},
  \bibinfo {author} {\bibfnamefont {D.}~\bibnamefont {Voigt}}, \ and\ \bibinfo
  {author} {\bibfnamefont {J.}~\bibnamefont {Woerdman}},\ }\href@noop {}
  {\bibfield  {journal} {\bibinfo  {journal} {Phys. Rev. A}\ }\textbf {\bibinfo
  {volume} {75}},\ \bibinfo {pages} {032319} (\bibinfo {year}
  {2007})}\BibitemShut {NoStop}%
\bibitem [{\citenamefont {Shaham}\ and\ \citenamefont
  {Eisenberg}(2011)}]{shaham2011realizing}%
  \BibitemOpen
  \bibfield  {author} {\bibinfo {author} {\bibfnamefont {A.}~\bibnamefont
  {Shaham}}\ and\ \bibinfo {author} {\bibfnamefont {H.}~\bibnamefont
  {Eisenberg}},\ }\href@noop {} {\bibfield  {journal} {\bibinfo  {journal}
  {Phys. Rev. A}\ }\textbf {\bibinfo {volume} {83}},\ \bibinfo {pages} {022303}
  (\bibinfo {year} {2011})}\BibitemShut {NoStop}%
\bibitem [{\citenamefont {Werner}(1989)}]{werner}%
  \BibitemOpen
  \bibfield  {author} {\bibinfo {author} {\bibfnamefont {R.~F.}\ \bibnamefont
  {Werner}},\ }\href@noop {} {\bibfield  {journal} {\bibinfo  {journal} {Phys.
  Rev. A}\ }\textbf {\bibinfo {volume} {40}},\ \bibinfo {pages} {4277}
  (\bibinfo {year} {1989})}\BibitemShut {NoStop}%
\bibitem [{\citenamefont {Myatt}\ \emph {et~al.}(2000)\citenamefont {Myatt},
  \citenamefont {King}, \citenamefont {Turchette}, \citenamefont {Sackett},
  \citenamefont {Kielpinski}, \citenamefont {Itano}, \citenamefont {Monroe},\
  and\ \citenamefont {Wineland}}]{wineland2000}%
  \BibitemOpen
  \bibfield  {author} {\bibinfo {author} {\bibfnamefont {C.~J.}\ \bibnamefont
  {Myatt}}, \bibinfo {author} {\bibfnamefont {B.~E.}\ \bibnamefont {King}},
  \bibinfo {author} {\bibfnamefont {Q.~A.}\ \bibnamefont {Turchette}}, \bibinfo
  {author} {\bibfnamefont {C.~A.}\ \bibnamefont {Sackett}}, \bibinfo {author}
  {\bibfnamefont {D.}~\bibnamefont {Kielpinski}}, \bibinfo {author}
  {\bibfnamefont {W.~M.}\ \bibnamefont {Itano}}, \bibinfo {author}
  {\bibfnamefont {C.}~\bibnamefont {Monroe}}, \ and\ \bibinfo {author}
  {\bibfnamefont {D.~J.}\ \bibnamefont {Wineland}},\ }\href@noop {} {\bibfield
  {journal} {\bibinfo  {journal} {Nature}\ }\textbf {\bibinfo {volume} {403}},\
  \bibinfo {pages} {269} (\bibinfo {year} {2000})}\BibitemShut {NoStop}%
\bibitem [{\citenamefont {Schindler}\ \emph {et~al.}(2013)\citenamefont
  {Schindler} \emph {et~al.}}]{Schindler}%
  \BibitemOpen
  \bibfield  {author} {\bibinfo {author} {\bibfnamefont {P.}~\bibnamefont
  {Schindler}} \emph {et~al.},\ }\href@noop {} {\bibfield  {journal} {\bibinfo
  {journal} {New J. Phys.}\ }\textbf {\bibinfo {volume} {15}},\ \bibinfo
  {pages} {123012} (\bibinfo {year} {2013})}\BibitemShut {NoStop}%
\bibitem [{\citenamefont {Bruzewicz}\ \emph {et~al.}(2019)\citenamefont
  {Bruzewicz}, \citenamefont {Chiaverini}, \citenamefont {McConnell},\ and\
  \citenamefont {Sage}}]{Bruzewicz}%
  \BibitemOpen
  \bibfield  {author} {\bibinfo {author} {\bibfnamefont {C.~D.}\ \bibnamefont
  {Bruzewicz}}, \bibinfo {author} {\bibfnamefont {J.}~\bibnamefont
  {Chiaverini}}, \bibinfo {author} {\bibfnamefont {R.}~\bibnamefont
  {McConnell}}, \ and\ \bibinfo {author} {\bibfnamefont {J.~M.}\ \bibnamefont
  {Sage}},\ }\href@noop {} {\bibfield  {journal} {\bibinfo  {journal} {Applied
  Phys. Rev.}\ }\textbf {\bibinfo {volume} {6}},\ \bibinfo {pages} {021314}
  (\bibinfo {year} {2019})}\BibitemShut {NoStop}%
\bibitem [{\citenamefont {Blais}\ \emph {et~al.}(2007)\citenamefont {Blais},
  \citenamefont {Gambetta}, \citenamefont {Wallraff}, \citenamefont {Schuster},
  \citenamefont {Girvin}, \citenamefont {Devoret},\ and\ \citenamefont
  {Schoelkopf}}]{blais2007}%
  \BibitemOpen
  \bibfield  {author} {\bibinfo {author} {\bibfnamefont {A.}~\bibnamefont
  {Blais}}, \bibinfo {author} {\bibfnamefont {J.}~\bibnamefont {Gambetta}},
  \bibinfo {author} {\bibfnamefont {A.}~\bibnamefont {Wallraff}}, \bibinfo
  {author} {\bibfnamefont {D.~I.}\ \bibnamefont {Schuster}}, \bibinfo {author}
  {\bibfnamefont {S.~M.}\ \bibnamefont {Girvin}}, \bibinfo {author}
  {\bibfnamefont {M.~H.}\ \bibnamefont {Devoret}}, \ and\ \bibinfo {author}
  {\bibfnamefont {R.~J.}\ \bibnamefont {Schoelkopf}},\ }\href@noop {}
  {\bibfield  {journal} {\bibinfo  {journal} {Phys. Rev. A}\ }\textbf {\bibinfo
  {volume} {75}},\ \bibinfo {pages} {032329} (\bibinfo {year}
  {2007})}\BibitemShut {NoStop}%
\bibitem [{\citenamefont {Blais}\ \emph {et~al.}(2020)\citenamefont {Blais},
  \citenamefont {Girvin},\ and\ \citenamefont {Oliver}}]{girvinNP}%
  \BibitemOpen
  \bibfield  {author} {\bibinfo {author} {\bibfnamefont {A.}~\bibnamefont
  {Blais}}, \bibinfo {author} {\bibfnamefont {S.~M.}\ \bibnamefont {Girvin}}, \
  and\ \bibinfo {author} {\bibfnamefont {W.~D.}\ \bibnamefont {Oliver}},\
  }\href@noop {} {\bibfield  {journal} {\bibinfo  {journal} {Nat. Phys.}\
  }\textbf {\bibinfo {volume} {16}},\ \bibinfo {pages} {247} (\bibinfo {year}
  {2020})}\BibitemShut {NoStop}%
\bibitem [{\citenamefont {Giovannetti}\ and\ \citenamefont
  {Fazio}(2005)}]{giovannettiPRA}%
  \BibitemOpen
  \bibfield  {author} {\bibinfo {author} {\bibfnamefont {V.}~\bibnamefont
  {Giovannetti}}\ and\ \bibinfo {author} {\bibfnamefont {R.}~\bibnamefont
  {Fazio}},\ }\href@noop {} {\bibfield  {journal} {\bibinfo  {journal} {Phys.
  Rev. A}\ }\textbf {\bibinfo {volume} {71}},\ \bibinfo {pages} {032314}
  (\bibinfo {year} {2005})}\BibitemShut {NoStop}%
\bibitem [{\citenamefont {Satoh}\ \emph {et~al.}(1999)\citenamefont {Satoh},
  \citenamefont {Susa},\ and\ \citenamefont {Matsuyama}}]{Satoh:99}%
  \BibitemOpen
  \bibfield  {author} {\bibinfo {author} {\bibfnamefont {S.}~\bibnamefont
  {Satoh}}, \bibinfo {author} {\bibfnamefont {K.}~\bibnamefont {Susa}}, \ and\
  \bibinfo {author} {\bibfnamefont {I.}~\bibnamefont {Matsuyama}},\ }\href@noop
  {} {\bibfield  {journal} {\bibinfo  {journal} {Appl. Opt.}\ }\textbf
  {\bibinfo {volume} {38}},\ \bibinfo {pages} {7080} (\bibinfo {year}
  {1999})}\BibitemShut {NoStop}%
\bibitem [{\citenamefont {Fisher}\ \emph {et~al.}(2012)\citenamefont {Fisher},
  \citenamefont {Prevedel}, \citenamefont {Kaltenbaek},\ and\ \citenamefont
  {Resch}}]{Fisher_2012}%
  \BibitemOpen
  \bibfield  {author} {\bibinfo {author} {\bibfnamefont {K.~A.~G.}\
  \bibnamefont {Fisher}}, \bibinfo {author} {\bibfnamefont {R.}~\bibnamefont
  {Prevedel}}, \bibinfo {author} {\bibfnamefont {R.}~\bibnamefont
  {Kaltenbaek}}, \ and\ \bibinfo {author} {\bibfnamefont {K.~J.}\ \bibnamefont
  {Resch}},\ }\href@noop {} {\bibfield  {journal} {\bibinfo  {journal} {New J.
  Phys.}\ }\textbf {\bibinfo {volume} {14}},\ \bibinfo {pages} {033016}
  (\bibinfo {year} {2012})}\BibitemShut {NoStop}%
\bibitem [{\citenamefont {{Campany}}\ and\ \citenamefont
  {{Fernández-Pousa}}(2012)}]{6358698}%
  \BibitemOpen
  \bibfield  {author} {\bibinfo {author} {\bibfnamefont {J.}~\bibnamefont
  {{Campany}}}\ and\ \bibinfo {author} {\bibfnamefont {C.~R.}\ \bibnamefont
  {{Fern\'{a}ndez-Pousa}}},\ }in\ \href@noop {} {\emph {\bibinfo {booktitle} {IEEE
  Photonics Conference 2012}}}\ (\bibinfo {year} {2012})\ pp.\ \bibinfo {pages}
  {473--474}\BibitemShut {NoStop}%
\end{thebibliography}

%

\end{document}